\documentclass{amsart}
\usepackage{amsthm,amsmath,amsxtra,amscd,amssymb,color,xspace,esint}
\def\l@subsection{\@tocline{3}{0.5pt}{0.5pc}{3pc}{}}
\usepackage[all]{xy}

\usepackage{latexsym,amsmath,amssymb,mathrsfs}
\usepackage{amsfonts,eucal,amsthm,graphicx,color}

\usepackage{cancel}
\usepackage{marginnote}
\usepackage{bm}

\numberwithin{equation}{section}
\newcommand\sdtimes{\mathbin{\hbox{\hskip2pt\vrule height 4.1pt depth -.3pt
width.25pt\hskip-2pt$\times$}}}
\newcommand{\ve}{{\varepsilon}}
\newcommand{\ep}{{\epsilon}}

\newcommand\beq{\begin{equation}}
\newcommand\eeq{\end{equation}}

\newcommand {\BC}   {\mathbb C}

\newcommand {\BR}   {\mathbb R}
\newcommand {\BP}   {\mathbb P}

\newcommand {\BT}   {\mathbb T}

\newcommand {\bR}   {\mathbf{R}}

\newcommand {\bx}{ \mathbf{x}}

\newcommand {\BZ}   {\mathbb Z}

\newcommand {\CalA} {\mathcal A}
\newcommand {\CalB} {\mathcal B}
\newcommand {\CalC} {\mathcal C}

\newcommand {\CalD} {\mathcal D}
\newcommand {\CalE} {\mathcal E}
\newcommand {\CalF} {\mathcal F}
\newcommand {\CalG} {\mathcal G}
\newcommand {\CalH} {\mathcal H}

\newcommand {\CalL} {\mathcal L}
\newcommand {\CalM} {\mathcal M}
\newcommand {\CalN} {\mathcal N}
\newcommand {\CalO} {\mathcal O}
\newcommand {\CalP} {\mathcal P}

\newcommand {\CalR} {\mathcal R}
\newcommand {\CalS} {\mathcal S}
\newcommand {\CalT} {\mathcal T}

\newcommand {\CalV} {\mathcal V}
\newcommand {\CalX} {\mathcal{X}}
\newcommand {\CalY} {\mathcal{Y}}

\newcommand {\fg} {\mathfrak{g}}

\newcommand{\fo}{\vert\kern -.03in\_}

\newcommand {\ii} {\mathrm{i}}

\newcommand{\pa}{{\partial}}

\begin{document}

\title{Fluid dynamics as an intersection problem}
\author{Nikita Nekrasov, Paul Wiegmann}
\address{$^{\mathsf{N}}$ Simons Center for Geometry and Physics, Yang Institute for Theoretical Physics, Stony Brook NY 11794-3636\\
nnekrasov@scgp.stonybrook.edu\footnote{corresponding author}\\
$^{\mathsf{W}}$ Leinweber Institute for Theoretical Physics, Kadanoff Center for Theoretical Physics and  Enrico Fermi Institute, University of Chicago, Chicago IL 60637}
\date{May 2026}

\setcounter{tocdepth}{2}

\begin{abstract}
The main result of the paper is a construction of an infinite-dimensional
symplectic manifold ${\mathscr P}_{M^4}$, a coisotropic submanifold
${\CalC}\subset{\mathscr P}_{M^4}$ defined as the zero locus of a moment map,
and a family of Lagrangian submanifolds ${\CalL}_{e,{\bf g}}$ determined by
the equation of state and the spacetime metric, such that
${\CalC}\cap{\CalL}_{e,{\bf g}}$ is precisely the covariant
Lichnerowicz--Carter system for a relativistic perfect fluid. On this basis, we clarify 
the notion of canonical velocity and
hydrodynamic invariants, such as the asymptotic Hopf invariant and the Ertel
invariant, and obtain a generalization of Kelvin's circulation theorem. We also
clarify the relation between the canonical velocity $\bf V$, the four-velocity $\bf u$, 
and other choices of frames. 
We point out a five-dimensional origin of the formalism underlying the covariant
formulation of hydrodynamics.

Our formalism extends naturally to fluids with additional degrees of freedom,
including multi-component fluids, charged fluids, and superfluids, and incorporates
the chiral anomaly and Onsager quantization. It also suggests a direction of
research connecting hydrodynamics to Poisson sigma models and topological field
theories.

Furthermore, we argue that a similar approach applies to the theory of self-dual
fields, such as chiral bosons in $1+1$ dimensions, tensor fields of $(2,0)$ theory
in $1+5$ dimensions, and the $4$-form field of type-IIB supergravity in $1+9$
dimensions.
\end{abstract}
\maketitle
\vskip 2cm
\tableofcontents
\vskip 2cm

\section{Introduction}
The motion of an ideal  fluid in a domain of Euclidean space is described by a system of five equations: Euler's equations for the velocity vector field  $\bf u$, the continuity equation for the mass density $\rho$, and  the adiabatic equation for the entropy density $s$
\beq
\begin{aligned}
& \rho \left( \partial_t+ {\bf u} \cdot\boldsymbol{\nabla} \right) {\bf u} = - \boldsymbol{\nabla} P \ ,  \\
& {\dot\rho} + { \boldsymbol\nabla} \left( {\rho} {\bf u} \right) = 0 \ ,\\
& {\dot s} +{ \boldsymbol\nabla}  \left(  s {\bf u} \right)  = 0 \ .\\
\end{aligned}
\label{eq:euler_eqs1}
\eeq
The right-hand side of \eqref{eq:euler_eqs1} is driven by the gradient of the pressure $P$, a function of $\rho$ and $s$ determined by the thermodynamics of the fluid at rest.

The equations \eqref{eq:euler_eqs1} are Galilean invariant. They are defined in a domain of ${\BR}^{3}$ with the flat metric. It is easy to extend them to a domain of a three-dimensional manifold $B^3$, endowed with a metric, see Section \ref{sec:epa}. It is also well known how to formulate relativistic fluid dynamics on a spacetime manifold $M^4$ with a metric ${\bf g} = g_{\mu\nu} dx^{\mu} dx^{\nu}$. 

We may view fluid mechanics as a field theory in which the metric is treated as a parameter (a background metric).
In a simple classical field theory, the equations of motion follow from a variational problem
\beq
\frac{{\delta} {\CalS}_{M,{\bf g}} [ {\Phi}]}{\delta \Phi} = 0
\label{eq:cfs}
\eeq
on the space ${\CalF}_{M}$ of fields $\Phi$ on the spacetime $M$ which transform as scalars, vectors, tensors, spinors, etc. 
Equivalently, solutions to \eqref{eq:cfs} can be viewed as the intersection of the graph of the derivative ${\delta}{\CalS}_{M,{\bf g}}$ of the action functional with the
zero section of the cotangent bundle ${\CalP}_{M} = T^{*}{\CalF}_{M}$ of the space of fields. Both the graph ${\sf{P}} = \frac{{\delta} {\CalS}_{M,{\bf g}} [ {\Phi}]}{\delta \Phi}$ of ${\delta}{\CalS}_{M,{\bf g}}$ and the zero section $\sf{P}  = 0$ are {\it Lagrangian} submanifolds of the symplectic manifold ${\CalP}_{M}$. 

In more complicated cases, the dynamics are described by Hamilton equations formulated using Poisson brackets, i.e., a vector field $V \in {\sf Vect}({\CalX})$ on a Poisson manifold $({\CalX}, {\pi})$ with a Poisson structure $\pi$ that is not necessarily invertible. The trajectories solving these equations are described, as reviewed in Section \ref{sec:cm}, as the intersection ${\CalC} \cap {\CalL}$ of two subvarieties of an infinite-dimensional symplectic manifold
${\CalM}_{\CalX}$. The manifold ${\CalM}_{\CalX}$ depends only on the smooth manifold $\CalX$, the subvariety ${\CalC}$ is canonically associated with the choice of $\pi$, the subvariety
${\CalL}$ depends on a different set of structures, including a choice of a function $H \in {\sf C}^{\infty}({\CalX})$. With respect to the symplectic structure on ${\CalM}_{\CalX}$ both subvarieties ${\CalC}$ and ${\CalL}$ are {\it coisotropic}, with ${\CalL}$ being {\it Lagrangian}. We recall the relevant notions from classical mechanics (symplectic geometry) in Section
\ref{sec:cm} below. 

{}This general setting is not covariant: the space, hidden in the structure of the phase space $\CalX$, and time $t$ are treated differently. 

An important special example of this construction in mechanics is provided by spinning tops\footnote{The aspects of the dynamics that spinning tops share with hydrodynamics were studied by L.~Euler, H.~Poincar{\'e}, and V.~Arnold. Following tradition, we refer to equations of this kind as Euler–Arnold spinning-top equations.},
reviewed in Section \ref{sec:epa}. In this case the Poisson manifold ${\CalX} = \mathfrak{g}^{*}$
is the dual space to the (possibly infinite-dimensional) Lie algebra $\mathfrak{g} = {\sf Lie}(G)$ of the Lie group $G$.

{}The hydrodynamics of a perfect fluid, whether Galilean-invariant or relativistic, falls into the same class of problems. We show that the Euler equations in the Lichnerowicz–Carter form \cite{Carter,L} can likewise be formulated in terms of intersection theory. More precisely, a fluid flow is represented by an inhomogeneous differential
form ${\boldsymbol{\Phi}} \in {\Omega}^{\bullet}(M^4)$, described in detail below:
\beq
{\boldsymbol{\Phi}} \in {\CalC} \cap {\CalL}   \subset {\mathscr{P}}_{M^4}\subset {\Omega}^{\bullet}(M^4)\,.
\label{eq:intersection}
\eeq
Thus, the flow lies in the intersection of two nearly middle-dimensional subvarieties
of the symplectic manifold ${\mathscr{P}}_{M^4}$ associated with spacetime.

The {\it coisotropic} subvariety ${\CalC}$ is defined in purely differential--topological terms; it is the zero locus 
\beq
{\CalC}    = {\boldsymbol{\mu}}^{-1}(0) \subset {\mathscr{P}}_{M^4}\, 
\label{eq:momcm}
\eeq
of the moment map
for the group
\beq
{\mathsf{Diff}}(M^4) \sdtimes {\mathsf{C}}^{\infty}(M^4)
\eeq
acting on ${\mathscr{P}}_{M^4}$ by symplectomorphisms. 

The \emph{Lagrangian} submanifold ${\CalL} \subset {\mathscr{P}}_{M^4}$ encodes the {\it equation of state}, characterized by a generating functional and the choice of natural thermodynamic and kinetic variables, such as momentum, mass, and entropy densities, as discussed in the next section. Apart from the spacetime metric, the submanifold ${\CalL}$ may also depend on a variety of background fields, such as Abelian and non-Abelian gauge fields. 
\begin{figure}
    \centering
    \includegraphics[width=0.8\linewidth]{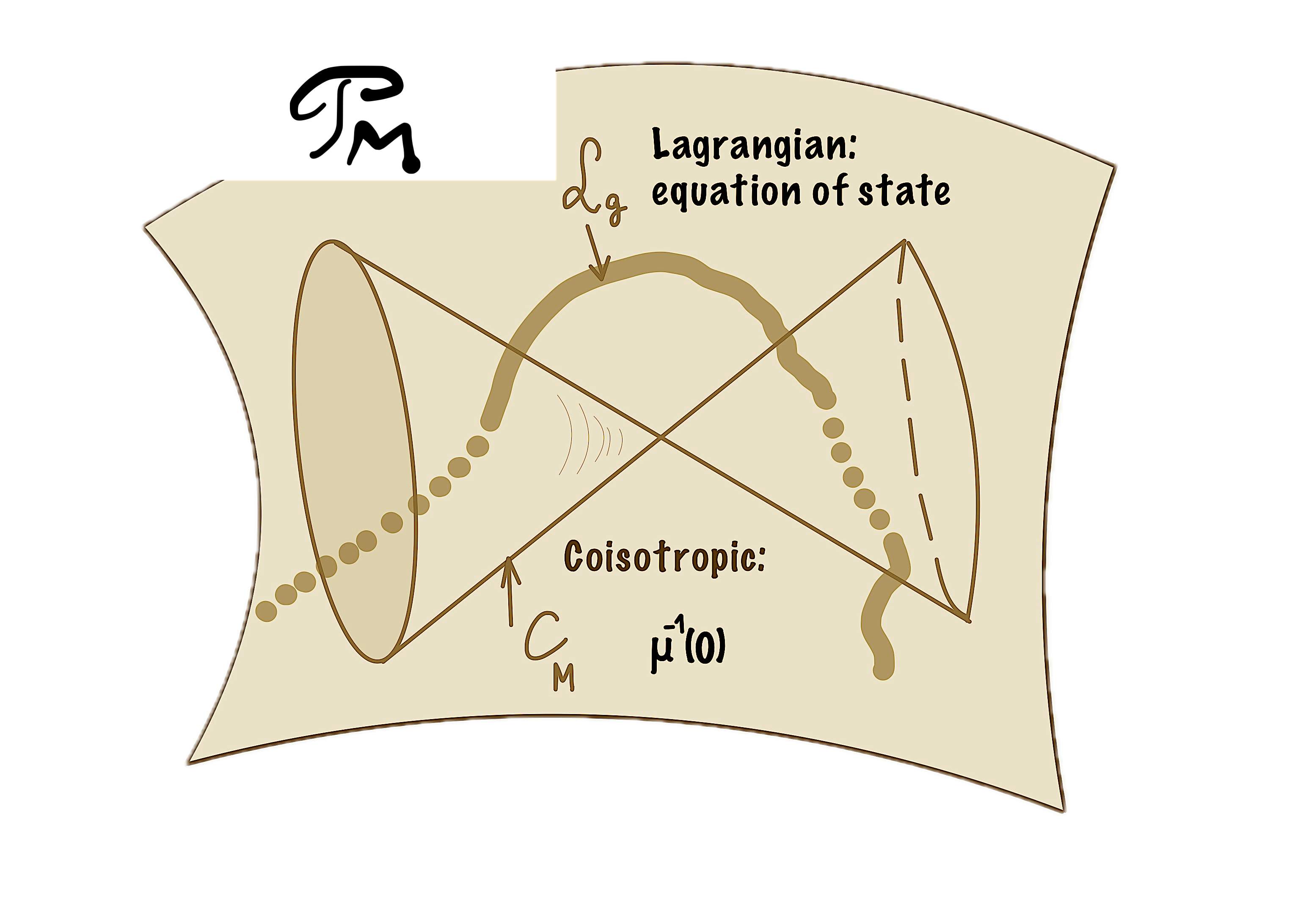}
    \caption{The coisotropic ${\CalC}$ and Lagrangian ${\CalL}$ inside ${\mathscr{P}}_{M^4}$}
    \label{fig:intersection}
\end{figure}

The simple geometric statement \eqref{eq:intersection}, illustrated in Fig.~\ref{fig:intersection}, summarizes the equations of Lichnerowicz and Carter \cite{Carter,L}.

Let  $B^3 \subset M^4$ be a smooth submanifold and choose a  vector field ${\ell} \in {\sf Vect}(M^4)$
defined in a tubular neighborhood  of $B^3$ and transverse to it. The flow of $\ell$ identifies this neighborhood locally with $B^3 \times I$, where $I$ is an open interval containing $0$. With this choice, the fields defining ${\mathscr{P}}_{M^4}$ can be described in three-dimensional terms. In this way, we recover a part of the geometric data defining the 
Euler-Arnold top on the dual of the Lie algebra of the group
\beq
{\mathsf{Diff}(B^3)} \sdtimes {\mathsf{C}}^{\infty}(B^3, {\BR}^{2})\,.
\eeq

{}
The conventional Euler equations correspond to the situation in which the spacetime metric ${\bf g}$ admits a Killing vector $K$; in this case one can describe
\eqref{eq:intersection} in the {\it evolution form}, locally representing
\beq
M^4 \approx B^3 \times {\BR}
\eeq
with $B^3$ being the Cauchy surface, transverse to the vector field ${\ell} = K$.

For a general ${\bf g}$ without isometries, the covariant formalism provides a non-stationary version of the Euler equations, incorporating relativistic effects such as gravitational drag.

Our formalism allows for a relatively simple incorporation of anomalous particle production arising from self-intersections of vortex spacetime surfaces (spacetime 2-surfaces swept out by evolving vortex lines), as well as other generalizations, including coupling to background gauge fields. In this way, we clarify and simplify some of the constructions of \cite{Abanov:2022zwm,Wiegmann:2024sqh}.

The paper is organized as follows. Section \ref{sec:epa} reviews the formulation of compressible flows with spatially varying entropy as Euler--Arnold equations (a kind of spinning-top equations), associated with a suitable Lie algebra. Section \ref{sec:cm} recalls the relevant notions from symplectic and Poisson geometry. In particular, we explain how Hamiltonian dynamics on a Poisson manifold ${\CalX}$ can be represented as an intersection problem in the symplectic manifold associated with the space of paths ${\CalP}_{\CalX} = {\rm Maps}(I,{\CalX})$; after this paper was ready for publication, we learned that this example was also considered in \cite{Cattaneo:2003dp}.

Section \ref{sec:covh} begins the main construction: the covariant formulation of relativistic hydrodynamics. We introduce the symplectic space ${\mathscr{P}}_{M^4}$ of fields
${\boldsymbol{\Phi}} = \boldsymbol{(S,p,n,\nu)} \in {\Omega}^{0\oplus 1 \oplus 3 \oplus 4}(M^4)$, representing the entropy per particle, fluid momentum per particle, particle flux, and thermal energy, respectively. We then define the coisotropic submanifold ${\CalC}$ as the zero locus of the moment map for the action of ${\sf Diff}(M^4)\sdtimes {\sf C}^{\infty}(M^4)$, and obtain the corresponding Lichnerowicz--Carter equations, together with Kelvin's and Ertel's theorems. Section \ref{sec:chemistry} introduces the second ingredient of the intersection problem, the Lagrangian submanifold ${\CalL}$ determined by the equation of state and the spacetime metric. 

Section \ref{sec:3d} explains how the four-dimensional covariant formalism reduces, after choosing a hypersurface and a transverse vector field, to the three-dimensional Euler--Arnold description reviewed earlier. Section \ref{sec:anomf} studies magnetic deformations of the symplectic form and interprets them as anomalous terms in hydrodynamics, thereby allowing particle production. Section \ref{sec:gut} discusses several extensions and related constructions: Euler--Maxwell systems, superfluids, particle creation through vorticity instantons, vortex spacetime surfaces, connections on gerbes and Onsager quantization, a five-dimensional parent formalism, non-abelian spectators, and barotropic models. In particular, the constructions suggest the pattern of symmetry reductions
\beq
{\sf Diff}(N^5) \longrightarrow {\sf Diff}(M^4) \sdtimes {\sf C}^{\infty}(M^4,{\BR}) \longrightarrow
{\sf Diff}(B^3) \sdtimes {\sf C}^{\infty}(B^3,{\BR}^{2})\,.
\eeq
Finally, the appendices collect more speculative connections to topological strings, Poisson sigma models, higher-dimensional topological field theories, viscosity, wavefronts, and Liouville theory viewed from the same intersection-theoretic perspective.

\subsubsection{Acknowledgments} We thank A.~Abanov, A. Cappelli, Ya.~Eliashberg, D.~Freed, M.~Kontsevich, R.~Suszek and D.~Sullivan for interesting discussions.

NN thanks Institut Mittag-Leffler (Stockholm), Uppsala University, IHES, University of Warsaw, and especially ICTS-TIFR (Bengaluru) for their hospitality during the preparation of the manuscript.  The results of this work were first presented at the IML program ``Cohomological Methods in QFT'' (January 2025, Stockholm), MaximFest (January 2025, University of Miami), CRC1624 opening conference  (April 2025, University of Hamburg), and Nag Memorial Lectures at IMSc and CMI (November 2025, Chennai).  P. W. gratefully acknowledges the hospitality of the Institute for Advanced Studies at  Tel-Aviv University where part of this work was reported.

\section{Euler-Arnold equations}\label{sec:epa}

Recall \cite{Arnold} that the equations of an incompressible fluid can be viewed as the Euler-Poincar\'e equations of a rigid body extended to the Lie algebra of divergence-free vector fields. This approach is not restricted to the incompressible case, as we recall here.

Associated with any Lie algebra $\mathfrak{g}$, the Euler-Poincar\'e equation describes a dynamical system on its dual space $\mathfrak{g}^{*}$. We stress that this is not a symplectic manifold: the phase space is merely Poisson. There are, of course, several ways of relating this dynamics to one on a symplectic manifold, which we shall review in the next section, but for now we remain purely at the level of $\mathfrak{g}^*$ dynamics.

Let ${\bf P} \in {\mathfrak{g}}^{*}$ and $\boldsymbol{\Omega} \in {\mathfrak{g}}$, and fix a map $ {\mathfrak{g}}^{*} \mapsto {\mathfrak{g}}$. We shall assume this map to be Lagrangian, in the sense that its graph is a Lagrangian submanifold of ${\mathfrak{g}} \times {\mathfrak{g}}^{*}$ endowed with the canonical symplectic structure. 

The equations of motion can be derived in two stages. The first stage establishes the relationship between the momentum and the angular velocity, given by
\beq
{\dot {\bf P}} = - {\rm ad}_{{\boldsymbol{\Omega}}}^{*} {\bf P}\ . 
\label{eq:top}
\eeq
This relation captures the purely kinematic aspects of the system, which arise exclusively from its Lie-Poisson structure.

The second step is to constrain $\left({\bf P}(t), {\boldsymbol{\Omega}}(t)\right)$ by a Lagrangian submanifold in ${\mathfrak{g}}^{*} \times {\mathfrak{g}}$. A simple way to do this is to choose a map ${\bf P} \mapsto {\boldsymbol{\Omega}(\bf P)}$ expressing the generalized \emph{angular velocity} $\boldsymbol{\Omega}$. We do so by choosing a function(al) ${\CalH}({\bf P})$ on $\mathfrak{g}^{*}$ and setting 
\beq
{\boldsymbol{\Omega}} ({\bf P}) = \frac{{\delta}{\CalH}}{\delta {\bf P}}\,.
\label{eq:1pol}
\eeq
{}
This step completes the equations of motion.

Alternatively, one picks a Legendre transform of ${\CalH}({\bf P})$,  the  functional ${\CalA}({\boldsymbol \Omega})$ on $\mathfrak{g}$ and defines the generalized momentum $ {\bf P}({\boldsymbol{\Omega}}) = \frac{{\delta}{\CalA}}{\delta {{\boldsymbol{\Omega}}}}$
 as the  inverse map (if it exists)  ${\mathfrak{g}} \mapsto {\mathfrak{g}}^*$, called the inertia map \cite{Arnold}. 
Other choices are also possible, with the generating 
function  depending on a part of ${\bf P}$ and a part of $\boldsymbol{\Omega}$. For example, the classical Euler top is associated with $\mathfrak{g} = {\mathfrak{so}}(3)$, the inertia map ${\mathfrak{g}} \mapsto {\mathfrak{g}}^*$ is associated with nondegenerate quadratic $\CalH$ or $\CalA$. 

We emphasize that Eq. \eqref{eq:top} alone is sufficient to derive the momentum conservation law.
Throughout, we will exploit the distinction between “kinematics” and “dynamics,” as well as the freedom in choosing the generating functional, in our description of hydrodynamics.

\subsection{Novikov groups}

\label{sec:phmng}

For a manifold $X$ and a Lie group $G$ define the group 
\beq
{\CalG}_{X}^{G} = {\sf Diff}(X) \sdtimes {\sf C}^{\infty}(X, G)
\eeq
which consists of pairs $(F,{\varphi})$, where $F: X \to X$ is a diffeomorphism, and ${\varphi}: X \to G$ is a smooth map. The product of pairs is defined via composition:
\beq
(F_{1}, {\varphi}_{1}) \circ (F_{2}, {\varphi}_{2}) = \left( F_{1} \circ F_{2}, ({\varphi}_{1} \circ F_{2} ) \times {\varphi}_{2} \right)
\eeq
where $\times$ stands for the group law in $G$. In the main part of this paper we shall use the groups
\beq
{\CalG}_{B^{3}}^{(2)} := {\CalG}_{B^3}^{{\BR}^{2}}\, , \ {\CalG}_{M^4}^{(1)} : = {\CalG}_{M^4}^{{\BR}}
\eeq
where ${\BR}^{2}, {\BR}$ are viewed as abelian groups with addition as the group law. The importance of the role of ${\CalG}_{B^3}^{(2)}$ in hydrodynamics of ideal 
compressible fluid was understood in \cite{Novikov}. The semi-direct product groups are extensively discussed in \cite{Holm}.

\subsection[Geometry vs equation of state]{Separating geometry from the equation of state}
\label{sec:uxit}

There are several ways to pursue an analogy between the equations \eqref{eq:euler_eqs1} describing fluids and equations describing spinning tops, e.g., \cite{Holm}. The formulation depends on the choice of independent variables — the natural variables. For illustration, we select the triple consisting of the densities of momentum, mass, and entropy (see \cite{Volovik}).

First, we treat 
the mass density and the entropy density as $3$-forms $\boldsymbol{\rho}= \frac{\rho}{3!}\, {\epsilon}_{ijk} dx^i \wedge dx^j \wedge dx^k$ and   $\boldsymbol{ s} = \frac{s}{3!} \,{\epsilon}_{ijk} dx^i \wedge dx^j \wedge dx^k$  in ${\Omega}^{3}(B^3)$, where ${\epsilon}_{ijk} $ is the Levi-Civita symbol on $B^3$.  Then we introduce the {\it momentum density} form
\beq
\boldsymbol{\Pi} = {\Pi}_{i} \, dx^i \otimes dx^1 \wedge dx^2 \wedge dx^3 \, 
 \in {\Omega}^{1}(B^3) \otimes_{\mathsf{C}^{\infty}(B^3)} {\Omega}^{3}(B^3) \, , 
\label{eq:mompervol}
\eeq
and write the fluid equations as the time evolution of the densities introduced above
\begin{align}
&\left(\partial_t + L_{u} \right)\boldsymbol {\rho} = 0 ,\quad\left(\partial_t + L_{u} \right)\boldsymbol{s} = 0\label{2.40} \,,\\
&\left(\partial_t + L_{u} \right)\boldsymbol{\Pi}+ d\xi \otimes \boldsymbol{\rho}+dT \otimes\boldsymbol{s}=0\, . 
  \label{eq:euler_eqs2}
\end{align}
These equations are driven by the velocity ${\bf u}$, the temperature $T$, and the potential $\xi$.

 These relations depend neither on the metric nor on the thermodynamics of the fluid. Together, Eqs.~\eqref{2.40}, \eqref{eq:euler_eqs2} cut out a coisotropic submanifold in the phase space ${\CalP}$ parametrizing all collections of fields  ${\Phi} = (u, \xi, T; \boldsymbol{\Pi}, \boldsymbol{\rho}, \boldsymbol{s})$; we review this notion in Sect.~\ref{sec:cm}. The remaining equations, expressing $(u, \xi, T)$ in terms of $(\boldsymbol{\Pi}, \boldsymbol{\rho}, \boldsymbol{s})$, are provided by the equation of state. This equation not only implies the existence of the pressure function, but also defines a second submanifold of phase space, which is Lagrangian. Under generic conditions, the coisotropic and Lagrangian submanifolds intersect transversely. In practical terms, this means that Eqs.~\eqref{2.40}, \eqref{eq:euler_eqs2}, together with the equation of state reviewed in the next subsection, form a closed system.

\subsection{Thermodynamic relations and the momentum-flux tensor}
\label{sec:thermodynamics}

From the definition of ${\boldsymbol{\Pi}}$, its transformation under infinitesimal diffeomorphisms of $B^3$ is
\beq
(L_{u}\boldsymbol{\Pi})_k=\partial_i(u^i\Pi_k)+\Pi_i\partial_k u^i\,.
\eeq
We may therefore rewrite Eq.~\eqref{eq:euler_eqs2} in the Newtonian form
\beq
{\partial}_t {\Pi}_{k} + {\partial}_{i} ( u^{i} {\Pi}_{k} ) =f_k\,,
\label{2.71}
\eeq
where the force connection components are given by
\beq
f_{k} \, = \,  -  
\left( \Pi_i\,{\partial}_{k} u^i+\rho\, {\partial}_{k} \xi+s\, {\partial}_{k} T\, \right) 
\label{F}
\eeq
The Lagrangian submanifold of $\CalP$ depends on the choice of a Riemannian metric $h$ on $B^3$. It can be described by a generating functional ${\CalH}(\boldsymbol{\Pi},\boldsymbol{\rho},\boldsymbol{s})$, which expresses $(u,\xi,T)$ in terms of the conjugate variables $(\boldsymbol{\Pi},\boldsymbol{\rho},\boldsymbol{s})$:
\beq
u^i=\frac{\delta\CalH}{\delta \Pi_i}, \qquad
\xi=\frac{\delta\CalH}{\delta {\boldsymbol{\rho}}}, \qquad
T=\frac{\delta\CalH}{\delta {\boldsymbol{s}}}\,.
\label{2.8}
\eeq
Once the generating functional is chosen, we refer to Eqs.~\eqref{2.8} as the equation of state, or more precisely, as the equation of state in the $(\boldsymbol{\Pi},\boldsymbol{\rho},\boldsymbol{s}\vert {u},\xi,T)$ polarization.

We only consider Lagrangians defined by the local generating functionals
\beq
{\CalH}\left(\boldsymbol{\Pi},\boldsymbol{\rho},\boldsymbol{s}\right)
=
\int_{B^3} e\left(\boldsymbol{\Pi},\boldsymbol{\rho},\boldsymbol{s}\right)\, {\rm vol}_{h} \, ,
\label{eq:locgf}\eeq
for some scalar function $e$, where
\beq
{\rm vol}_{h} = \sqrt{{\rm det}h}\, d^{3}x\ . 
\eeq
If the field configuration ${\Phi}\in{\CalP}$ belongs to a Lagrangian submanifold of ${\CalP}$, that is, if the relation between $(\boldsymbol{\Pi},\boldsymbol{\rho},\boldsymbol{s})$ and $({u},\xi,T)$ is encoded by a generating functional of the form \eqref{eq:locgf}, then the work form \eqref{F} is exact: $f=-d\left( P \sqrt{{\rm det}h} \right)$, where
\beq
P:=
\frac{\Pi_i\, u^i
+\rho\, {\xi}
+s\, T}{\sqrt{{\rm det}{h}}}
- e\,.
\label{2.6}
\eeq
The function $P$ is identified with the fluid pressure, and Eq.~\eqref{2.6} may be viewed as a hydrodynamic form of the Gibbs-Duhem relation.

Equation \eqref{2.71} may therefore be rewritten as
 the momentum conservation law
\beq
\partial_t\Pi_k+\partial_i\Pi^i_k=0\,,
\label{Pi}
\eeq
where
\beq
\Pi^i_k=u^i\Pi_k+\delta^i_k P
\label{2.91}
\eeq
is the momentum-flux tensor density.

Together with the continuity equations \eqref{2.40}, this implies the Euler equations in their original form, Eq.~\eqref{eq:euler_eqs1}.

The generating functional is naturally identified with the energy of the fluid. Consider a specific Galilean-invariant choice of $e({\bf\Pi}, {\boldsymbol{\rho}}, {\boldsymbol{s}})$: 
\beq
{\CalH}\left({\boldsymbol{\Pi}},{\boldsymbol{\rho}},{\boldsymbol{s}}\right)
=
\int_{B^3}\left(
\frac{h^{ij}\Pi_i\Pi_j}{2\rho}
+\ve(\rho,s) \sqrt{{\rm det}(h)}
\right)\,d^3 x\,.
\label{2.7}
\eeq
Here $\varepsilon(\rho,s)$ is the fluid energy density. 
Evaluating \eqref{2.6} for \eqref{2.7} yields the Gibbs-Duhem relation in its thermodynamic form:
\beq
dP=\rho\,d\mu+s\,dT\,.
\eeq

In this way, we recover the Galilean relation between the velocity field ${\bf u}\in{\rm Vect}(B^3)$, which enters \eqref{eq:euler_eqs2} through the Lie derivative $L_{\bf u}$, and the momentum:
\beq
\Pi_i=h_{ij}\,\rho\,u^j\,,
\label{eq:momentum_per_volume}
\eeq
as well as the scalar function
\beq
\xi=-\frac{h^{ij}}{2\rho^2}\Pi_i\Pi_j+\frac{\mu}{m}\,.
\label{eq:chempot}
\eeq
Here $\mu/m=\partial {\ve}/\partial\rho$ is the chemical potential per atomic mass $m$, while ${\ve}(\rho,s)$ denotes the fluid energy density at rest, regarded as a function of the mass density and entropy density.

Similarly, the temperature is given by
$$
T=\frac{\partial {\ve}}{\partial s}\,.
$$

The relations \eqref{2.8} are the analogue of the inertia map \eqref{eq:1pol}. In the hydrodynamic setting, we refer to them as the equation of state. As Eq.~\eqref{2.7} shows, the equation of state depends on the metric $h$, whereas the relations \eqref{eq:euler_eqs2} and \eqref{Pi} do not.

As in the case of the spinning top, Eqs.~\eqref{eq:euler_eqs2} constitute five relations on ten functional variables; that is, they describe the system off shell. They define a metric-independent coisotropic submanifold of the symplectic manifold to be discussed below. On shell, once the equation of state is specified, \eqref{eq:euler_eqs2} becomes a complete system of five functional equations for five functional variables. This is the intersection-theoretic problem referred to in the title of the paper.

Of course, the triples $(\boldsymbol{\Pi},\boldsymbol{\rho},\boldsymbol{s})$ and $({u},\xi,T)$ may be interchanged. In that case, the generating functional is the negative Legendre transform of \eqref{2.7}:
\beq
{\CalA}({u},\xi,T)
=
{\rm Crit}_{{\boldsymbol{\Pi}},{\boldsymbol{\rho}},{\boldsymbol{s}}}
\left\{
-\int_{B^3}\left(
\iota_{u}\boldsymbol{\Pi}
+\xi\,{\boldsymbol{\rho}}
+T\,{\boldsymbol{s}}
\right)
+{\CalH}\left(
{\boldsymbol{\Pi}},{\boldsymbol{\rho}},{\boldsymbol{s}}
\right)
\right\}.
\label{2.9}
\eeq
The on-shell value of this functional is the volume integral of the pressure, cf.~\cite{Whitham}:
\beq
{\CalA}=-\int_{B^3} P\,{\rm vol}_h\,.
\eeq

Finally, one may work in the polarization in which the momentum and velocity are exchanged, organizing the variables into the triples $({u},\boldsymbol{\rho},\boldsymbol{s})$ and $(\boldsymbol{\Pi},\xi,T)$. In this case one obtains the generating functional\footnote{The functional below is often called the fluid Lagrangian. One should keep in mind, however, that the fluid motion does not follow from the Euler-Lagrange equations derived from it.}
\beq
{\CalS}({ u},\boldsymbol{\rho},\boldsymbol{s})
=
\left( \int_{B^3}\iota_{u}\boldsymbol{\Pi}\right) \,-\CalH
=
\int_{B^3}
\left(
\frac{1}{2}\rho \|{\bf u}\|^2-\ve(\rho,s)
\right)\,{\rm vol}_h\,.
\label{eq:fluid_lagrangian}
\eeq
Each of these formulations is useful in different applications. Among them, the choice
$$
(\boldsymbol{\Pi},\boldsymbol{\rho},\boldsymbol{s}),\qquad
({u},\xi,T)
$$
provides the most direct connection to the spinning top.

\subsection{Poisson and symplectic manifolds}
\label{sec:cm}

Let $\CalX$ be a Poisson manifold, i.e. a smooth manifold endowed with the bivector
\beq
{\pi} = \frac 12 {\pi}^{ij} (x) \frac{\partial}{\partial x^i} \wedge \frac{\partial}{\partial x^j}\, , \ 
\eeq
obeying the Jacobi identity:
\beq
[ {\pi}, {\pi} ] := \frac 16\, 
\left( {\pi}^{lk} {\partial}_{l} {\pi}^{ij} + {\pi}^{li} {\partial}_{l} {\pi}^{jk} + {\pi}^{lj} {\partial}_{l} {\pi}^{ki} \right) \frac{\partial}{\partial x^{i}} \wedge  \frac{\partial}{\partial x^{j}} \wedge \frac{\partial}{\partial x^{k}} = 0 .  
\label{eq:jacobi}
\eeq
It defines the Poisson bracket
\beq
\{ f, g \}_{\pi} = {\pi}^{ij} \frac{\partial f}{\partial x^i} \frac{\partial g}{\partial x^j} = 
\iota_{\pi} df \wedge dg
\eeq
and Hamiltonian vector fields: $H \in C^{\infty}({\CalX}) \mapsto V_{H} \in {\sf Vect}({\CalX})$
\beq
V_{H} = \iota_{\pi} dH = {\pi}^{ij} \frac{\partial H}{\partial x^i} \frac{\partial}{\partial x^j}\,.
\label{eq:hamvect}
\eeq
Recall that a Poisson manifold is not, in general, symplectic. If $\pi$ is invertible, then 
\beq
{\omega} = {\pi}^{-1}
\label{eq:sf}
\eeq
is a closed nondegenerate, i.e. symplectic, $2$-form. 

An important notion in Poisson geometry is that of a {\it coisotropic submanifold}. 
${\CalC} \subset {\CalX}$ is coisotropic, if the ideal $I_{\CalC} \subset C^{\infty}({\CalX})$
of functions, vanishing on $\CalC$ is closed under the Poisson bracket, in other words:
\beq
f, g \Bigr\vert_{\CalC} = 0 \Longrightarrow \{ f, g \} \Bigr\vert_{\CalC} = 0\,.
\label{eq:coiso}
\eeq
In the physics literature, such submanifolds are usually described by {\it first class constraints}. 
If $\CalX$ is symplectic, then the minimal dimension a coisotropic submanifold may have is half of ${\rm dim}\,{\CalX}$. In this case, such a submanifold is called {\it Lagrangian}. It is characterized as a maximal-dimensional submanifold on which the symplectic form vanishes, in other words, it is {\it isotropic} for $\omega$. 

A Lagrangian submanifold ${\CalL} \subset{\CalX}$ can be locally described as the graph of the differential of a function, called the \emph{generating function}. Indeed, let $(p_i, x^i)$
be the local Darboux coordinates on $\CalX$, so that
\beq
{\omega} = \sum_{i=1}^{k} dp_i \wedge dx^i \, , \ {\rm dim}{\CalX} = 2k\, , 
\eeq
then, for any $I \subset \{ 1, \ldots, k \}$ , the $1$-form 
\beq
\theta_{I} = \sum_{i \in I}  p_i dx^i  - \sum_{j \notin I} x^j dp_j
\eeq
is closed, when restricted to $\CalL$, 
\beq
d{\theta}_{I} \vert_{\CalL} = 0\, , 
\eeq
and therefore locally is exact
\beq
{\theta}_{I} = dS_{I}\, , 
\eeq
with the function $S_I$ on $\CalL$. If the subset $\{ x^i \, | \, i \in I\} \cup \{ p_j \, | \, j \notin I \}$ of $k$ local coordinates on $\CalX$ can be used as a local coordinate system on $\CalL$ (if $\CalL$ is smooth, then there is at least one choice of $I$ for which it is true), then $S_I$ is called the \emph{generating function} of $\CalL$ in the \emph{polarization}
\beq
\left( x^i\, , \ p_j \, | \ p_i \, , \ x^j \right)_{i \in I\, , \, j \notin I}
\eeq

If $\CalC \subset \CalX$ is a coisotropic submanifold of a symplectic manifold, then ${\omega} \vert_{\CalC}$ is a degenerate closed two-form. Its kernel forms an integrable foliation $\CalF$. The quotient ${\CalC}/{\CalF}$ is not, in general, a well-behaved topological space, but locally it is a symplectic manifold. 

A good example of such a quotient is provided by a {\it symplectic quotient}. Assume that symplectic $\CalX$ is endowed with a Hamiltonian action of a Lie group $G$, so that
the infinitesimal $G$-action is described by a homomorphism:
\beq
{\CalV}: {\mathfrak{g}} \to {\sf Vect}({\CalX})\,  , \ {\xi} \mapsto {\CalV}_{\xi} \ .
\eeq
It comes with the moment map ${\mu}: {\CalX} \to {\fg}^{*}$, 
so that 
\beq
{\CalV}_{\xi} = \iota_{\pi} d{\mu}({\xi})  = V_{{\mu}({\xi})}
\label{eq:mom1}\eeq
for any $\xi \in \mathfrak{g}$, cf. \eqref{eq:hamvect}. The equivariant moment map obeys, in addition to \eqref{eq:mom1}, 
\beq
\left\{ {\mu}({\xi}_{1}), {\mu}({\xi}_{2}) \right\} = {\mu}( [{\xi}_{1}, {\xi}_{2}] )
\eeq
for any ${\xi}_{1}, {\xi}_{2} \in \mathfrak{g}$. In this case 
\beq
{\CalC} = {\mu}^{-1}(0)
\label{eq:coisfirst}
\eeq
is coisotropic, while 
\beq
{\CalX}//G = {\mu}^{-1}(0)/G
\eeq
is symplectic, as the foliation of kernels of 
$\omega \vert_{\CalC}$ is formed by the orbits of $G$. 
The geometric fact \eqref{eq:coisfirst} is prominently featured in our main construction. 

{}
For ${\CalX}$ a coadjoint orbit, the moment map is simply the embedding ${\mu}: {\CalX} \hookrightarrow {\mathfrak{g}}^{*}$. Another useful example is that of ${\CalX} = T^{*}{\mathfrak{g}} = \{ ( {\bf \Omega}, {\bf P}) \}$, with $G$ acting on $\mathfrak{g} \ni {\bf\Omega}$ in the adjoint and on ${\bf P}\in {\mathfrak{g}}^{*}$ in the coadjoint representation. The moment map is ${\rm ad}_{\bf\Omega}^{*}({\bf P})$. 

An important example of a Poisson manifold, which is not, in general, symplectic, is the dual space $\mathfrak{g}^{*}$ to a Lie algebra.  Given two functions $f_1, f_2 \in C^{\infty}({\fg}^{*})$,   the value
of the Poisson bracket $\{ f_{1}, f_{2} \}$ at a point ${\xi} \in {\fg}^{*}$ is given by:
\beq
\{ f_{1} , f_{2} \} ({\xi}) = {\xi} \left( \left[ \frac{\partial f_1}{\partial \xi} , \frac{\partial f_2}{\partial \xi} \right] \right)\,,
\label{eq:kkf}
\eeq
where $df \vert_{\xi} \in T^{*}{\fg}^{*} \approx {\fg}$ (the isomorphism requires care in an infinite-dimensional case). In linear coordinates $l_{a}$, $a = 1, \ldots , {\rm dim}{\fg}^{*}$, 
\beq
\{ l_{a} , l_{b} \} \ = \ f_{ab}^{c} \, l_{c}\,,
\label{eq:kirillov-kostant}
\eeq
where $f_{ab}^{c}$ are the structure constants of $\fg$ in the associated basis $T_{a} \in {\fg}$. This structure is featured, e.g., in the Euler top.

\subsubsection{Variational principle}

On a symplectic manifold $({\CalX}, {\omega})$, the parametrized trajectories 
\beq
{\gamma}: I \to {\CalX}
\eeq
of a Hamilton vector field, ${\pi} = {\omega}^{-1}$, 
\beq
{\dot {\bf x}} = V_{H} = {\omega}^{-1} \frac{\partial H}{\partial \bf x}
\eeq
are  extremizers ${\delta}{\CalS} = 0$ of a (multi-valued) action functional
\beq
{\CalS} [ {\gamma} ] = \int_{\gamma} d^{-1} {\omega} - \int_{I} {\gamma}^{*} H \, d{\tau}\,,
\label{eq:actionsympl}
\eeq
where $\tau$ is the coordinate on $I$. 

\subsection{Symplectic manifold out of Poisson pointwise}

\subsubsection{Symplectic out of Poisson: the hard way, part I}

The general theory says that $\CalX$ is foliated 
\beq
{\CalX} = \bigcup\limits_{\bf c} X_{\bf c}
\eeq
with symplectic leaves $X_{\bf c}$. 
Near a regular point, a leaf $X_{\bf c}$ is a level set of a collection of Casimir functions $c_{1}, \ldots, c_{k}$, $k = {\rm codim}X_{\bf c}$, which obey $V_{c_{i}} = 0$, $i = 1, \ldots, k$.

In the example $\CalX = {\fg}^*$,
the Casimir functions are the $G$-invariant functions on ${\fg}^{*}$, while  symplectic leaves
are the coadjoint orbits $X_{\bf c} = {\CalO}_{\bf c} = \{ \, Ad_{g}^{*}({\bf c}) \, | \, g \in G \, \}$, for some ${\bf c} \in {\fg}^{*}$. 

In applications of Poisson geometry to fluid dynamics, selecting a symplectic leaf seems like an insurmountable problem: often, Casimir functions are non-local. One exception is provided by the Ertel invariant, or potential vorticity, see, e.g.,  \cite{LL}: for any smooth function ${\psi} \in C^{\infty}({\BR})$
\beq
\int_{B^3}\, \boldsymbol{\rho}\, {\psi} \left( \frac{dS \wedge d{\bf p}}{\boldsymbol{\rho}} \right) \, , 
\label{eq:nonrelertel}
\eeq
with $S$ being  the entropy per particle (a specific entropy), 
is a ${\CalG}^{(2)}_{B^3}$-invariant. Notice that the Hopf invariant, the asymptotic linking number \cite{ArnoldKhesin} of vortex lines,  $\int_{B^{3}} {\bf p} \wedge d{\bf p}$, where $\bf p$ is the fluid momentum per particle 1-form, is not an invariant of ${\CalG}^{(2)}_{B^3}$, nor a Casimir function on $\left( {\CalG}^{(2)}_{B^3} \right)^{*}$. It is a Casimir of ${\CalG}^{(1)}_{B^3}$, the group governing the barotropic flows. 

\subsubsection{Symplectic out of Poisson: the hard way, part II}

Another way to associate a symplectic manifold to the Poisson manifold $\CalX$ is 
the symplectic covering: add degrees of freedom, conjugate to Casimirs. For example, for
$\CalX = {\fg}^{*}$ that would be $\CalY = T^* G$ with the standard Liouville symplectic form. 
The group $G$ acts on $T^* G$ by the lift of the left action on $G$. The space of
functions $C^{\infty}(T^*G)^G$, invariant under this action, forms a subalgebra with 
respect to the Poisson bracket on $T^*G$. Hence, the quotient $T^*G/G \approx {\fg}^{*}$ inherits the Poisson structure which is precisely \eqref{eq:kirillov-kostant}. 

Again, in practice, this construction is not very convenient, because now the set of dummy variables has doubled. 

\subsubsection{Physical meaning of Novikov groups}

Had we followed that route, we would have realized the physical meaning of 
 Novikov's group ${\CalG}^{(2)}_{B^3}$: its elements  act as advection of particle density and entropy. If we interpret the entropy as a density of {\it spectator} particles, namely passive
particles carrying no inertia, as suggested in \cite{Carter}, then the Novikov group keeps track of the `initial' positions of both the fluid particles and the spectators, as well as the two types of individual clocks carried by both inert particles 
and the spectators. We will see more evidence of this interpretation in the four-dimensional formalism discussed in Section \ref{sec:covh} below. 

\subsection{Symplectic manifold out of Poisson via pathways}

{}Consider the space ${\CalP}_{\CalX}$ of
parametrized paths in $\CalX$, i.e., maps ${\bf x}: I \to {\CalX}$, where the domain $I$ could be an interval, the real line $\BR$   or a circle $S^1$.  Now consider the cotangent bundle ${\mathscr{M}}_{\CalX} = T^{*}{\CalP}_{\CalX}$, i.e., the space of pairs $$ \left(\, {\bf x} = \left(  x^i(t) \right)  \in {\CalP}_{\CalX} \, , \,  {\boldsymbol{\xi}} = \left( {\xi}_i(t) dt \right) \in T^{*}_{\bf x}{\CalP}_{\CalX}\, \right) \ , $$
where ${\boldsymbol{\xi}} \in {\Gamma} \left( {\bf x}^{*} T^{*}{\CalX} \otimes {\Omega}^{1}(I) \right)$. The space ${\mathscr{M}}_{\CalX}$ carries the canonical symplectic form:
\beq
{\Omega}_{{\mathscr{M}}_{\CalX}} = \int_{\BR} {\delta}x^{i} (t) \wedge {\delta}{\xi}_{i} (t) \, dt\,.
\label{eq:lf}
\eeq
Note that the natural action of ${\sf Diff}({\BR})$ on ${\mathscr{M}}_{\CalX}$ preserves ${\Omega}_{{\mathscr{M}}_{\CalX}}$. This action is generated by the moment map
\beq
{\mu}_{{\sf Diff}({\BR})} =  {\xi}_{i} \frac{dx^{i}}{dt} = : T_{tt} \, 
\eeq
valued in quadratic differentials on the timeline. 

\subsection{Hamiltonian dynamics as intersection problem}

We can now present the integral trajectories of the Hamiltonian vector field $V_{H}$ on $\CalX$
as the solutions to the intersection problem:
\beq
{\rm trajectories} \in {\CalC}  \cap {\CalL}  
\label{eq:intersect}
\eeq 
described below. The submanifold ${\CalC}$ encodes the choice of Poisson structure on $\CalX$, it is going to be coisotropic, cf. Section \ref{sec:cm}. The submanifold  ${\CalL}$ encodes the choice of the Hamiltonian and the choice of time parameterization, it is going to be Lagrangian, cf. Section \ref{sec:cm}. 

\subsubsection{The kinematic information}

The first ingredient of our intersection  perspective \eqref{eq:intersect} is 
the subvariety ${\CalC}  \subset {\mathscr{M}}_{\CalX}$. It encodes the Poisson structure on $\CalX$:
\beq
 {\CalC}   = \Biggl\{ \, ( {\bf x}, {\boldsymbol{\xi}} ) \, \Bigr| \, \frac{dx^{i}}{dt}  = {\pi}^{ij} (x(t)) \, {\xi}_{j} \, \Biggr\}\,.
 \label{eq:cois1}
 \eeq
We note that $T_{tt}= 0$ on ${\CalC} $. Moreover, ${\CalC} $ is preserved by the ${\sf Diff}({\BR})$ action on ${\mathscr{M}}_{\CalX}$. 

 {}
 Let us now show $\CalC$ is coisotropic. To this end we compute the Poisson brackets of the equations defining ${\CalC} $. Fix a point
 ${\bx} = (x(t)) \in {\CalP}_{\CalX}$. Let
 ${\zeta}_{i}(t)$ be a test function valued in ${\bx}^{*}T^{*}{\CalX}$. 
 Denote by 
 \beq
 {\mathfrak{M}}_{\zeta} = \int_{I} {\zeta}_{i}(t) \left( {\dot x^{i}}(t)  - {\pi}^{ij}(x(t)) {\xi}_{j}(t)\right) dt \ .
 \label{eq:testeq}
 \eeq
 Then
 \beq
 \left\{ 
  {\mathfrak{M}}_{\zeta}\, ,\, {\mathfrak{M}}_{\tilde\zeta} \right\}^{{\Omega}_{{\mathscr{M}}_{\CalX}}} = 
 {\mathfrak{M}}_{[[{\zeta}, {\tilde\zeta}]]} \, , 
 \label{eq:poissbrm}
 \eeq
 where the double bracket stands for
 \beq
 [[{\zeta}, {\tilde\zeta}]]_{j} = {\partial}_{j}{\pi}^{ik} (x(t)) {\zeta}_{i}(t) {\tilde\zeta}_{k}(t) \ .
\label{eq:dbracket}
\eeq
In deriving \eqref{eq:poissbrm} we used \eqref{eq:jacobi}. Thus, ${\CalC} $ is a \emph{coisotropic} submanifold in ${\mathscr{M}}_{\CalX}$, it is given by first-class constraints, i.e., the Poisson brackets of equations defining ${\CalC}$ vanish on ${\CalC} $, cf. \cite{Cattaneo:2003dp}.

{}
A standard exercise in symplectic geometry, given a coisotropic subvariety, e.g., ${\CalC} \subset {\mathscr{M}}_{\CalX}$, is to measure how much it differs from a Lagrangian (i.e., a maximal ${\Omega}_{{\mathscr{M}}_{\CalX}}$-isotropic
submanifold). To this end, we need to compute the distribution of the kernels of 
\beq
{\omega}_{\CalX, \pi} = {\Omega}_{{\mathscr{M}}_{\CalX}} \bigr\vert_{{\CalC} }\,.
\eeq
Ignoring for the moment the issue with boundary conditions at $t \to \pm \infty$, the vector
\beq
{\bf u} = \left( \, u^{i}_{x}(t) \, , \, u^{\xi}_{i}(t) dt \right) \in T_{({\bf x}, {\boldsymbol{\xi}})} {\mathscr{M}}_{\CalX}
\eeq
is tangent to ${\CalC}  \ni ({\bf x}, {\boldsymbol{\xi}})$ if
\beq
{\dot u}^{i}_{x} = {\pi}^{ij}\left( {\bf x}(t) \right) u^{\xi}_{j}(t) + {\partial}_{l}{\pi}^{ij}\left( {\bf x}(t) \right)  {\xi}_{j}(t) u^{l}_{x}(t) \, , 
\label{eq:tangp}
\eeq
and it furthermore belongs to ${\rm ker} \left( {\omega}_{\CalX, \pi} \right)$ if
\beq
\begin{aligned}
& u^{i}_{x} (t) = {\pi}^{ij}\left( {\bf x}(t) \right) {\zeta}_{j}(t)  = \frac{\delta  {\mathfrak{M}}_{\boldsymbol{\zeta}}}{\delta {\xi}_{i}}\, , \\
& u^{\xi}_{i} (t) = {\dot \zeta}_{i}(t) + {\partial}_{i} {\pi}^{jk}\left( {\bf x}(t) \right) 
{\xi}_{j}(t) {\zeta}_{k}(t)  = - \frac{\delta  {\mathfrak{M}}_{\boldsymbol{\zeta}}}{\delta x^{i}}\\
\end{aligned}
\label{eq:kerom}
\eeq
for some section ${\boldsymbol{\zeta}} \in {\Gamma}\left( {\BR}, {\bf x}^{*} T^{*}{\CalX} \right)$. The compatibility of \eqref{eq:kerom} and \eqref{eq:tangp} is a consequence of the Jacobi identity \eqref{eq:jacobi}. The quotient 
\beq
{\CalT}_{({\bf x}, {\boldsymbol{\xi}})} = T_{({\bf x}, {\boldsymbol{\xi}})}{\CalC} /{\rm ker}\left( {\omega}_{\CalX, \pi} \right)
\eeq
of the vector space of solutions to \eqref{eq:tangp} by the vector space
of solutions to \eqref{eq:kerom} is a symplectic vector space. If it is zero, then ${\CalC} $ 
is actually Lagrangian. In any case, the distribution of the kernels ${\rm ker}\left( {\omega}_{\CalX, \pi} \right)$ is integrable, i.e., is tangent to a foliation ${\CalF}_{\pi} \subset {\CalC} $. If this foliation is regular, the space ${\CalM}_{\CalX, \pi} = {\CalC} /{\CalF}_{\pi}$ of leaves is a symplectic manifold, with ${\CalT}_{({\bf x}, {\boldsymbol{\xi}})}$ being the tangent space to ${\CalM}_{\CalX, \pi}$ at the point $[({\bf x}, {\boldsymbol{\xi}})]$ representing the leaf passing through the point $({\bf x}, {\boldsymbol{\xi}})$. 

\subsubsection{Symplectic case}

If $\pi$ is invertible, i.e., there exists a closed two-form $\omega_{\CalX}$ such that
${\pi} = {\omega}_{\CalX}^{-1}$, then ${\CalC} $ could be Lagrangian, described
by the generating function, cf. \eqref{eq:actionsympl}
\beq
\int_{\gamma} d^{-1} {\omega}_{\CalX}
\label{eq:geomac}
\eeq
once a polarization on $\CalX$ is chosen. More precisely, if the domain of our paths is $S^1$, then the generating function of ${\CalC} $ is a multi-valued function, whose differential is well-defined and given by the integral of $\omega$ along the loop
\beq
{\delta} S_{{\CalC} } ( {\bf u} ) = \int_{S^1} {\omega} ( {\dot\gamma}, {\bf u} ) dt\,.
\eeq
If the domain is an interval $I$, then one should restrict the endpoints of the path to lie on fixed submanifolds $L_s, L_t \subset {\CalX}$, Lagrangian w.r.t ${\omega}$. 

More general boundary conditions on paths deserve a separate analysis.

\subsubsection{The dynamical information}

The second ingredient in our story, the subvariety ${\CalL}$, encodes the information about the Hamiltonian driving the dynamics.  The submanifold ${\CalL}$ is Lagrangian, with generating function given by
\beq
{\CalH}\left[ {\bf x} \right] = \int_{\BR}\, dt\, {\bf x}^{*}H \ \Leftrightarrow \ {\CalL}   = \left\{ \, ({\bf x}, {\boldsymbol{\xi}}) \, \Biggr| \, {\xi}_{i} = \frac{\partial H}{\partial x^i} \, \right\}\,.
\label{eq:vh}
\eeq
We emphasize that \eqref{eq:vh} depends on the specific parametrization of the path. The group ${\sf Diff}({\BR})$ does not leave ${\CalL}$ invariant: a diffeomorphism $t \mapsto {\tilde t}(t)$ changes ${\CalH}$. In what follows, we shall slightly change the viewpoint.  The data defining the Lagrangian ${\CalL}$ is really the metric ${\bf g} = dt^2$ on the worldline, cf. Section \ref{sec:covh}. 

The important fact, reviewed in Section \ref{sec:cm}, is that ${\CalL}$  can be described by numerous
generating functions relative to numerous choices of polarization on ${\mathscr{M}}_{\CalX}$.

{}
In the next section  we generalize ${\mathscr{M}}_{\CalX}$ so that the fluid dynamics is formulated in spacetime. In this way we vindicate the intersection problem view on Poisson dynamics.

\section{Hydrodynamics in four dimensions, I: geometry}\label{sec:covh}

Let $M^4$ be a four-manifold, interpreted as spacetime. We associate to $M^4$ the symplectic manifold ${\mathscr{P}}_{M^4} = {\Omega}^{0 \oplus 1 \oplus 3 \oplus 4}(M^4)$:
\beq
{\mathscr{P}}_{M^4} = \left\{ \, {\bf S}, {\bf p}, {\bf n}, {\boldsymbol{\nu}} \, | \, {\bf S} \in C^{\infty}(M^4), {\bf p} \in {\Omega}^{1}(M^4), {\bf n} \in {\Omega}^{3}(M^4), {\boldsymbol{\nu}} \in {\Omega}^{4}(M^4) \, \right\}\,.
\label{eq:forms}
\eeq
We choose the canonical symplectic form on $\CalP_{M^4}$. It  is given by:
\beq
{\Omega}_{\CalP_{M^4}} = \int_{M^4} {\delta} {\bf S} \wedge {\delta} {\boldsymbol{\nu}} + {\delta} {\bf p} \wedge {\delta} {\bf n}\,.
\label{eq:symf}
\eeq
The fields  $({\bf S}, {\bf p}, {\bf n}, {\boldsymbol{\nu}})$ are treated as independent. Their physical meanings are as follows: the scalar ${\bf S}$ and the single component of the top form $\boldsymbol{\nu}$ are thermodynamic fields at local equilibrium. Respectively, they are the entropy per particle and a ``thermal energy'' --- temperature times particle density. The other two are the four-fluid momentum (momentum per particle) and the particle four-flux.  They are subject to a set of regularity conditions, which we implicitly assume. In particular, the particle density, entropy density, and temperature are strictly positive (nowhere zero on $M^4$). In addition, we assume that $d{\bf p} \wedge d{\bf p}$ is nowhere zero.
Together, these fields have ten components related to ten components of the triples $ (\boldsymbol{\Pi}, \boldsymbol{\rho},\boldsymbol{s}),\ ({u}, \xi, T)$  used in Section \ref{sec:uxit}.  The relation is made explicit in Section \ref{sec:3d} below. In particular,  $\boldsymbol{s}=(\boldsymbol{\rho}/m) {\bf S},\ \boldsymbol{\nu}=(\boldsymbol{\rho}/m) T$.  For the purposes of this section, we organize the variables into two doublets $({\bf n},{\bf S})$ and $({\bf p},\boldsymbol{\nu})$. In the four-dimensional formalism, densities per particle are more natural than densities per volume, which were used in the three-dimensional formalism of Section \ref{sec:epa}.

As in the previous section, we shall study the intersections of two subvarieties in ${\mathscr{P}}_{M^4}$, a coisotropic subvariety ${\CalC}   $, defined through the action of the group ${\sf Diff}(M^4) \sdtimes {\sf C}^{\infty}(M^4)$ 
on ${\mathscr{P}}_{M^4}$, and a Lagrangian submanifold ${\CalL}_{e, \bf g}$. The latter is determined by the equation of state. We begin with a set of geometric observations insensitive to the equation of state. 
\subsection{Flow lines, fluid velocity, and frames}
\subsubsection[First pencil of flow lines]{$({\bf S}, {\bf p})$-pencil of flow lines: }
\label{sec:sp}

A pair $({\bf S}, {\bf p}) \in {\Omega}^{0 \oplus 1}(M^4)$ defines a system of \emph{parametrized} lines $\gamma$ as the extrema of the functional, cf.~\eqref{eq:actionsympl}, 
\beq
{\CalS}_{{\bf S}, {\bf p}} [{\gamma}] = \int_{\gamma} {\bf p} + \int_{I} \, d{\tau} \, {\gamma}^{*} {\bf S}\,,
\label{eq:ssp}
\eeq
where\footnote{To be precise, one should specify the boundary conditions. One possible choice, for $I$ being an interval $[{\tau}_{-}, {\tau}_{+}]$ is to fix two surfaces ${\Sigma}_{-}, {\Sigma}_{+} \subset M^4$, such that
$d{\bf p} \vert_{\Sigma_{\pm}} = 0$, and to require ${\gamma}({\tau}_{\pm}) \subset {\Sigma}_{\pm}$.}
{}
$I$ is the domain of the parameterization (an interval or a real line $\BR$), ${\gamma}: I \to M^{4}$ is the line, and $\tau$ is the coordinate on $I$. 
{}
The curves ${\gamma}$ that extremize ${\CalS}_{{\bf S}, {\bf p}}$ obey the Hamiltonian vector field equation
\beq
\iota_{\bf V}(\bf dp)=\bf dS\, ,  \label{eq:hamvps}
\eeq
with  Hamiltonian  $-\bf S$ and (pre-)symplectic form $\bf dp$. This equation defines an Eulerian velocity 
\beq
{\bf V} = \frac{d{\gamma}}{d{\tau}}\, , 
\label{eq:flow1}
\eeq
thereby justifying the term \emph{flow line} for ${\gamma} ({\tau})$. We refer to $\bf V$ as a {\it canonical} velocity.
\footnote{The notion of velocity depends on the choice of parametrization, also referred to as {\it fluid frame}.  A function similar to  \eqref{eq:ssp}  and the parametrizations 
\eqref{eq:flow1} were considered by Carter  \cite{Carter} (see also \cite{Markakis:2016udr} and references therein).  Note that $\bf V$ differs from $u$ considered in Section \ref{sec:thermodynamics} and the  $4$-velocity $\bf u$,  defined as the unit tangent vector to a particle’s worldline. The latter  corresponds to yet another parametrization discussed in Section \ref{sec:landau}.}

The functional ${\CalS}_{{\bf S}, {\bf p}}$ is invariant under certain transformations of the $({\bf S}, {\bf p})$ pair.
For any compactly supported $f \in {\sf C}^\infty(M^4)$, the transformation by a shift of the fluid momentum by an exact form
\begin{equation}
 {\bf p}\;\longmapsto\;{\bf p} + {\bf d}f\, , 
\label{eq:1stacp}
\end{equation}
is a symmetry of ${\CalS}_{{\bf S}, {\bf p}}$ modulo the boundary terms:
\beq
{\CalS}_{{\bf S}, {\bf p}+{\bf d} f} [ {\gamma}] = {\CalS}_{{\bf S}, {\bf p}} [ {\gamma}]  + {f}\left( {\gamma}({\tau}_{+}) \right) - { f}\left( {\gamma}({\tau}_{-}) \right) \,.
\eeq
Thus the extrema of ${\CalS}_{{\bf S}, {\bf p}+{\bf d}f}$ and those of ${\CalS}_{{\bf S}, {\bf p}}$ are the same. 

Now let us observe that the functional \eqref{eq:ssp} is invariant under the transformation
\begin{equation}
 {\bf p}\;\longmapsto\;{\bf p} + {\bf S}{\bf d} f\,
\label{eq:2ndacp}
\end{equation}
supplemented by a change of the parametrization of the line $\gamma$ \footnote{Here, to preserve the interval $I$ the function ${\bf f}$
 should also obey ${\bf f} \vert_{{\Sigma}_{\pm}} = 0$.} 
 \beq
 {\tau} \mapsto  {\tau} + { f}\left({\gamma}({\tau}) \right)\, , 
 \eeq
 $\bf S$ is kept fixed under both transformations.
 Subsequently, the velocity  is transformed under the change of a frame as
 \beq
{\bf V}\;\longmapsto\;\left( 1 - L_{\bf V} {f} \right)^{-1} {\bf V}\,.
\label{eq:renormv}
 \eeq
We will see in Section \ref{sec:coisotropic} that either of the two transformations \eqref{eq:1stacp}, \eqref{eq:2ndacp} is responsible for the continuity equations and the adiabatic equation. 

For future use define the function
\beq
{\boldsymbol{\zeta}} = {\bf S} + \iota_{\bf V} {\bf p}
\label{eq:zeta}
\eeq
which can be interpreted as the Lagrangian for the action  \eqref{eq:ssp} on the flow line passing through a point in spacetime. 
The importance of \eqref{eq:zeta} is clear from the  Lagrangian version of \eqref{eq:hamvps}:
\beq
L_{\bf V} {\bf p} = {\bf d}{\boldsymbol{\zeta}}\ .
\label{eq:lvpdz}
\eeq
Equation \eqref{eq:lvpdz}
implies Kelvin's theorem:
\beq
\frac{d}{d\tau} \oint_{C_{\tau}} {\bf p} = \oint_{C_{\tau}} {\bf d} {\boldsymbol{\zeta}} = 0
\eeq
where the family of closed contours $C_{\tau}$ is transported by the $\bf V$-flow according to \eqref{eq:flow1}. In the domain where ${\bf dp} \wedge {\bf dp} \neq 0$ Kelvin's theorem is nothing but the statement about the Poincare-Cartan invariants of Hamiltonian mechanics.

It is not difficult to generalize Kelvin's theorem. Let $g \in {\sf C}^{\infty}({\BR})$ be a smooth function, such that the function 
\beq
{\Upsilon}_{g}({\bf S}, {\bf p}) := \, \left( {\bf S} g ({\bf S}) \right)^{\prime} \, - \,
{\boldsymbol{\zeta}} \, g^{\prime}({\bf S}) 
\eeq
does not vanish on $M^4$. Consider the $1$-form $g({\bf S}) {\bf p}$. It obeys:
\beq
\iota_{\bf V} {\bf d}\left( g({\bf S}) {\bf p} \right) = {\Upsilon}_{g}({\bf S}, {\bf p})  d{\bf S} 
\eeq
where we used the {\it adiabatic equation}
\beq
L_{\bf V} {\bf S} = 0
\label{eq:adiabat}
\eeq
following from \eqref{eq:hamvps}. Now, define the collinear vector field
\beq
{\bf V}_{g} = {\Upsilon}_{g}^{-1}  {\bf V}
\eeq
so that
\beq
L_{{\bf V}_{g}} \left( g({\bf S}) {\bf p} \right) =  {\bf d}\left( {\bf S}   +  {\Upsilon}_{g}^{-1} g({\bf S}) \iota_{\bf V} {\bf p} \right)
\eeq
implying
\beq
\frac{d}{dt} \oint_{C_{t}} \, g({\bf S}) {\bf p} = 0
\label{eq:genken}
\eeq
 where now $C_{t} = \{ ({\bf x}_{t}(s)\sim {\bf x}_{t}(s+1)) \} \subset M^4$ is the family of closed contours transported by the ${\bf V}_{g}$-flow
\beq
\frac{{\pa}{\bf x}_{t}}{{\pa}t} = {\bf V}_{g}( {\bf x}_t)
\eeq

\subsubsection[Second pencil of flow lines]{$({\bf n}, {\boldsymbol{\nu}})$-pencil of flow lines} 
\label{sec:nnu}

A particle flux $3$-form ${\bf n}$ on $M^4$ defines the distribution of kernels ${\rm ker}({\bf n}) \subset TM^4$. This kernel is one-dimensional. The canonical velocity ${\bf { V}} \in {\rm ker}({\bf n})$ is normalized by the condition
\beq
\iota_{{\bf { V}}} {\boldsymbol{\nu}} = {\bf n}\, . 
\label{eq:flow2}
\eeq
Then the  flow lines as integral lines of $\bf V$ are  parametrized  
by a residue\footnote{At any point of  $M^4$ choose any basis $e_1,\dots,e_{4}$ of $T_m M^4$ with $\iota_{ e_1} {\bf n} = 0$. Then $d{\tau} (e_{1}) = {\boldsymbol{\nu}}(e_1, e_2, e_3, e_4) / {\bf n}_{m} (e_2, e_3, e_4)$.}
\beq
d{\tau} = {\boldsymbol{\nu}}/{\bf n}  \vert_{{\rm ker}({\bf n})}\label{eq:tau}\,.
\eeq
If, in addition ${\bf dn}=0$, then the flow of $\bf V$ preserves both ${\boldsymbol{\nu}}$ and $\bf n$: 
\beq
L_{\bf V} {\bf n} = 0
\label{eq:invn}
\eeq
\subsection[Coisotropic subvariety]{$\CalC$-subvariety}\label{sec:coisotropic}

We now define a submanifold
$$
{\CalC}\subset {\mathscr{P}}_{M^4}
$$
first, geometrically, as
 the set of $\left({\bf S},{\bf p},{\bf n},{\boldsymbol{\nu}}\right)$ for which the parametrized flow lines defined
in Section \ref{sec:sp} coincide with the parametrized flow lines defined in Section \ref{sec:nnu}, and preserve $\boldsymbol{\nu}$, and second, algebraically, by
\beq
{\CalC}
=
\left\{\, 
\left({\bf S},{\bf p},{\bf n},{\boldsymbol{\nu}}\right)
\;\middle|\;
\iota_{\bf V}{\boldsymbol{\nu}}={\bf n},
\ \iota_{\bf V}({\bf dp})={\bf dS}\, , \ {\bf dn} = 0\,
\right\}.
\label{eq:cois0}
\eeq
The equation
\beq
{\bf dn}=0\, 
\label{eq:continuity_n}
\eeq
is the continuity equation.

\subsubsection{Invariants and generalized Ertel theorems}

We can now combine
\eqref{eq:adiabat}, \eqref{eq:invn}
and \eqref{eq:hamvps} 
to conclude: 
the
$3$-form, 
{\it Ertel current}, cf. \eqref{eq:nonrelertel},
\beq
{\bf J}_{E} = {\bf dS} \wedge {\bf dp}\, , 
\label{eq:closedj}
\eeq
is closed and $\bf V$-equivariant:
\beq
{\bf dJ}_{E} = 0, \ 
\iota_{\bf V} {\bf J}_{E} = - {\bf dS} \wedge {\bf dS} = 0\, . 
\eeq
We can also define (away from the locus ${\bf n} \wedge {\bf p} = 0$) the scalar function
\beq
\mathsf{Q}=\frac{{\bf p} \wedge  {\bf J}_{E}}{\bf p\wedge n}
\label{eq:ertelscalar}
\eeq
So far it makes sense on all of ${\mathscr P}_{M^4}$, away from the ${\bf p} \wedge {\bf n} = 0$ locus on $M^4$. Now let us restrict to ${\CalC}$. Being annihilated by $\bf V$, ${\bf J}_{E}$
must be proportional to $\bf n$. It is easy to see that the coefficient
\beq
{\bf J}_{E} = \mathsf{Q} \, {\bf n}\, 
\label{eq:ertelsc}
\eeq
is given by \eqref{eq:ertelscalar}. 
The Lagrangian scalar  $\mathsf{Q}$ is called the {\it specific potential  vorticity} \cite{Katz}. 
Applying $L_{\bf V}$ to both sides of \eqref{eq:ertelsc}, using \eqref{eq:adiabat}, \eqref{eq:invn} and \eqref{eq:hamvps},  we derive
\beq
L_{\bf V} \mathsf{Q} = 0\, , 
\label{eq:advecte}
\eeq
the {\it advectivity} of $\mathsf{Q}$. The statements \eqref{eq:closedj}, \eqref{eq:advecte} hold also for
the generalized Ertel currents ${\bf J}_{E}^{g} = g({\bf S}) {\bf J}_{E}$, for any smooth function 
$g \in {\sf C}^{\infty}({\bR})$.

\subsubsection{${\CalC}$ is a coisotropic variety}

Equations \eqref{eq:hamvps}, \eqref{eq:flow2} are  Lichnerowicz--Carter equations. We now explain the algebraic definition from the viewpoint of symmetry.

Recall that the group ${\CalG}_{M^4}^{(1)}$ is the semidirect product
\beq
{\CalG}_{M^4}^{(1)}={\sf Diff}(M^4)\sdtimes {\sf C}^{\infty}(M^4)\,.
\eeq
There are two natural actions of ${\CalG}^{(1)}_{M^4}$ on ${\mathscr{P}}_{M^4}$. The group ${\sf Diff}(M^4)$ acts in the obvious way by spacetime diffeomorphisms, but its extension to ${\CalG}^{(1)}_{M^4}$ is not unique: one may use either \eqref{eq:1stacp} or \eqref{eq:2ndacp}.

Both transformations \eqref{eq:1stacp} and \eqref{eq:2ndacp} lift to symplectomorphisms of ${\mathscr{P}}_{M^4}$. The transformation \eqref{eq:1stacp} lifts to a shift of the fluid momentum:
\beq
\left({\bf S},{\bf p},{\bf n},{\boldsymbol{\nu}}\right)
\mapsto
\left({\bf S},{\bf p}+{\bf d}f,{\bf n},{\boldsymbol{\nu}}\right),
\label{eq:1staction}
\eeq
whereas \eqref{eq:2ndacp} lifts to the linear symplectic transformation
\beq
\left({\bf S},{\bf p},{\bf n},{\boldsymbol{\nu}}\right)
\mapsto
\left(
{\bf S},
{\bf S}({\bf p}/{\bf S}+{\bf d}f),
{\bf n},
{\boldsymbol{\nu}}-{\bf d}f\wedge{\bf n}
\right).
\label{eq:2ndaction}
\eeq
In the second realization, the ratio ${\bf p}/{\bf S}$ plays the role of the fluid momentum. If ${\bf S}$ is interpreted as the ratio of the density of spectator particles to the density of fluid particles, then ${\bf p}/{\bf S}$ is the fluid momentum per spectator particle.

We now compute the ${\CalG}^{(1)}_{M^4}$ moment map defined by \eqref{eq:mom1}:
\beq
{\boldsymbol{\mu}}
=
{\mu}_{\sf Diff}\oplus {\mu}_{{\sf C}^{\infty}},
\label{eq:decommom}
\eeq
where
\beq
\begin{aligned}
& 1^{\rm st}\ {\rm action}:\qquad
{\mu}_{{\sf C}^{\infty}}={\bf dn}\in {\Omega}^{4}(M^4)
=
\bigl({\sf C}^{\infty}(M^4)\bigr)^{*},\\
& 2^{\rm nd}\ {\rm action}:\qquad
{\mu}_{{\sf C}^{\infty}}={\bf ds}\in {\Omega}^{4}(M^4)
=
\bigl({\sf C}^{\infty}(M^4)\bigr)^{*},\\
& \langle {\mu}_{\sf Diff},\epsilon\rangle
=
\int_{M^4}
\iota_{\epsilon}{\bf dS}\,{\boldsymbol{\nu}}
+
\iota_{\epsilon}{\bf dp}\wedge{\bf n}
+
{\bf d}\iota_{\epsilon}{\bf p}\wedge{\bf n}
\end{aligned}
\label{eq:momg1}
\eeq
for a test vector field $\epsilon\in {\sf Vect}(M^4)$.

One can rewrite \eqref{eq:momg1} as
\beq
{\mu}_{\sf Diff}
=
\left(\iota_{\bf V}{\bf dp}-{\bf dS}\right)\otimes{\boldsymbol{\nu}}
+
{\bf p}\otimes{\bf dn}
\in
{\Omega}^{1}(M^4)\otimes{\Omega}^{4}(M^4),
\label{eq:momdiff4}
\eeq
or, equivalently,
\beq
{\mu}_{\sf Diff}
=
\iota_{{\bf n}^{\vee}}{\bf dp}
+
{\bf p}\otimes{\bf dn}
-
{\bf dS}\otimes{\boldsymbol{\nu}}\,,
\label{eq:momdiff41}
\eeq
where ${\bf n}^{\vee}$ denotes ${\bf n}$ viewed as a $4$-form-valued vector field on $M^4$.
We can now state the algebraic definition of ${\CalC}$ as follows:
\beq
{\CalC} = {\boldsymbol{\mu}}^{-1}(0)
\label{eq:coismom}
\eeq
We have the identity, cf. \eqref{eq:zeta}
\beq
\iota_{\bf V}{\mu}_{\sf Diff}
=
{\boldsymbol{\zeta}}\,{\bf dn}
-
{\bf ds}
\in {\Omega}^{4}(M^4),
\eeq
relating the ${\mu}_{{\sf C}^{\infty}}$ for the $1^{st}$ and the $2^{nd}$ ${\sf C}^{\infty}$-actions. 
We show below that on the Lagrangian submanifold representing the equation of state 
$\zeta$ becomes the reduced chemical potential $\mu/T$.

Setting the moment map ${\mu}_{\sf Diff}$ to zero, we obtain
\beq
\iota_{{\bf n}^{\vee}}{\bf dp}
+
{\bf p}\otimes{\bf dn}
=
{\bf dS}\otimes{\boldsymbol{\nu}}\,.
\label{eq:diffeq}
\eeq
As a consequence, the divergences of the particle flux and entropy flux are related by
\beq
{\boldsymbol{\zeta}}\,{\bf dn}= {\bf ds}\,.
\eeq
If, in addition, we impose the Abelian constraint ${\mu}_{{\sf C}^{\infty}}=0$, then the definition   \eqref{eq:cois0} follows. In that case, Eq.~\eqref{eq:diffeq} reduces to the Carter--Lichnerowicz equation,
\beq
\iota_{{\bf n}^{\vee}}{\bf dp}
=
{\bf dS}\otimes{\boldsymbol{\nu}}\,,
\label{eq:CL}
\eeq
which is equivalent to \eqref{eq:cois0}.

\subsection{Symmetries of phase space}
\label{sec:symmphsp}

There are canonical transformations of ${\mathscr{P}}_{M^4}$ compatible with the ${\CalG}^{(1)}_{M^4}$-action. 
These transformations preserve ${\CalC}$ and suggest additional constraints that may be imposed when formulating the intersection problem.   

{}Let $F \in {\sf Diff}({\BR})$, and $\varphi$ a smooth function on $\BR$. 
The map 
\beq
\left({\bf S},{\bf p},{\bf n},{\boldsymbol{\nu}}\right) \mapsto
\left( \, F({\bf S}),\, e^{{\varphi}({\bf S})} {\bf p},\, e^{-{\varphi}({\bf S})} {\bf n}, \, \frac{{\boldsymbol{\nu}} - {\varphi}^{\prime}({\bf S}) {\bf p} \wedge {\bf n}}{F^{\prime}({\bf S})}   \right)
\label{eq:diffcov}
\eeq
preserves ${\Omega}_{{\mathscr{P}}_{M^4}}$. The transformations \eqref{eq:diffcov} are generated
by the Hamiltonians
\beq
{\CalH}_{a} = \int_{M^4} a({\bf S})  {\boldsymbol{\nu}}
\label{eq:aham}
\eeq
and
\beq
{\CalH}_{b} = \int_{M^4} b({\bf S}) {\bf p} \wedge {\bf n}
\label{eq:bham}
\eeq
for some functions $a, b \in {\sf C}^{\infty}({\BR})$. 

The vector field $\bf V$ transforms, under \eqref{eq:diffcov}, 
\beq
{\bf V} \mapsto \frac{F^{\prime}({\bf S}) e^{-{\varphi}({\bf S})}}{1 + ( {\bf S} - {\boldsymbol{\zeta}} ) {\varphi}^{\prime}({\bf S}) }
\, {\bf V}
\label{eq:vtransfr}
\eeq
This transformation is related to the generalization \eqref{eq:genken} of Kelvin's theorem. 

In addition to \eqref{eq:aham},\eqref{eq:bham} one can define the following ${\CalG}_{M^4}^{(1)}$-invariants:
\beq
{\CalH}_{c} = \int_{M^4} c({\bf S}) {\bf d n}
\label{eq:cham}
\eeq
and
\beq
{\CalH}_{d} = \int_{M^4} d({\bf S}) {\bf dp} \wedge {\bf dp}
\label{eq:dham}
\eeq
for some functions $c, d \in {\sf C}^{\infty}({\BR})$. These transformations shift $\bf p$ by differentials ${\bf dS}$, mix $\bf n$ with ${\bf p}\wedge{\bf dp}$, and mix $\boldsymbol{\nu}$ with ${\bf dp}\wedge{\bf dp}$ and ${\bf p}\wedge{\bf n}$, multiplied by functions of $\bf S$. 

Note that the Ertel scalar $\mathsf{Q}$ \eqref{eq:ertelsc} times any function of $\bf S$ can be defined on all ${\mathscr P}_{M^4}$ as the ratio of densities of ${\CalH}_{d}$ and
${\CalH}_{b}$, for appropriate $b,d$ functions. Also note that the restrictions of \eqref{eq:aham}, \eqref{eq:bham}, \eqref{eq:cham}, \eqref{eq:dham} on $\CalC$ are not independent. 

\subsection{Coisotropic vs Lagrangian}

As recalled above, coisotropic submanifolds of a symplectic manifold $\mathscr P$ may have any dimension between the minimal, middle-dimensional case of a Lagrangian submanifold and the maximal case $\mathscr P$ itself. The zero level
\beq
{\CalC} = {\boldsymbol{\mu}}^{-1}(0)
\eeq
of the moment map $\boldsymbol{\mu}$ for ${\CalG}^{(1)}_{M^4}$ has functional dimension five; in this sense, it lies much closer to the Lagrangian end of the spectrum than a typical coisotropic submanifold.

The restriction of ${\Omega}_{{\mathscr P}_{M^4}}$ to ${\CalC}$ is nonzero, however. It has a kernel, spanned
by the orbits of the ${\CalG}^{(1)}_{M^4}$-action. 
In this way the quotient
\beq
{\mathscr{M}}_{M^{4}} = {\CalC}   /{\CalG}^{(1)}_{M^{4}}
\eeq
is a symplectic variety (a singular symplectic manifold), which 
measures how far ${\CalC}$ is from being a Lagrangian submanifold. Since $\mathscr{P}_{M^4}$ is the cotangent bundle to ${\Omega}^{0} (M^4) \oplus {\Omega}^{1}(M^4)$, the symplectic quotient ${\mathscr{M}}_{M^{4}}$ can be viewed as a refined version of the cotangent bundle to the quotient
\beq
\left( {\Omega}^{0} (M^4) \oplus {\Omega}^{1}(M^4) \right)/{\CalG}^{(1)}_{M^{4}} \ .
\eeq
We therefore want to study the space 
parametrized by the ${\CalG}^{(1)}_{M^{4}}$-invariants built from ${\bf S}$ and ${\bf p}$. From our list
of local invariants only \eqref{eq:dham} is a suitable functional.

\subsection{Asymptotic Hopf invariants}
One way to think about the collection of invariants \eqref{eq:dham} is as a
 distribution $H_{s}$ of helicity integrals over the 3-manifolds of the level sets ${\bf S}(x)=s$
\beq
H_{s} = \int_{{\bf S}^{-1}(s)} {\bf p} \wedge d{\bf p}\, . 
\label{eq:hofsinv}
\eeq
These integrals are invariant under both the \eqref{eq:1staction} and \eqref{eq:2ndaction} actions of ${\CalG}^{(1)}_{M^{4}}$.

The invariants \eqref{eq:hofsinv} are familiar in the context of $2+1$ dimensional hydrodynamics as
asymptotic Hopf invariants  \cite{ArnoldKhesin}. Even though our problem is four dimensional, the level sets of $\bf S$
are three dimensional invariant submanifolds of the $\bf V$-flow. 

For future use, define the helicity $3$-form
\beq
{\bf h} = {\bf p} \wedge {\bf dp}
\label{eq:helicity}
\eeq

\section{Hydrodynamics in four dimensions II: equation of state}
\label{sec:chemistry}
Let ${\bf g}$ be a Lorentzian metric on $M^4$, and define the scalar particle density $n_{\bf g}$ via
 \beq
n_{\bf g}^{2} = \frac{{\bf n} \wedge \star_{\bf g} {\bf n}}{{\rm vol}_{\bf g}} = 
n_{\mu\nu\lambda} \, n^{\mu\nu\lambda} \, , 
\label{eq:ng}
\eeq
where ${\rm vol}_{\bf g} = \sqrt{-{\rm det}({\bf g})} \, d^{4}x$
is the volume form, and the indices are raised by the metric.  Let 
$e= e(n_{\bf g},S)$, called the \emph{relativistic energy density}, be a function of two variables, $n_{\bf g}$ and $S$, such that
\beq
w = \left( \frac{{\partial} {e}}{{\partial n_{\bf g}}} \right)_{S} \, , \ T = \frac{1}{n_{\bf g}} \left( \frac{{\partial} {e}}{{\partial} S} \right)_{n_{\bf g}}
\label{eq:esf}
\eeq
which are called the \emph{specific relativistic enthalpy} and the \emph{temperature}, respectively, and are assumed to be positive.

\subsection{Lagrangian submanifolds}

Recall that a submanifold  $L \subset {\CalP}$ of a symplectic manifold $({\CalP}, {\omega})$
is Lagrangian, if it is \emph{isotropic}, i.e., ${\omega} \vert_{L} = 0$, and \emph{maximal}, i.e., 
${\rm dim}(L) = \frac 12 {\rm dim}({\CalP})$. The latter definition can be extended to the infinite-dimensional case as follows: any isotropic $L' \subset {\CalP}$ which contains $L$, $L \subset L'$ is equal to $L$. 

Now let us choose a set of local Darboux coordinates, i.e., $(p_i, q^i)$, such that
\beq
{\omega} = \sum_{i=1}^{n}  dp_{i} \wedge dq^i\, , \qquad {\rm dim}({\CalP}) = 2n\, , 
\eeq
and assume $q^1, \ldots, q^n$ are also good local coordinates on $L$. Then there exists a function $S(q)$, such that
\beq
p_{i} = \frac{{\partial}S}{{\partial}q^{i}}\, , \ i = 1, \ldots, n
\label{eq:spq}
\eeq
for $(p,q) \in L$. We refer to the function $S$ as the \emph{generating function} of $L$ \emph{in the $(q|p)$ polarization}. Viewed as a function of $q$, it need not be single-valued, since the projection $L \to {\BR}^{n}_{q}$ may be many-to-one. In fact,  $S$ should not be regarded as a function of $q$, but rather a function defined on $L$. 

The Darboux coordinates are not unique. One may, for instance, choose $p$ in place of $q$ and $-q$ in place of $p$; we refer to this choice as the $(p|q)$ polarization. In this polarization, the Lagrangian submanifold $L$ is described by the generating function ${\tilde S}(p)$ obeying
\beq
q^i =  \frac{{\partial}{\tilde S}}{{\partial}p_{i}}\,,
\label{eq:sqp}
\eeq
which is related to $S(q)$ given by \eqref{eq:spq}. The relation is the  Legendre transform:
\beq
{\tilde S}(p) = \sum_{i=1}^{n} p_i q^i -S(q) \, , 
\eeq
where one solves \eqref{eq:spq} to express $q$ in terms of $p$.
 {}
 We are now ready to introduce the second ingredient of our intersection viewpoint on fluid dynamics, 
  a Lagrangian submanifold ${\CalL} \subset{\mathscr{P}}_{M^4}$. 
 A purely geometric characterization of ${\CalL}$, analogous to the characterization of ${\CalC}$ in terms of the $({\bf S}, {\bf p})$ and $({\bf n}, {\boldsymbol{\nu}})$ flows, remains desirable. 
 
 For the time being, we use the formalism of generating functions \cite{Arnold}. Specifically, we define 
\beq
{\CalL} = \Biggl\{ \, ( {\bf S}, {\bf p}, {\bf n}, {\boldsymbol{\nu}} )\ \Bigr| \ {\bf p} \, = \, \frac{w}{n_{\bf g}}   \cdot \star_{\bf g} {\bf n}\, , \ {\boldsymbol{\nu}} \, = \, T n_{\bf g}  \cdot {\rm vol}_{\bf g} \, \Biggr\}\,,
\label{eq:pol1}
\eeq
where $w, T$ are given by \eqref{eq:esf}.

The vector field ${\bf V}$ defined by \eqref{eq:flow2} is therefore related to ${\bf p}$ via
\beq
{\bf p} = T  w \cdot {\bf V}^{\flat}\,,
\eeq
where ${\bf V}^{\flat} : = {\bf g}( {\bf V}, \cdot )$.
We can thus relate ${\bf V}$ to $T$: 
\beq
1 = T \Vert {\bf V} \Vert_{\bf g}\,,
\label{eq:carter1}
\eeq
where
\beq
{\bf g}({\bf V}, {\bf V}) = \Vert {\bf V} \Vert_{\bf g}^{2}\,.
\eeq
In the framework of classical mechanics and symplectic geometry, the representation \eqref{eq:pol1} is associated with the generating function $S(q) = {\CalA} ({\bf S}, {\bf n})$ in the polarization $({\bf S}, {\bf n} \mid {\bf p}, {\boldsymbol{\nu}}) = (q \mid p)$:
\beq
{\CalA}({\bf S}, {\bf n}) = -\int_{M^4} {e} (n_{\bf g}, {\bf S} ) \, {\rm vol}_{\bf g}\,
\label{eq:genes}
\eeq

\subsubsection{The $4$-velocity $\bf u$}
\label{sec:landau}
Often, it is convenient to use another vector field ${\bf u} \in {\sf Vect}(M^4)$, the $4$-velocity, the unit tangent vector to a particle’s worldline, cf. \cite{LL}.
It is collinear with $\bf V$, but its normalization uses the metric, not the $4$-form $\boldsymbol{\nu}$: 
\beq
{\bf u} = \frac{\bf V}{||\bf V||_{\bf g}}\,.
\label{eq:Lfield}
\eeq
The flows generated by $\bf V$ and $\bf u$  are different, 
in particular, the Kelvin theorem does not hold \cite{LL} for a material contour comoving along the $\bf u$-vector field.

\subsubsection{Equation of state in other polarizations}

Instead of ${\bf S}, {\bf n}$ we can use, e.g., ${\bf S}, {\bf p}$ as coordinates, and
${\bf n}, {\boldsymbol{\nu}}$ as momenta. The generating function in the new polarization
$( {\bf S}, {\bf p} | {\bf n} , {\boldsymbol{\nu}})$ is computed by the
(partial) functional Legendre transform, e.g.,
\beq
{\CalL}= \{ \, \left( {\bf S}, {\bf p}, {\bf n}, {\boldsymbol{\nu}} \right) \, | \,  {\bf n} = w^{-1}{\partial}_{w} P
\, \star_{\bf g} {\bf p}\, , \ {\boldsymbol{\nu}} = -{\partial}_{S} P
\cdot {\rm vol}_{\bf g}\,
\}\,,
\label{eq:pol2}
\eeq
where now
\beq
p_{\bf g} = \sqrt{\frac{{\bf p} \wedge \star_{\bf g} {\bf p}}{{\rm vol}_{\bf g}}} = w 
\eeq
and
\beq
P=
P(w,{\bf S})
\eeq
is the pressure, given by the negative Legendre transform of ${e}( n_{\bf g},{\bf S})$ with respect to the particle-density argument: 
\beq
P(w,S)=-\varepsilon(n_{\bf g},S)+W ,
\eeq
where $W=n_{\bf g}w$ is the enthalpy density.

The generating function ${\tilde S} ({p}) 
$ of ${\CalL}, {\bf g}$ in the $({\bf S}, {\bf p}| {\bf n} , {\boldsymbol{\nu}}) = ({p}| {q})$ polarization is \cite{Schutz}: 
\beq
{\CalH}
( {\bf p},{\bf S}) = - \int_{M^4} P
( p_{\bf g},{\bf S})\, {\rm vol}_{\bf g}\ .
\label{eq:genes2}
\eeq
We emphasize that the Lagrangian submanifolds ${\CalL}$ as defined
by \eqref{eq:pol1} and \eqref{eq:pol2} are identical; they only differ in their descriptions. They
 describe the same set of fields.

\subsection{Deformations of Lagrangian submanifolds}

\subsubsection{Stress-energy tensor}

We emphasize the explicit dependence of the Lagrangian submanifold ${\CalL}$ representing the equation of state, on the metric $\bf g$. In other words, changing $\bf g$ changes $\CalL$. 

Recall that Hamiltonian dynamics in classical mechanics can be studied not only in terms of the motion of individual points on phase space $\CalP$, but also through the motion of submanifolds, specifically Lagrangian submanifolds \cite{Arnold}. This is justified a posteriori as the quasiclassical limit of the evolution of states in quantum mechanics.

Now imagine that our Lagrangian submanifold belongs to a multiparametric family $\left( {\CalL}_{b} \right)_{b \in B}$, parametrized by some space $B$. Let $0 \in B$ be a marked point. Let us assume ${\CalL}_{0}$ is simply-connected. Then, to each 
direction $i$ in $T_{b}B$ we associate a function $h_{i}\in C^{\infty}({\CalL}_{0})$
describing the deformation in the corresponding direction. We can write a multi-component Hamilton-Jacobi equation
\beq
\frac{\partial S}{\partial b^{i}} = h_{i} \left(\, \frac{\partial S}{\partial q}, q\, ; \, b \right)\,.
\label{eq:HJb}
\eeq
The right-hand side of \eqref{eq:HJb} uses some smooth extrapolation of $h_i$ from $\CalL_{0}$ to its neighborhood in $\CalP$. Different extrapolations lead to different equations \eqref{eq:HJb}; however, they can be mapped to one another by a reparametrization of $\CalL_0$. 

{}
In the present context, the analogue of $B$ is the space ${\rm Met}(M^4)$ of metrics $\bf g$ on $M^4$. The corresponding functions $h_i$ are the components of the \emph{stress-energy tensor}. In the $({\bf S}, {\bf n}|{\bf p}, {\boldsymbol{\nu}})$ polarization \eqref{eq:genes}: 
\beq\label{eq:tmn}
T_{\mu\nu} \,  :=  \, \frac{2}{{\rm vol}_{\bf g}}  \frac{\delta \CalA}{\delta g^{\mu\nu}} \,.\eeq
Without much computation, we can bring \eqref{eq:tmn} to the familiar form: \beq
T^{\mu\nu} = W u^{\mu} u^{\nu} + P \, g^{\mu\nu}\, , 
\label{eq:tmncan}
\eeq
in the normalization $u^{\mu} u_{\mu} = -1$. 
{}
In the  $({\bf S}, {\bf p} | {\bf n} , {\boldsymbol{\nu}} )$ polarization \eqref{eq:genes2} we obtain the same stress-tensor
\beq
T_{\mu\nu} \,  =  \, \frac{2}{{\rm vol}_{\bf g} }\frac{\delta \CalH}{\delta g^{\mu\nu}} \ ,  
\eeq
once the equation of state is imposed. 

{}
We can now illustrate that the \emph{equations of motion of hydrodynamics are conservation laws}. The group ${\CalG}^{(1)}(M^4)$ acts on ${\CalP}(M^4)$ preserving the coisotropic submanifold ${\CalC}   $. It does not preserve ${\CalL}$, rather, a transformation $g \in {\CalG}^{(1)}(M^4)$ consisting of a
diffeomorphism and an abelian transformation moves ${\CalL}$ by moving the metric.
 Let us consider an infinitesimal diffeomorphism $g = {\exp}({\ep})\in {\sf Diff}(M^4) \subset {\CalG}^{(1)}(M^4)$, generated by a vector field ${\ep} \in {\sf Vect}(M^4)$.  The
corresponding change in the metric --- the parameter ${\bf g}$ of ${\CalL}$ --- is given by
\beq
{\delta}g^{\mu\nu} = {\nabla}^{\mu} {\ep}^{\nu} + {\nabla}^{\nu} {\ep}^{\mu} \ .
\eeq
Now we recall the Hamilton-Jacobi equation \eqref{eq:HJb} which tells us that 
the change in the generating function \eqref{eq:genes} ${\CalA}$ associated with the deformation of  ${\CalL}$ is equal to the associated Hamiltonian \eqref{eq:momg1}
\beq
{\CalA}_{ {\bf g} + {\nabla}\cdot {\ep}} - 
{\CalA}_{ {\bf g}} = \langle {\mu}_{\sf Diff} ,  {\ep} \rangle\,.
\label{eq:hjep}
\eeq
Since the right-hand side of \eqref{eq:hjep} vanishes on ${\CalC}    \cap {\CalL}_{e, {\bf g}}$, we arrive at the familiar conservation law
\beq
{\nabla}^{\mu} T_{\mu\nu} = 0 \ .
\label{eq:nabtmn}
\eeq
 Of course, for generic $\bf g$ without isometries
no conserved quantities follow from \eqref{eq:nabtmn}. On the other hand, the Killing vectors of $\bf g$
lead to conservation laws, cf. \cite{Markakis:2016udr}. 

Thus the intersection ${\CalC}\cap{\CalL}_{e,{\bf g}}$ imposes, on the one hand, the Carter--Lichnerowicz equation and the continuity equation through the moment-map constraint defining ${\CalC}$, and, on the other hand, the equation of state through the Lagrangian submanifold ${\CalL}_{e,{\bf g}}$. The usual conservation of the stress-energy tensor is then the Hamilton-Jacobi expression of the covariance of the family ${\CalL}_{e,{\bf g}}$ under spacetime diffeomorphisms.

\subsubsection{Backgrounds and Currents}

There is a more general class of Lagrangian submanifolds ${\CalL} \subset {\mathscr P}_{M^4}$ of physical significance. 
Let $a_{1} \in {\Omega}^{1}(M^4)$ and $a_{3} \in {\Omega}^{3}(M^4)$ be fixed differential forms on $M^4$. 
We may call $a_1$ a {\it background gauge field}, and $a_3$ a {\it background charge flux}. 
Define ${\CalL}_{{\bf g}, a_{1}, a_{3}}$ as a translation of ${\CalL}_{\bf g}$ by $(a_1, a_3)$. In terms of the generating functional, this amounts to taking:
\beq
{\CalA}_{{\bf g}, a_{1}, a_{3}} = \int_{M^4} \, e \left( \sqrt{\frac{({\bf n}- a_{3}) \wedge \star_{\bf g} ({\bf n}- a_{3})}{{\rm vol}_{\bf g}}} \, , \, {\bf S} \right) \, {\rm vol}_{\bf g} + \int_{M^4} a_{1} \wedge {\bf n}
\label{eq:genfunl1}
\eeq
The corresponding conservation laws are obtained by applying the Hamilton-Jacobi argument above to the action of ${\CalG}^{(1)}_{M^4}$ on the $({\bf g}, a_{1}, a_{3})$-backgrounds. 

{}There is a larger class of Lagrangian submanifolds, defined by analogues
of \eqref{eq:genfunl1} in which $a_1$ and $a_3$ are promoted to $\bf S$-dependent $1$- and
$3$-forms, respectively. The symmetries of Section \ref{sec:symmphsp} preserve this class.
An even larger class is obtained by allowing the energy density to depend on
derivatives of $\bf n$ and $\bf S$, for example through the invariants
\beq
\frac{{\bf dn} \wedge \star_{\bf g} {\bf dn}}{{\rm vol}_{\bf g}}\, , \ \frac{{\bf d} \star_{\bf g} {\bf n} \wedge \star_{\bf g} {\bf d} \star_{\bf g} {\bf n}}{{\rm vol}_{\bf g}}\, , \ \frac{{\bf dS} \wedge \star_{\bf g} {\bf dS}}{{\rm vol}_{\bf g}}\ .
\eeq

\section{Reduction to three dimensions}\label{sec:3d}

Let $B^3 \subset M^4$ be a smooth submanifold, and $\ell \in {\sf Vect}(M^4)$ be a vector field
defined in a neighborhood $U_{B^3} \subset M^4$ of $B^3$, which is transverse to $B^3$. 

For any $({\bf S}, {\bf p}, {\bf n}, {\boldsymbol{\nu}}) \in {\mathscr{P}}_{M^4}$ define the 
 Eulerian fields on $B^3$ by: \beq
\begin{aligned} 
& {\boldsymbol{\rho}}= {\bf n} \vert_{B^3} \, , \\
& {\boldsymbol{s}}= {\bf S} {\bf n} \vert_{B^3} \, , \\
& \boldsymbol{\Pi} = \left( {\bf p} \vert_{B^3} \right) \otimes \left( {\bf n} \vert_{B^3} \right) \, , \\
& {\bf u} \in {\sf Vect}(B^3): \iota_{\bf u} {\boldsymbol{\rho}} =  - \left( \iota_{\ell} {\bf n} \right) \vert_{B^3} \, , \\
& \boldsymbol{\Theta}={\boldsymbol{\rho}} T = \left( \iota_{\ell} {\boldsymbol{\nu}} \right) \vert_{B^3} \, , \\
& \tilde{\mu}= \left( \iota_{\ell} {\bf p} \right) \vert_{B^3} - T\left( {\bf S} \vert_{B^{3}} \right) - \iota_{\bf u} \left( {\bf p} \vert_{B^{3}} \right)\,.
\end{aligned}\label{eq:Eulerpacket}
\eeq

\subsection{Coisotropic side}
Using the flow generated by $\ell$ identify the neighborhood $U_{B^3}$ with the product
\beq
U_{B^3} \approx B^{3} \times I \, , 
\eeq
where an interval $I \subset {\BR}$ is parametrized by $t$, such that
\beq
{\ell}  = {\pa}_{t}
\eeq
We  decompose:
\beq
{\bf p} ={\bf m}+ {\bf p}_0\, dt\, , \ {\bf n} = {\boldsymbol{\rho}} + {\bf j}  \wedge dt\, , \ {\boldsymbol{\nu}} = dt \wedge \boldsymbol{\Theta} 
\eeq
where all fields are now $t$-dependent forms on $B^3$. Specifically, ${\bf m}$ is a $3$-momentum per particle
and ${\bf p}_0$ is the energy per particle, $\bf j$ is the 2-form particle flux, and ${\boldsymbol{\Theta}}$ is the  thermal energy (particle density times temperature).  The symplectic structure \eqref{eq:symf} reads, up to unimportant sign changes:
\beq
{\Omega}_{\mathscr{P}_{M^4}} = \int\, dt\, \int_{B^3} \left(
 {\delta}{\bf S} \wedge {\delta} {\boldsymbol{\Theta}}   + {\delta} {\bf m}\wedge \delta {\bf j}  + {\delta} {\boldsymbol{\rho}} \wedge {\delta} {\bf p}_0 
\right)\,.
\label{eq:3from4}
\eeq

{}
Now we rewrite Eqs. \eqref{eq:cois0} in a way, which will make the analogy with 
\eqref{eq:cois1} transparent:
\beq
\begin{aligned}
& {\dot {\boldsymbol{\rho}}} + d \iota_{\bf u} {\boldsymbol{\rho}} = 0 \\
& {\dot {\bf S}} + \iota_{\bf u} d{\bf S} = 0 \\
& \dot{\bf m}+ d{\bf p}_0+ \iota_{\bf u} d{\bf m} +T \, d{\bf S} = 0\\
\end{aligned}
\label{eq:tderalg}
\eeq
where we decompose the $4$-dimensional vector field $\bf V$ as
(cf. \eqref{eq:Eulerpacket}):
\beq
{\bf V} = T^{-1} \left( {\partial}_{t} - {\bf u} \right)
\eeq
We now recognize in \eqref{eq:tderalg} the equations \eqref{eq:euler_eqs2}. 
In other words,  the phase space ${\mathscr{P}}_{B^3 \times {\BR}}$ is identified with 
$T^{*} {\sf Paths}\left({\sf Lie}^{*}\left({\CalG}^{(2)}_{B^3}\right) \right)$, the cotangent bundle to the space of paths in the dual Lie algebra of Novikov's group associated with the three-manifold $B^3$.

To summarize, we have reformulated the covariant equations \eqref{eq:cois0} as the three-dimensional evolution equations \eqref{eq:tderalg}, equivalent to \eqref{2.40} and \eqref{eq:euler_eqs2}.

This also clarifies the relation of the kinematic equations to coadjoint orbits.
Let $\mathfrak{g}={\sf Lie}({\CalG}^{(2)}_{B^3})$ and identify
${\bf P}=(\boldsymbol{\Pi},\boldsymbol{\rho},\boldsymbol{s})\in\mathfrak{g}^{*}$ and
${\boldsymbol{\Omega}}=({\bf u},\xi,T)\in\mathfrak{g}$ using the natural pairing.
Equations \eqref{2.40} and \eqref{eq:euler_eqs2} then have the form
\eqref{eq:top}, so their velocity is tangent to the coadjoint orbit through
${\bf P}$. Formally, wherever the time-dependent infinitesimal action integrates,
the evolution remains on this orbit, independently of the equation of state.
The orbit carries the Kirillov-Kostant symplectic structure induced by
\eqref{eq:kkf}, whereas the equation of state selects the Hamiltonian
and hence a particular motion on the orbit. Thus the coisotropic formulation
encodes motion within coadjoint orbits without requiring us to classify or
choose an orbit in advance. This is the distinction discussed in Section
\ref{sec:cm} and illustrated by the Ertel invariants \eqref{eq:nonrelertel}.

\subsection{Change in polarization}

Now assume the metric $\bf g$ is static, ${\ell}$ is the Killing vector field ${\partial}_{t}$, 
\beq
{\bf g} = e^{2f(x)} dt^2 - e^{-\frac{2f(x)}{3}} h
\label{eq:ngh}
\eeq
with a dilation factor determined by $f:B^3 \to {\BR}$ and three-dimensional metric $h = h_{ij}(x) dx^i dx^j$.

We observe that the polarizations used in describing the Lagrangian ${\CalL}_{{e}, \bf g}$
in four and three-dimensional formalisms are different. To go from the $({\bf S}, {\bf n} | {\bf p}, {\boldsymbol{\nu}})$-polarization to $(S, {\pi}, {\boldsymbol{\rho}} | c, b,a )$ polarization we perform the partial Legendre transform:
\beq
{\tilde\CalL}(S, {\pi}, {\boldsymbol{\rho}}) = \int_{B^3} {\pi} \wedge b + {\boldsymbol{\rho}} a - {\varepsilon} ( n_{\bf g}, S) \, {\rm vol}_{h}
\eeq
where
\beq
\begin{aligned}
& n_{\bf g}^2 = \left( \frac{\boldsymbol{\rho}}{{\rm vol}_{h}} \right)^2 - \frac{b \wedge \star_{h} b}{{\rm vol}_{h}} \\
& n_{\bf g} = \frac{\boldsymbol{\rho}}{{\rm vol}_{h}}  \sqrt{1-\Vert v \Vert_{h}^{2}} \approx
\frac{\boldsymbol{\rho}}{{\rm vol}_{h}} - \frac{\Vert \Pi \Vert_{h}^{2}}{2 \boldsymbol{\rho}} \\
\end{aligned}
\eeq
Thus, in the static situation, the four-dimensional Lagrangian submanifold ${\CalL}_{e,{\bf g}}$ reduces, after the change of polarization above, to the three-dimensional Lagrangian submanifold generated by the corresponding energy functional, e.g. \eqref{2.7}, up to the gravitational redshift factors determined by $f$. This completes the comparison between the covariant intersection problem and the Euler-Arnold formulation reviewed in Section \ref{sec:epa}.

\section{Anomalies through magnetic terms}
\label{sec:anomf}

Let us recap what we have done so far. We identified 
the hydrodynamics  of the perfect fluid with  the intersection ${\CalC}    \cap {\CalL}  $ of two coisotropic subvarieties, of which one is Lagrangian, in an auxiliary
symplectic manifold $\mathscr{P}_{M^4}$ associated with spacetime $M^4$. 

Actually, 
\beq
\mathscr{P}_{M^4} = T^{*}\mathscr{X}_{M^4}\ , \ {\CalC}    = 
{\boldsymbol{\mu}}^{-1}(0) \, , 
\label{eq:pxhydro}
\eeq
where ${\boldsymbol{\mu}}$ is the ${\CalG}^{(1)}_{M^4}$-moment map
for its action 
on $T^{*}\mathscr{X}_{M^4}$ which is the canonical lift of ${\CalG}^{(1)}_{M^4}$-action 
on
the base $\mathscr{X}_{M^4} = {\sf C}^{\infty}(M^4) \times {\Omega}^{1}(M^4)$, with $f \in {\sf C}^{\infty}(M^4)$
acting by
\beq
{\bf f}: ({\bf S}, {\bf p} ) \mapsto ({\bf S} , {\bf p} + {\bf d}f)
\label{eq:gachydro}
\eeq
\subsection{A comment on geometry}

In this and the next section we denote the de Rham differential on $\mathscr{X}$, ${\mathscr{P}}$ by $\delta$ in order not to confuse it
with de Rham differential $d$ on $M^4$. 

Let us discuss this geometry more abstractly. Suppose $\mathscr{X}$ is a smooth manifold with local coordinates $(x^i)$. Suppose $\mathscr{X}$ is endowed with a smooth action of a Lie group $G$. Let
\beq
{\xi} \mapsto V_{\xi} \in {\sf Vect} ({\mathscr{X}})\, , \ {\xi} \in \mathfrak{g} = {\sf Lie}(G)
\eeq
be the associated homomorphism of Lie algebras:
\beq
V_{\xi}^{j} {\partial}_{j} V^{i}_{\xi'} - V_{\xi'}^{j} {\partial}_{j} V^{i}_{\xi} = V_{[{\xi}, {\xi}']}^{i} \ . 
\eeq
Let $(p_i)$ be the linear coordinates on $T^{*}_{x}\mathscr{X}$, so that the canonical Liouville symplectic form on $T^{*}\mathscr{X}$ is
\beq
{\Omega}_{0} = \delta p_i \wedge \delta x^i
\eeq
Then
\beq
{\boldsymbol{\mu}}_{\xi} (x,p) = p \cdot V_{\xi} (x) : = p_i V^{i}_{\xi}(x)
\eeq
is the Hamiltonian of the associated vector field ${\CalV}_{\xi}$ on $T^{*}{\mathscr{X}}$:
\beq
{\CalV}_{\xi} = V_{\xi}^{i} {\partial}_{x^i}  - {\partial}_{i} V^{j}_{\xi} p_{j}  {\partial}_{p_{i}}\,.
\label{eq:liftv}
\eeq
Just like $V_{\xi}$'s the vector fields ${\CalV}_{\xi}$'s form the Lie algebra $\mathfrak{g}$:
\beq
[{\CalV}_{\xi}, {\CalV}_{\xi'} ] = {\CalV}_{[{\xi}, {\xi}']}\,.
\eeq
Moreover
\beq
{\boldsymbol{\mu}}(x,p) = p_{i} V^{i}_{\cdot}(x) \, : \, T^{*}{\mathscr{X}} \to {\mathfrak{g}}^{*}
\eeq
is the equivariant moment map. 

Our problem is to find the intersection
\beq
{\Phi} = (x,p)  \in \left( {\CalC} = {\boldsymbol{\mu}}^{-1}(0) \right)  \cap {\CalL}
\label{eq:intloc}
\eeq
of the zero locus of the moment map with some Lagrangian submanifold ${\CalL}$, 
possibly taken from some family, in our story parametrized by spacetime metrics and other 
backgrounds. If ${\CalL}$ is described in the $(x|p)$-polarization with the help
of the generating function $S = S(x)$  as
\beq
p_{i} = {\partial}_{i}S
\eeq
then \eqref{eq:intloc} reduces to the search of $x$, such that
\beq
V^{i}_{\xi}(x) {\partial}_{i} S = 0\, , 
\label{eq:oldx}
\eeq
for all $\xi \in \mathfrak{g}$. In other words, \eqref{eq:intloc} is a weak form of the variational principle.

\subsection{Geometry with magnetic field}

Let us assume the $G$-action on
$\mathscr{X}$ preserves a closed $2$-form ${\CalB} \in {\Omega}^{2}({\mathscr{X}})$: 
\beq
{\delta} {\CalB} = 0\, , \ {\CalB} = \frac 12 B_{ij}(x) {\delta}x^i \wedge {\delta}x^j  \ .
\eeq
We further assume the group action
 to be $\CalB$-Hamiltonian (we do not assume $\CalB$ to be non-degenerate): for any ${\xi} \in \mathfrak{g}$,  
\beq
\iota_{V_{\xi}} {\CalB} = - {\delta}m_{\xi}\, , \ m_{\xi} \in {\sf C}^{\infty} ({\mathscr{X}})
\eeq
We also assume that there is a choice of constants in  defining $m_{\xi}$ such that the map ${\bf m} : \mathscr{X} \to {\mathfrak{g}}^{*}$ is $G$-equivariant:
\beq
L_{V_{\xi}} m_{\xi'} = - L_{V_{\xi'}} m_{\xi} = m_{[{\xi},{\xi}']}
\eeq
For any $\xi \in \mathfrak{g}$ the vector field \eqref{eq:liftv} preserves the deformed symplectic form
\beq
{\Omega}_{\CalB} = {\delta} p_i \wedge {\delta} x^i + \frac{1}{2} B_{ij}(x) {\delta} x^i \wedge {\delta}x^j
\label{eq:omgk}
\eeq
with the deformed moment map given by
\beq
{\boldsymbol{\tilde\mu}}_{\xi} = {\boldsymbol{\mu}}_{\xi} + m_{\xi}
\label{eq:mudef}
\eeq
obeying
\beq
\left\{ {\boldsymbol{\tilde\mu}}_{\xi} , {\boldsymbol{\tilde\mu}}_{\xi'} \right\}^{{\CalB}} = {\boldsymbol{\tilde\mu}}_{[{\xi}, {\xi}']}
\eeq
where
\beq
\{ p_{i}, x^{j} \}^{{\CalB}} = {\delta}_{i}^{j} \, , \, \{ x^i , x^j \}^{\CalB} = 0 \, , \, \{ p_i , p_j \}^{\CalB} = B_{ij}
\label{eq:pbk}
\eeq
So, \emph{the old symmetry preserving the deformed symplectic structure defines the deformed moment map} and the deformed \emph{coisotropic} submanifold 
\beq
{\tilde{\CalC}} = \left\{ \, (x,p) \, | \,  {\boldsymbol{\tilde\mu}} (x,p) = 0 \, \right\}
\label{eq:ncois}
\eeq
Now we need to understand the fate of the Lagrangian submanifold ${\CalL}$, which, for $\CalB = 0$ we defined
using the generating function $S$.
The (local) functions 
\beq
{\sigma}_{i}(x,p) \equiv p_{i} - {\partial}_{i}S \, , \ i = 1, \ldots, {\rm dim}({\mathscr{X}}) 
\eeq
no longer Poisson commute
\beq
\{ {\sigma}_{i}, {\sigma}_{j} \}^{\CalB} =  B_{ij}
\eeq
thus the equations ${\boldsymbol{\sigma}} = ({\sigma}_i) = 0$ no longer form first-class constraints, so ${\boldsymbol{\sigma}}^{-1}(0)$ is not Lagrangian. Suppose $\CalB$ is exact
\beq
{\CalB} = {\delta}{\CalA} \, , \ B_{ij} = {\partial}_{i} A_{j} - {\partial}_{j} A_{i}
\eeq
with some $1$-form 
\beq
{\CalA} = A_{i}(x) {\delta}x^i \ . 
\eeq
Then, 
\beq
{\sigma}_{i}(x,p) + A_{i} (x) = 0
\eeq
are first-class constraints, defining the deformed family of Lagrangians
\beq
{\CalL}_{\bf t}^{({\CalB})} = \left\{ \, (x,p) \, | \, p = - {\CalA} + {\delta}S\, \right\}
\label{eq:deflbtk}
\eeq
The ${\CalB}$-deformed intersection problem 
\beq
{\Phi}_{{\CalB}} \in {\tilde{\CalC}} \cap {\CalL}^{(B)}
\eeq
is thus solved by the deformed system of equations: for any $\xi \in \mathfrak{g}$
\beq
V^{i}_{\xi}{\partial}_{i} S = A_{i} V^{i}_{\xi} - m_{\xi} \ .
\label{eq:newx}
\eeq
The difference between \eqref{eq:oldx} and \eqref{eq:newx} is
the \emph{rocket} term
\beq
{\alpha}_{\xi} = - \iota_{V_{\xi}} {\CalA} + m_{\xi}
\label{eq:rocket}
\eeq
We have:
\beq
{\delta}{\alpha}_{\xi} = - L_{V_{\xi}} {\CalA}
\eeq
Of course, $\CalA$ is not uniquely defined by $\CalB$, as one can shift ${\CalA} \to {\CalA} + {\delta}{\phi}$, with ${\phi} = {\phi}(x)$ a function on $\mathscr{X}$. Such a shift
is equivalent to changing the original generating function $S \mapsto S - {\phi}$. Thus, the term \eqref{eq:rocket}
can be eliminated by the gauge transformation of $\CalA$ if for any $\xi \in {\mathfrak{g}}$ the function ${\alpha}_{\xi}$ can be represented as $L_{V_{\xi}} {\phi}$, for some $\phi \in {\sf C}^{\infty}({\mathscr{X}})$.

In the context of field theory, the choices of $\phi$ are constrained by the requirements of locality, general covariance etc.

\subsection{Anomaly in hydrodynamics}
\label{sec:anomhydro}

The abstract discussion above refers to the problem of interest
as (\ref{eq:pxhydro}, 
\ref{eq:gachydro}), with 
\beq
\left( x = ({\bf p}, {\bf S})\, | \,   p = ({\bf n}, {\boldsymbol{\nu}}) \right) \, , 
\eeq
and 
the closed $2$-form given by:
\beq
{\CalB} = \frac{k}{2} \int_{M^4} d{\bf p} \wedge {\delta} {\bf p} \wedge {\delta} {\bf p}\,,
\label{eq:bhydro}
\eeq
where the level $k$ is a parameter. We refer to this modification as a \emph{current anomaly}.
The two-form $\CalB$ is exact
\beq
{\CalB} = {\delta}{\CalA}\, , \ {\CalA} = \frac{k}{2} \int_{M^4}{\bf h} \wedge {\delta} {\bf p}\,,
\eeq
where we used 
 the helicity 3-form ${\bf h}={\bf p}\wedge {\bf d p}$ from \eqref{eq:helicity}. 

The corresponding components of the moment map $\boldsymbol{m} = (m_{\sf Diff}, m_{\sf C^{\infty}})$ are given by, e.g. \cite{Fock:1991dt}:
\beq
m_{\sf Diff}=-\frac{k}{2}\, {\bf p} \otimes   {\bf d}{\bf h}  \, , \ m_{\sf C^{\infty}} = -\frac{k}{2}\, {\bf d}{\bf h} 
\eeq
Note that the one-form $\CalA$ is ${\sf Diff}(M^4)$-invariant
but not ${\sf C}^{\infty}$-invariant. 
This conclusion agrees with  \cite{Wiegmann:2024sqh}.

{}Explicitly,
\beq
{\tilde\mu}_{{\sf C}^{\infty}} = {\bf d} {\bf n} - \frac{k}{2}  {\bf d}{\bf h} \, .
\label{eq:anomab1}
\eeq
The deformed coisotropic subvariety ${\tilde\CalC}$, cf. \eqref{eq:mudef} is now defined by ${\tilde\mu}_{{\sf C}^{\infty}} = 0$,   hence
\beq
d {\bf n} = \frac{k}{2}  {\bf d}{\bf h}\ .
\label{eq:anomab}
\eeq
Define
\beq 
{\bf n}_k= {\bf n} - k {\bf h} 
\eeq
the ${\sf C}^{\infty}(M^4)$-invariant 3-form representing the particle flux. Then \eqref{eq:anomab} 
reads
\beq
{\bf d} {\bf n}_k= -\frac{k}{2} {\bf  dh}\,.\label{6.38}
\eeq
Thus, the coisotropic subvariety ${\tilde\CalC}$ corresponding to the level $k$ is
the space of $({\bf S}, {\bf p}, {\bf n}, {\boldsymbol{\nu}})$ solving
\beq
\iota_{{\bf n}_k^\vee} {\bf d}{\bf p}={\bm \nu}\otimes {\bm d}{\bf S} -  {\mathfrak{r}}_{k} 
\label{eq:kcois}
\eeq
with the \emph{rocket term} $\mathfrak{r}_k$ given by
\beq
{\mathfrak{r}}_{k} = {\bf p}\,\otimes  {\bf dn}_k \,.
\eeq
Using the identity 
\beq
{\bf p }\otimes {\bf dh}=2\iota_ {\bf h^\vee} ({\bf dp})\ , 
\eeq
where ${\bf h}^\vee$ 
is the helicity 3-form  ${\bf h}={\bf p}\wedge {\bf d}{\bf p} $ viewed as a  vector field, 
we can hide the rocket term by redefining velocity. Then the Eq. \eqref{eq:kcois} remains  Hamiltonian
\beq
\iota_{{\bf V}_k} {\bf d}{\bf p}={\bf d}{\bf S} 
\eeq
but now with the velocity ${\bf V}_k$
defined by 
\beq
\iota_{{\bf V}_k} \boldsymbol{\nu} = {\bf n}_{k} +  k{\bf h}\,.
 \label{6.44}
 \eeq
 The vector field ${\bf V}_k$  continues to advect the entropy as 
 \beq
 L_{{\bf V}_k}{\bf S}=0\,.
 \eeq
Hence, the second law of thermodynamics and Kelvin's circulation theorem continue to hold.
We may therefore view the anomaly as a modification of the hydrodynamic flow which preserves these fundamental properties while allowing for particle production, as in \eqref{6.38}. The price one pays is that the velocity ${\bf V}_k$ defined by \eqref{6.44} is no longer collinear with the particle flux ${\bf n}_k$:
\beq
\iota_{{\bf V}_{k}} {\bf n}_{k}\neq 0\, .
\eeq

\section{Generalizations and related topics}\label{sec:gut}

Our formalism allows several immediate generalizations. Some are briefly  listed in this section. 
\subsection{Euler-Maxwell fluid system}

Consider a charged fluid interacting with an electromagnetic field, itself governed by Maxwell's equations with sources. This provides, for example, a limiting description of a poorly conducting charged fluid, usually referred to as the Euler-Maxwell system.

We start by defining the extended phase space 
\beq
{\widetilde{\mathscr P}}_{M^4} = \left\{ \left( {\bf S}, {\bf p}, {\bf B}, {\bf F}, {\bf n}, {\boldsymbol{\nu}} \right) \right\} = {\Omega}^{\bullet}(M^4) \oplus {\Omega}^{2}(M^4)
\label{eq:allforms}
\eeq
where we added a pair $({\bf B}, {\bf F})$ of two-forms on $M^4$ to the quadruple $({\bf S}, {\bf p}, {\bf n} , {\boldsymbol{\nu}})$ defined in \eqref{eq:forms}. The functional dimension of ${\widetilde{\mathscr P}}_{M^4}$
is equal to $22$. We endow ${\widetilde{\mathscr P}}_{M^4}$ with the symplectic form
\beq
{\Omega}_{{\widetilde{\mathscr P}}_{M^4}} = \int_{M^4} {\delta} {\bf S} \wedge {\delta} {\boldsymbol{\nu}}
 + {\delta}{\bf p} \wedge {\delta}{\bf n} + {\delta} {\bf B} \wedge {\delta} {\bf F}
 \label{eq:omgtildep}
 \eeq
 We now define the larger group ${\widetilde{\CalG}}_{M^4} = {\sf Diff}(M^4) \sdtimes {\Omega}^{1}(M^4) \oplus B^{2}(M^4)$: the group of diffeomorphisms, extended by the abelian additive group
of one-forms and exact $2$-forms. The latter is isomorphic to ${\Omega}^{1}(M^4)/Z^1(M^4)$, the space of $1$-forms modulo
closed $1$-forms. The functional dimension of this group is $4+4+3 = 11$, i.e. half of the functional dimension of ${\widetilde{\mathscr P}}_{M^4}$. Let us now define the action of ${\widetilde{\CalG}}_{M^4}$
on ${\widetilde{\mathscr P}}_{M^4}$. The diffeomorphisms act in the natural way, while the abelian group acts as follows: for $a \in {\Omega}^{1}(M^4), b \in {\Omega}^1(M^4)$ set
\beq
(a,b) \cdot \left( {\bf S}, {\bf p}, {\bf B}, {\bf F}, {\bf n}, {\boldsymbol{\nu}} \right) =
\left( {\bf S}, {\bf p}+a, {\bf B}+{\bf d}b, {\bf F}+da, {\bf n}, {\boldsymbol{\nu}} \right)
\label{eq:bbaction}
\eeq
The action  \eqref{eq:bbaction} preserves \eqref{eq:omgtildep} and comes with the moment map:
\beq
\begin{aligned}
& {\mu}_{{\Omega}^{1}} = {\bf n} - {\bf d B}\, , \\
& {\mu}_{{\Omega}^{1}/Z^{1}} = {\bf d F}
\end{aligned}
\label{eq:maxwell}
\eeq
The action of ${\sf Diff}(M^4)$ also preserves \eqref{eq:omgtildep}, with the corresponding moment map
\beq
{\mu}_{\sf Diff} = -{\bf dS} \otimes {\boldsymbol{\nu}} + {\bf p} \otimes {\bf dn} + 
\iota_{{\bf n}^{\vee}} {\bf dp} + \iota_{{\bf dB}^{\vee}} {\bf F} - \iota_{{\bf dF}^{\vee}} {\bf B}
\eeq
Setting ${\boldsymbol{\mu}} : = \left( {\mu}_{\sf Diff} , {\mu}_{{\Omega}^{1}}, {\mu}_{{\Omega}^{1}/Z^{1}} \right)=0$ we get
\beq
 \iota_{\bf V} ( {\bf dp + F} )={\bf dS}  \, , 
\label{eq:Lichnem}
\eeq
with the $\iota_{\bf V}{\bf F}$ term representing the Lorentz force, 
as well as
\beq
{\bf n }= {\bf d B}\, , 
\label{eq:maxwell1}
\eeq
and
\beq
{\bf dF} = 0
\label{eq:maxwell2}
\eeq
The next step is to introduce the Lagrangian submanifold ${\CalL}$, which depends on the spacetime metric $\bf g$ and is characterized by the energy density $e$. The minimal choice would be $e=e(f, n, S)$, so that the generating functional in the $({\bf S}, {\bf n}, {\bf F} | {\bf B}, {\bf p} , {\boldsymbol{\nu}})$-polarization reads as
\beq
{\CalA} \left( {\bf S}, {\bf n}, {\bf F} \right) = -\int_{M^4} e \left( \frac{{\bf F} \wedge \star_{\bf g} {\bf F}}{2{\rm vol}_{\bf g}} \, , \, {\bf n}_{\bf g}\, , \, {\bf S} \right) \, {\rm vol}_{\bf g}
\eeq
Thus, on $\CalL$, 
\beq 
{\bf B}\, =\, {\mu}_{\sf em}\star_{\bf g} {\bf F}
\eeq
where $\mu_{\sf em}= -\frac{\pa e}{\pa f}$ is a permeability, 
so that \eqref{eq:maxwell1} and \eqref{eq:maxwell2} become Maxwell's equations with $\bf n$ being  the electric current 3-form.
There exist more general forms of energy density, including, e.g.
$e=e(f_{+}, f_{-} , {\bf n}_{\bf g} , {\bf S})$, with
\beq
f_{+} = \frac{{\bf F} \wedge \star_{\bf g} {\bf F}}{2{\rm vol}_{\bf g}}\, , \ f_{-} = \frac{{\bf F} \wedge  {\bf F}}{2{\rm vol}_{\bf g}}
\eeq
The charge conservation law ${\bf dn} =0$ follows from \eqref{eq:maxwell1} \footnote{The symmetry \eqref{eq:bbaction} forbids anomalous magnetic terms such as
$$
\int_{M^4} {\bf dp} \wedge {\delta}{\bf p} \wedge {\delta}{\bf p}
$$
in the symplectic form, unlike the symmetry considered in Section \ref{sec:anomhydro}.}. 

\subsection{Superfluid} A superfluid provides another generalization. In this case,
we consider another extended phase space 
\beq
{\widetilde{\mathscr P}}_{M^4} = \left\{ \left( {\bf S}, {\boldsymbol{\theta}}, {\bf p}, {\bf n}, {\boldsymbol{\nu}} , {\boldsymbol{\Gamma}} \right) \right\} = {\Omega}^{0  \oplus 1 \oplus 3 \oplus 4 \oplus 4}(M^4) \times {\sf C}^{\infty}( M^4, S^1 )
\label{eq:001344forms}
\eeq
where the variable ${\boldsymbol{\theta}}$, called the {\it superfluid phase}, cf. \cite{Volovik}, is periodic and describes the map $M^4 \to S^1$. We endow the phase space 
with the symplectic form
\beq
{\Omega}_{{\widetilde{\mathscr P}}_{M^4}} = \int_{M^4} {\delta} {\boldsymbol{\theta}} \wedge {\delta} {\boldsymbol{\Gamma}} + {\delta} {\bf S} \wedge {\delta} {\boldsymbol{\nu}} + {\delta} {\bf p} \wedge {\delta} {\bf n}
\eeq
and the symmetry ${\CalG}^{(2)}_{M^4}$ acting via the usual diffeomorphisms and 
\beq
(a, b) \cdot \left( {\bf S}, {\boldsymbol{\theta}}-a, {\bf p}+da+{\bf S} db , {\bf n}, {\boldsymbol{\nu}} - {\bf n} \wedge db , {\boldsymbol{\Gamma}} \right)
\eeq
for $a,b \in {\sf C}^{\infty}(M^4)$. The moment map equations defining ${\CalC}$ read
\beq
{\mu}_{\sf Diff} = -{\bf dS} \otimes {\boldsymbol{\nu}} + {\bf p} \otimes {\bf dn} + 
\iota_{{\bf n}^{\vee}} {\bf dp} + {\bf d}{\boldsymbol{\theta}} \otimes {\boldsymbol{\Gamma}} = 0
\eeq
as well
 as
 \beq
{\bf dn} = {\boldsymbol{\Gamma}}\, , \ {\bf d} ({\bf S}{\bf n} ) = 0
 \eeq
 which imply
 \beq
 {\bf dS} = \iota_{\bf V} {\bf dp} + \left( \frac{\boldsymbol\Gamma}{\boldsymbol\nu} \right) \, \left( {\bf p} + {\bf d} {\boldsymbol{\theta}} \right)
 \eeq
 as well as
 \beq
 L_{\bf V} {\boldsymbol{\theta}} = - {\boldsymbol{\zeta}}
 \eeq

\subsection{Particle creation}
\label{sec:partcreation}

Let us turn to a geometric mechanism of particle creation, arising from the topology of the vorticity two-form.

In the discussion of anomalous fluid dynamics of Section \ref{sec:anomhydro}
the Gauss law for the abelian ${\sf C}^{\infty}(M^4)$ -factor of ${\CalG}^{(1)}_{M^4}$ yields a modification of the continuity equation
\beq
{\bf dn}_k + \frac{k}{2} {\bf dp} \wedge {\bf dp} = 0\,.
\label{eq:gaussanom}
\eeq
Eq. \eqref{eq:gaussanom}
states that the number of particles in a volume $U^3$, as measured by the integral
$$ N(U^3) = \int_{U^3} {\bf n}_k $$
can change if $U^3$ crosses a \emph{vorticity instanton}, a point of a self-intersection of the vortex spacetime surfaces, Poincar{\'e} dual to a closed vorticity 2-form ${\bf dp}$. This produces a nontrivial effect when ${\bf dp}$ is closed but not exact.

This requires a modification of the formalism, in which we relax the global well-definiteness of $\bf p$. We now describe two such modifications. 

\subsection{Vortex spacetime surfaces}
\label{sec:vs}

One possibility  is to allow for codimension-two singularities
\beq
{\bf dp} = \sum_{l=1}^{n} {\Gamma}_{l} \, {\delta}^{(2)}_{\Sigma_{l}}
\label{eq:sing}
\eeq
where ${\delta}^{(2)}_{\Sigma}$ is a two-form with support on a 2-surface $\Sigma$, which we assume smooth. We can either think of $\Sigma$'s
as fixed. In this case the space of fields can remain the same, up to a redefinition of $\bf p$, but the group ${\CalG}^{(1)}_{M^4}$ is reduced (so that the diffeomorphisms preserve the collection ${\Sigma}_{1}, \ldots, {\Sigma}_{n}$ of surfaces). 

If ${\Sigma}_{1}, \ldots, {\Sigma}_{n}$ are not fixed and their spacetime position is determined by the fluid dynamics, then we modify 
\beq {\mathscr{P}}_{M^4} \to {\widetilde{\mathscr{P}}}_{M^4} = \bigsqcup_{n=0}^{\infty} \ {\widetilde{\mathscr{P}}}_{M^4} [n] \eeq 
by replacing the set of smooth fields $({\bf S}, {\bf p}, {\bf n}, {\boldsymbol{\nu}})$ defined on $M^4$, by the set
of fields, smooth on $M^4 \backslash \left( {\Sigma}_{1} \cup \ldots \cup {\Sigma}_{n} \right)$, for some $n$ and some two dimensional surfaces ${\Sigma}_{1}, \ldots, {\Sigma}_{n}$. The space ${\widetilde{\mathscr{P}}}_{M^4} [n]$ is endowed with the closed two-form:
\begin{multline}
{\Omega}_{{\widetilde{\mathscr{P}}}_{M^4} [n]} = {\Omega}_{{\mathscr{P}}_{M^4}} + \\
\frac{k}{2} \sum_{l=1}^{n} \int_{\Sigma_{l}} \left( \iota_{{\delta}{\sigma} \wedge {\delta} {\sigma}} \left( d{\bar {\bf p}} \wedge d{\bar {\bf p}} \right) + 2\left( \iota_{{\delta}{\sigma}} d{\bar{\bf p}} \right) \wedge {\delta}{\bar{\bf p}} + {\delta}{\bar{\bf p}} \wedge {\delta}{\bar{\bf p}} \right) \, 
\end{multline}
where ${\delta}{\sigma} \in {\Gamma}\left( {\Sigma}_{l}, TM^4/T{\Sigma}_{l} \right)$ represent the deformations of ${\Sigma}_{l} \subset M^4$, and ${\bf {\bar p}}$ is the smooth part of ${\bf p}$, 
$$ {\bf {\bar p}} = {\bf p} - \sum_{l} \frac{\Gamma_{l}}{2\pi} d{\varphi}_{l} $$ 
where ${\varphi}_{l}$ are the angular coordinates on $M^4$, defined locally near $\Sigma_l$'s.
We note that the $\Sigma_l$'s contribution to the symplectic form is similar in form to Weinstein's symplectic form on the space of unparametrized loops in a $3$-manifold endowed with the volume form: a coadjoint orbit of the group of volume-preserving diffeomorphisms \cite{ArnoldKhesin}. In our case, the
$3$-manifold has become $M^4$, the unparametrized loops are the submanifolds ${\Sigma}_{1}, \ldots, {\Sigma}_{n}$, and the role of the volume form is played by 
$\frac{k}{2} {\bf dp}\wedge {\bf dp}$. 
This spacetime symplectic structure should be compared with the
Hamiltonian structures on vortex filaments and vortex membranes studied in 
\cite{Khesin}. In the localized induction approximation, the latter lead to the binormal,
or more generally skew mean-curvature, flow of codimension-two vortex membranes.

The anomalous particle creation can now take place:
\beq
{\Delta} N(U^3) \, = \, k \sum_{a, b}  \Gamma_{a} \Gamma_{b}\, 
{\#}  {\Sigma}_{a} \cap {\Sigma}_{b} \cap  U^3 
\eeq

\subsection{Connections on gerbes and Onsager quantization}

Another possible generalization of our formalism is to allow $\bf p$ and $\bf n$ to be locally, but not necessarily globally, well-defined. In this way they become connections. The simplest such generalization is to view $\bf p$ as a connection $1$-form. This means that we take $M^4$ to be the base of a principal $G$-bundle $P$, with $G = {\BR}$ or $G = U(1)$.
The space $P$ carries a free $G$-action, with $M^4 = P/G$. The form $\bf p$ is then a $G$-invariant $1$-form on $P$,
\beq
{\bf p} \in {\Omega}^{1}(P)^{G}
\eeq
obeying 
\beq
\iota_{U} {\bf p} = 1\, , \ L_{U} {\bf p} = 0\, , 
\eeq
where $U \in {\sf Vect}(P)$ is the vector field generating the $G$-action. The group ${\CalG}^{(1)}_{M^4}$ generalizes to the group ${\sf Aut}_{G}(P)$ of diffeomorphisms of $P$, commuting with the $G$-action. This group fits into the exact sequence:
\beq
1 \longrightarrow {\CalG}(P) \longrightarrow {\sf Aut}_{G}(P) \longrightarrow {\sf Diff}(M^4)_{[P]} \longrightarrow 1
\eeq
where ${\CalG}(P)$ is the gauge group, 
\beq
{\CalG}(P) = {\Gamma}\left( M^4, P \times_{\rm Ad} G \right)\, , 
\eeq
and ${\sf Diff}(M^4)_{[P]} \subset {\sf Diff}(M^4)$ is the subgroup of all diffeomorphisms preserving all the Chern classes of $P$. In our paper we only discuss diffeomorphisms in the connected component of the identity. All of them belong to ${\sf Diff}(M^4)_{[P]}$:
\beq
{\sf Diff}(M^4)_{0} \subset
{\sf Diff}(M^4)_{[P]}
\eeq
Although ${\bf p}$ lives on $P$, its variation ${\delta}{\bf p}$ lives on $M^4$, since the variation is horizontal and invariant. Thus the terms $\int_{M^4} {\delta}{\bf p} \wedge {\delta} {\bf n}$ and $\int_{M^4} {\bf dp} \wedge {\delta}{\bf p} \wedge {\delta}{\bf p}$ continue to make sense in the connection setting. The novelty is that ${\bf dp}$ is now a closed $2$-form on $M^4$, but for $G = U(1)$ it need not be exact; consequently, the level $k$ becomes an observable integer.

The dual construction, in which $\bf n$ is promoted to the analogue of a connection, involves $2$-gerbes over $M^4$. 
We will not discuss it further, except to mention that it naturally leads to
Onsager quantization of circulation  \cite{Onsager}. The latter reads 
\beq
\Gamma_{l} = 2\pi \hbar n_l\, , \ n_l \in {\BZ}
\eeq
In this formulation, the vortex spacetime surfaces become analogous to gauge-theory surface defects, appearing in the BPS/CFT correspondence \cite{NN2004, NekBPSCFT}, cf. Section \ref{sec:BPSCFT}.

{}
When the Onsager quantization condition
 is imposed, the value of the level $k$ is quantized in units of the Planck constant. 

 When both $\bf p$ and $\bf n$ are connections on gerbes, there is another quantized integer, the coefficient in front of the 
 \beq
 \int {\delta}{\bf p} \wedge {\delta}{\bf n}
 \eeq
 term in the symplectic form. One can then expect some braiding relations between the
 line operators
 \beq
 e^{{\kappa} \oint {\bf p}}
 \eeq
 associated to loops and membrane operators
 \beq
 e^{{\kappa}' \oint {\bf n}}
 \eeq
 associated to $3$-dimensional closed submanifolds.

\subsection{Five-dimensional formalism}

Let $N^5$ be a five-dimensional closed manifold. Define
\beq
{\CalR}_{N^5} = {\Omega}^{1 \oplus 4} (N^5) = \Biggl\{ ( {\bf P}, {\bf N} ) \, | \, {\bf P} \in {\Omega}^{1}(N^5)\, , \ {\bf N} \in {\Omega}^{4}(N^5) \Biggr\}
\eeq
It is again a symplectic manifold with the natural action of the group 
\beq
{\CalG}_{N^5} = {\sf Diff}({N^5})\ .
\eeq
The moment map for the natural ${\CalG}_{N^5}$-action on ${\CalR}_{N^5}$ is given by:
\beq
{\mu}_{{\CalG}_{N^5}} = {\bf P} \otimes {\bf D N} + \iota_{{\bf N}^{\vee}} {\bf D P}
\eeq
where ${\bf N}^{\vee}$ is the same thing as ${\bf N}$ but viewed as a section of
$Vect(N^5) \otimes {\Omega}^{5}(N^5)$. Setting ${\mu}_{{\CalG}_{N^5}} = 0$ implies, away from the hypersurface ${\bf N} \wedge {\bf P} = 0$, that
\beq
{\bf D}{\bf N} = 0\, , \ \iota_{\CalV} {\bf DP} = 0\, , \ \iota_{\CalV} {\bf N} = 0
\label{eq:5moment}
\eeq
where we denote the de Rham differential on $N^5$ by $\bf D$ in order not to confuse it with the four-dimensional de Rham differential $\bf d$ acting on forms on $M^4$ to be defined below. The vector field ${\CalV} \in {\sf Vect}(N^5)$ in \eqref{eq:5moment} is simply ${\bf N}^{\vee}$ divided by some non-zero $5$-form. ${\CalV} $ is not uniquely defined; it can be multiplied by any non-zero function on $N^5$. Define  
\beq
{\CalC}_{N^5} = {\mu}_{{\CalG}_{N^5}}^{-1}(0) \ , 
\label{eq:coisn}
\eeq
which is again coisotropic. 
Given any metric $\bf G$ on $N^5$ and a reasonable function ${\ve} = {\ve}(n)$, $n \in {\BR}$, 
with Legendre transform 
\beq
{\mathsf{p}}(p) = n {\partial}_{n} {\ve} - {\ve}\ ,
\label{eq:5dpressure}
\eeq
$p = {\partial}_{n}{\ve}$,  we can define a Lagrangian ${\CalL}_{{\ve},{\bf G}} \subset {\CalR}_{N^5}$ by the generating function 
\beq
{\CalE}_{{\ve},{\bf G}}  ({\bf N}) = \int_{N^5} {\ve} \left( \sqrt{\frac{{\bf N} \wedge \star_{\bf G} {\bf N}}{{\rm vol}_{\bf G}}} \right) \, {\rm vol}_{\bf G}
\label{eq:genfunn}
\eeq
in the $({\bf N}|{\bf P})$ polarization, or
\begin{multline}
{\tilde\CalE}_{{\mathsf{p}},{\bf G}}  ({\bf P}) = \int_{N^5} {\bf P} \wedge {\bf N} - {\CalE} ({\bf N}) = \int_{N^5} {\mathsf{p}} \left( \sqrt{\frac{{\bf P} \wedge \star_{\bf G} {\bf P}}{{\rm vol}_{\bf G}}} \right) \, {\rm vol}_{\bf G} \, , \\
{\bf P} = \frac{\delta {\CalE}_{{\ve},{\bf G}}}{\delta {\bf N}} = \frac{1}{n} {\partial}_{n}{\ve} \star_{\bf G} {\bf N}\, ,
\end{multline}
in the $({\bf P} | {\bf N} )$ polarization, with $\mathsf{p}$, defined by the Eq. \eqref{eq:5dpressure}, playing the role of the pressure. 

{}With ${\CalL}_{{\ve},{\bf G}}$ in place we define 
\beq
{\mathsf{5d\ flows}}\ = \ {\CalC}_{N^5} \cap {\CalL}_{{\ve},{\bf G}}  \subset {\CalR}_{N^5}
\eeq
In analogy with \cite{Bilal:1990wn}, we can impose the moment map equations for ${\CalG}_{N^5}$ but use additional geometric structures on $N^5$ in defining the Lagrangian submanifold ${\CalL}$. For example, assume we have a distinguished vector field $\bf K \in {\sf Vect}(N^5)$ (this is similar to a choice of parabolic subgroup in $SL(n, {\BC})$ in \cite{Bilal:1990wn} in defining the $W$-projective structures on a Riemann surface). 

Having an extra structure such as $\bf K$ we can work with more general Lagrangians, in other words, less symmetric equations of state. For example, given a reasonable function ${\ve}(n,s)$ of two variables, with the partial Legendre transform
${\mathsf{P}}(W,S) = n {\partial}_{n} {\ve} - {\ve}$, $W = {\partial}_{n}{\ve}$, we define
\beq
{\CalP}_{\mathsf{p}, {\bf G}, {\bf K}}  ({\bf N}) = \int_{\CalN}\, {\mathsf{p}}( p_{\bf g, \bf K}, \iota_{\bf K} {\bf P} ) \, {\rm vol}_{\bf G}
\eeq
where we assume ${\bf G}$ to be $K$-invariant, and
\beq
p_{\bf g, \bf K} = \sqrt{\frac{{\bf P} \wedge \star_{\bf G} {\bf P} - {\bf G}({\bf K}, {\bf K}) (\iota_{\bf K} {\bf P} )^2}{{\rm  vol}_{\bf G}} }
\eeq

\subsubsection{The uses of five dimensions}

Suppose $N^5 = M^4 \times {\BR}$ with the $\BR$ factor parametrized by the $\theta$ coordinate. 
We can write
\beq
{\bf P} = {\bf p} + {\bf S} d\theta \, , \ {\bf N} = {\boldsymbol{\nu}} + {\bf n} \wedge d\theta
\eeq
where for each $\theta$, $( {\bf S} , {\bf p}, {\bf n}, {\boldsymbol{\nu}}) \in {\mathscr{P}}_{M^4}$.  In other words the $({\bf P}, {\bf N}) \in {\CalP}_{N^5}$ is a path in ${\mathscr{P}}_{M^4}$ parametrized by $\theta$. Moreover, the  flows solving \eqref{eq:5moment}
obey
\beq
\frac{\partial}{\partial \theta} \left( \begin{matrix} {\bf p} \\  {\boldsymbol{\nu}} \end{matrix} \right)  = \left( \begin{matrix} {\bf dS} + \iota_{\bf V} {\bf dp} \\  {\bf dn} \end{matrix} \right) \ , 
\label{eq:euler5}
\eeq
where we normalized the representative  for $\CalV$ by $d{\theta} ({\CalV}) = 1$:
\beq
{\CalV} = \frac{\partial}{\partial \theta} - {\bf V}
\eeq
with some $\theta$-dependent ${\bf V} \in {\sf Vect}(M^4)$, still 
related to $\bf n$ and
$\boldsymbol{\nu}$ via \eqref{eq:flow2}. The substitution
${\Pi} = {\bf p} \otimes {\boldsymbol{\nu}}, {\rho} = {\boldsymbol{\nu}}$ maps \eqref{eq:euler5} to the equation
\beq
{\partial}_{\theta} {\xi} = {\mu}
\label{eq:5dto4d}
\eeq
for $$ {\xi} = \left( {\Pi}, {\rho} \right) \in  {\sf Lie}\left(  {\CalG}^{(1)}_{M^4} \right)^{*} $$
and
$$ {\mu} = {\mu}_{\sf Diff} \oplus {\mu}_{{\sf C}^{\infty}} : {\mathscr{P}}_{M^4} \to {\sf Lie}\left(  {\CalG}^{(1)}_{M^4} \right)^{*}$$ defined in \eqref{eq:momg1}, \eqref{eq:momdiff4}.
Although Eqs. \eqref{eq:5dto4d} are not the spinning-top equations for the four-dimensional Novikov group ${\CalG}^{(1)}_{M^4}$, the solutions to \eqref{eq:5dto4d} which stay within ${\CalL}$ are also solutions to the Euler-Poincar{\'e}-Arnold equations.
Indeed, imposing that the path remains in ${\CalL}$ converts the moment-map evolution in \eqref{eq:5dto4d} into the Euler-Poincaré-Arnold evolution associated with the induced Hamiltonian on the dual of ${\CalG}^{(1)}_{M^4}$. In this sense, the five-dimensional construction gives a parent system whose constrained trajectories reproduce the four-dimensional top.

But what if we stayed in four dimensions? One option is to declare the $\theta$-independence. 
A physical way to do so is to first compactify $N^5 = M^4 \times S^1$, then send the circumference of $S^1$ to zero. 

{} 
The compactification route opens new possibilities. One can study the five-dimensional manifolds $N^5$ which nontrivially fiber over $M^4$ with generic $U(1)$ fibers. Let $\bf K$ denote the vector field generating the $U(1)$ action on $N^5$, and assume $\bf G$ is $U(1)$-invariant. 

The $1$-form ${\bf P}$ on $N^5$ can then be decomposed as
\beq
{\bf P} = {\bf S} \, {\Theta} + {\bf p}
\eeq
where $\iota_{\bf K} {\Theta} = 1$, $\iota_{\bf K} {\bf p} = 0$, where
\beq
{\Theta} = \frac{{\bf G} ( {\bf K} , {\cdot} )}{{\bf G}({\bf K}, {\bf K})}
\eeq
is the connection $1$-form on $N^5$ (not on $M^4 = N^5/U(1)$!), well-defined outside the zeroes of $\bf K$. It obeys
\beq
L_{\bf K} {\Theta} = 0 \ .
\eeq
Likewise, 
\beq
{\bf N} =  {\Theta} \wedge  {\bf n}+ {\boldsymbol{\nu}}
\eeq
and 
\beq
{\CalV} = {\bf K} - {\bf V}
\eeq
We impose the $U(1)$-invariance:
\beq
L_{\bf K} {\bf N} = 0\, , \ L_{\bf K} {\bf P} = 0
\eeq
so that $({\bf S}, {\bf p}, {\bf n}, {\boldsymbol{\nu}}) \in {\mathscr{P}}_{M^4}$, and derive
\beq
{\bf dn} = 0\, , \ {\bf dS} + \iota_{\bf V} {\bf dP} = {\bf S} \left( \iota_{\bf V} {\bf F} \right) 
\label{eq:abelian}
\eeq
i.e., the curvature of the $N^5 \to M^4$ bundle exerts an additional force on the fluid. 

\subsubsection{Non-abelian spectators}

As explained in Section \ref{sec:phmng}, 
the group ${\CalG}^{(2)}_{B^3}$ is a particular case of a more general class of semi-direct product groups ${\CalG}_{B^3}^{G}$, based on the Lie group $G$. The corresponding
Euler-Arnold top describes the flow in three dimensions (cf. \cite{Bistrovic:2002jx,Jackiw:2004nm})
\beq
\begin{aligned}
& {\dot\Pi} + L_{\bf u} {\Pi} = - d{\mu}^{a} \otimes {\rho}_{a} \, \\
& {\dot\rho}_{a} + L_{\bf u} {\rho}_{a} + f_{ab}^{c} {\rho}_{c} {\mu}^{b} = 0
\end{aligned}
\label{eq:nonabelian}
\eeq
with ${\mathfrak{g}}^{*}$-valued density ${\boldsymbol{\rho}} = ( {\rho}_{a} ) \in {\Omega}^{3}(B^3)$, $a = 1, \ldots, {\rm dim}(G)$, 
and $\mathfrak{g}$-valued chemical potential ${\boldsymbol{\mu}} = ( \mu^{a} )$. The velocity vector field ${\bf u} \in {\sf Vect}(B^3)$
and the momentum per volume $\Pi \in {\Omega}^{1}(B^3) \otimes {\Omega}^{3}(B^3)$
are as in \eqref{eq:euler_eqs2}. The equation of state
determines the map ${\bf\Omega}: ({\boldsymbol{\rho}}, {\Pi}) \mapsto ({\boldsymbol{\mu}}, {\bf u})$ making \eqref{eq:nonabelian} a self-consistent system of evolution equations. For a simple Lie algebra $\mathfrak{g}$ the basic invariant polynomials $C_{1}, \ldots, C_{r} \in (S^{*}{\mathfrak{g}}^{*})^{G}$ of degrees
$d_1, \ldots, d_r$ define the generalized entropy
\beq
{\bf S} = \left( C_{1}({\boldsymbol{\rho}}):  \ldots :  C_{r} ({\boldsymbol{\rho}}) \right) \in W{\BP}^{d_1, \ldots, d_r}
\eeq
where $d_1 = 2$, 
which is transported passively
\beq
{\dot{\bf S}} + L_{\bf u} {\bf S}  = 0
\eeq
with ${\rho} = \left( C_{1}({\boldsymbol{\rho}}) \right)^{\frac{1}{2}}$ playing the role of the physical fluid density. 

Is there a four-dimensional covariant formulation? It would appear that
the higher dimensional version of the five dimensional formalism with $N^5$ replaced by
a $G$-bundle $X^{d}$ over $M^4$, with ${\bf P} \in {\Omega}^{1}(X^d)$ and ${\bf N} \in {\Omega}^{d-1}(X^d)$ should reduce to \eqref{eq:nonabelian} by an appropriate Kaluza-Klein reduction, generalizing the construction leading to \eqref{eq:abelian}. We leave a detailed analysis of this reduction to future work. 

\subsection{Barotropic models}
\label{sec:barotropic}

The paradigm of our paper is to realize the solution to the equations of motion as the intersection ${\CalC} \cap {\CalL} \subset {\mathscr P}$ of a coisotropic variety defined by the symmetries, and a Lagrangian subvariety defined by the equation of state. 
When the symmetry group includes the group of spacetime diffeomorphisms, the corresponding system is called a {\it fluid}. 
The minimal realization has 
the phase space ${\mathscr P}_{M^4}^{\rm min}$:
\beq
{\mathscr P}_{M^4}^{\rm min} = {\Omega}^{1}(M^4) \times {\Omega}^{3}(M^4)
\eeq
with the symplectic form given by
\beq
{\Omega}_{k}^{\rm min} = \int_{M^4} {\delta}{\bf p} \wedge {\delta} {\bf n} + k \int_{M^4} {\bf dp} \wedge
{\delta}{\bf p} \wedge {\delta}{\bf p}
\eeq
The coisotropic manifold ${\CalC}^{\rm min}$ is defined as the zero locus of the moment map for ${\sf Diff}(M^4)$ cf. \cite{Abanov:2022zwm}
\beq
{\mu}_{\sf Diff} = \iota_{{\bf n}^{\vee}_{k}} {\bf dp} + {\bf p} \otimes {\bf dn}_{k} = 0
\label{eq:diffbt}
\eeq
with
\beq
{\bf n}_{k} = {\bf n} + \frac{k}{2} {\bf p} \wedge {\bf dp}
\eeq
Equation \eqref{eq:diffbt} implies ${\bf dp}$ is degenerate,
\beq
{\bf dp} \wedge {\bf dp} = 0\, ,
\eeq
thus the helicity is a constant and  ${\bf dn} = 0$. Moreover, ${\bf dp}$ shares its kernel with ${\bf n}_{k}$:
\beq
{\bf dn}_{k} = 0\, , \ \iota_{\bf V} {\bf dp} = 0
\eeq
for any $\bf V$, such that $\iota_{\bf V} {\bf n}_{k} =0$. In the absence of $\boldsymbol{\nu}$, we cannot normalize the distribution of kernels to define the canonical vector field ${\bf V}_{k}$, but we can still define the flow lines. 
The same formalism works in $1+1$ dimensions, with $\bf n$ being a $1$-form, just like $\bf p$ is. 

The barotropic model described above embeds into our more general model with ${\mathscr P}_{M^4} = \{ ( {\bf S}, {\bf p}, {\bf n}, {\boldsymbol{\nu}} ) \}$ as the flows with vanishing Ertel scalar ${\mathsf{Q}} = 0$. 
Among those there are the isentropic flows, where ${\bf dS} = 0$, 
and the potential, or irrotational flows, where ${\bf dp} = 0$, e.g. ${\bf p} = {\bf d}{\boldsymbol{\psi}}$, for some ${\boldsymbol{\psi}} \in {\sf C}^{\infty}(M^4)$.

\appendix
\section{Outlook material}

Here we briefly discuss topics both related to and extending beyond fluid dynamics. 
They all share the same geometry discussed in the main part of the paper.

\subsection{Six dimensional viewpoint and Poisson sigma-model}

Our presentation of the equations of covariant relativistic hydrodynamics suggests an interesting connection
of fluid dynamics and topological string theory. Recall that the $A$-model \cite{Witten:1991zz} describes a simplified version
of string theory based on a sigma model with symplectic target space ${\CalP}^{2m}$, whose path integral
localizes onto the finite-dimensional moduli space of pseudoholomorphic maps of the worldsheet
Riemann surface to the target space. The mathematical counterpart of this theory
is provided by the theory of Gromov-Witten classes. The open string version of this theory involves the 
maps of Riemann surfaces with boundaries, conditioned on the boundaries landing on some Lagrangian submanifolds
$L_1, \ldots, L_k \subset {\CalP}^{2m}$ -- the $D$-branes of topological string theory. The mathematical counterpart of this theory is the theory of Fukaya categories. 

It was discovered some time ago that for special ${\CalP}^{2m}$ the category of $D$-branes must be enhanced by the inclusion
of additional branes, associated to coisotropic submanifolds \cite{Kapustin:2001ij}. These additional branes
play an important role in the modern approaches to quantization, the geometric Langlands program, and the analytic
continuation of path integrals. Upon string duality, one such brane becomes the brane of opers, which plays an important role in Liouville theory, as 
reviewed in Section \ref{sec:BPSCFT}. 

The intersections of coisotropic and Lagrangian branes provide the low-energy
approximations to the open string ground states describing the morphisms in the Fukaya category. These intersections, in general, receive worldsheet instanton corrections.

In our story, the symplectic manifold ${\CalP}^{2m}$ in question is the infinite dimensional symplectic vector space ${\mathscr{P}}_{M^4}$
associated with the four-dimensional spacetime $M^4$. Unfortunately for the comparison to \cite{Kapustin:2001ij}
${\mathscr{P}}_{M^4}$ does not seem to have a natural complex structure. The Hodge star $\star$ squares to $-1$ for Lorentzian metric when acting on ${\Omega}^{0 \oplus 4}$, but then it squares to $+1$ on ${\Omega}^{1 \oplus 3}$. For a Riemannian metric on $M^4$ these signs are reversed. 

Fortunately, there is another topological string theory, the so-called $C$-model \cite{Baulieu:2001fi}, based on the AKSZ Poisson-sigma model \cite{Alexandrov:1995kv} (see also \cite{Schaller:1994es}), which was used by M.~Kontsevich in his deformation quantization program \cite{Kontsevich:1997vb, Cattaneo:1999fm, Cattaneo:2001ys}. 
The $C$-model is based on a sigma model on a real Poisson manifold. Remarkably, it has an open string version with
coisotropic submanifolds as $D$-branes \cite{Cattaneo:2003dp}. We are therefore in the right context. 

Taking this approach seriously, we conclude that relativistic four-dimensional hydrodynamics is described
by a six dimensional hybrid topological theory on $M^4 \times I \times {\BR}$ or $M^4 \times D^2$. 
The BV action of the Poisson sigma model on ${\mathscr{P}}_{M^4}$ reads as the six dimensional theory:
\beq
\int_{M^4 \times D^2} {\bf P} \wedge {\rm d} {\bf Q} + {\bf P} \wedge {\bf P} + {\rm gauge\ fixing}
\eeq
where ${\rm d}$ is the de Rham differential on $D^2$, the fields ${\bf P} = {\bf P}^{(0)}+{\bf P}^{(1)} + {\bf P}^{(2)}$ and ${\bf Q} = {\bf Q}^{(0)}+{\bf Q}^{(1)} + {\bf Q}^{(2)}$ are inhomogeneous differential
forms on $M^4 \times D^2$ of bi-degrees $(m,d)$ with $m=0,1,3,4$ and $d = 0,1,2$, and the bosonic/fermionic parity assigned as follows: the even $d$ components ${\bf Q}^{(d)}/{\bf P}^{(d)}$ of ${\bf Q}/{\bf P}$ are bosons/fermions, while
the odd $d$ components ${\bf Q}^{(d)}/{\bf P}^{(d)}$ of ${\bf Q}/{\bf P}$ are fermions/bosons. 

Now imagine the boundary ${\partial}D^2 = S^1$ of the disk partitioned into the intervals $I_{+} \cup I_{-}$, with the boundary condition on $I_{\pm}$ being
\beq
{\bf Q}^{(0)} \Biggr\vert_{I_{+}} \in {\CalC}   \, , \ {\bf Q}^{(0)}\Biggr\vert_{I_{-}} \in {\CalL}_{\ve, {\bf g}}
\eeq
The intersection points $I_{+}\cap I_{-}$ would have to map to the intersection ${\CalC}    \cap {\CalL}_{\ve, {\bf g}}$, i.e., on-shell hydrodynamical flows on $M^4$. 

One can study the correlation functions
\beq
\langle {\CalO}_{1}(t_{1}, +) \ldots {\CalO}_{k}(t_{k}, +)\, {\tilde\CalO}_{1}({\tilde t}_{1}, -) \ldots {\tilde\CalO}_{\tilde k}({\tilde t}_{\tilde k}, -) \rangle
\label{eq:corrf}
\eeq
with ${\CalO}_{i}, {\tilde\CalO}_{\tilde i}$ being some functionals on ${\CalC}   $ and ${\CalL}_{{\ve},{\bf g}}$, respectively. 
The diagram technique of \cite{Cattaneo:2003dp} for computing \eqref{eq:corrf} generalizes the deformation quantization formulas of \cite{Kontsevich:1997vb} and provides a possible tool for studying these amplitudes.  

There is another use of the Poisson-sigma model in connection with non-relativistic hydrodynamics. In Section \ref{sec:2dym}
we discussed the traditional spinning tops and mentioned a connection to the two-dimensional Yang-Mills theory. The latter can also be viewed as a Poisson sigma model with the target space $\mathfrak{g}^{*}$, our familiar Poisson example. The authors of \cite{Losev:2019xip} studied this Poisson sigma model not as a gauge theory, but as a full quantum field theory allowing gauge non-invariant observables, working in Lorenz gauge. This suggests that the equation of state represented by the Lagrangian submanifold ${\CalL}_{\bf\Omega}$ may be viewed as another gauge choice\footnote{What is called $B$ in the Ref. \cite{Losev:2019xip} we call $E$} in $EF_{A}$  theory. For the Euler top based on a finite-dimensional Lie algebra, this gives a conventional-looking two-dimensional theory. For three-dimensional hydrodynamics, it gives a five-dimensional theory. 

\subsection{$M$-theory of vortex spacetime surfaces}

Adding the surfaces ${\Sigma}_{1}, \ldots, {\Sigma}_{n}$ as dynamical degrees of freedom, as we did in the discussion of anomalous hydrodynamics,  
is analogous to adding the $M2/M5$-branes to eleven-dimensional supergravity in defining $M$-theory. The supergravity background sourced by the $M$-branes is generically singular, yet allowing it has the significance of adding fundamental degrees of freedom
to the effective field theory. 

A natural problem is to identify a place for the components 
${\tilde{\CalP}}_{M^4}[n]$ of the extended phase space
in the framework of the Poisson sigma model on ${\mathscr{P}}_{M^4}$. One possibility we can envision is to associate to the collection $\Sigma_1,\ldots,\Sigma_n$ a coisotropic submanifold of ${\mathscr P}_{M^4}$, and therefore yet another boundary condition in the topological sigma model.

\subsection{Topological field theories in five and six dimensions}

The $2$-form  \eqref{eq:bhydro} is the (pre)symplectic form
one finds in the canonical formulation of the five-dimensional Chern-Simons theory \cite{Fock:1991dt}
\beq
S_{CS5} = k \int_{N^5} {\bf p} \wedge {\bf dp} \wedge {\bf dp}
\label{eq:cs5}
\eeq
In such a theory, there are two types of order observables: Wilson loops and Chern-Simons bodies:
\beq
W_{q}({C}) = e^{\ii q \oint_{C} {\bf p}}\, , \ CS_{l} (B) = e^{\ii l \int_{B} {\bf p} \wedge {\bf dp}}
\eeq
When $N^5 = M^4 \times {\BR}$, the analogues of charges are the points and surfaces. The linking number of loops in three dimensions, which shows up as
a braiding phase in the expectation values of Wilson loops in three-dimensional Chern-Simons theory, is now replaced by the linking number of loops and bodies, and by the triple linking of triple bodies \cite{Fock:1991dt}. The interesting feature of the 
theory \eqref{eq:cs5} observed in \cite{Fock:1991dt} was the emergence of diffeomorphism symmetries from the gauge symmetry Gauss law. Thus, even though the theory \eqref{eq:cs5} is a gauge theory, the kernel of the restriction of the closed two-form \eqref{eq:bhydro} on the zero level of the Gauss law
consists of both the infinitesimal gauge transformations and diffeomorphisms. Unfortunately, the action of these groups is rarely free, so the classical phase space of \eqref{eq:cs5} is highly singular.

Of course, our theory has more fields, and we impose both the diffeomorphism and the gauge, i.e., ${\sf C}^{\infty} (M^4)$-symmetry constraints. We can nevertheless ask which Chern-Simons-like theory these constraints would correspond to. 

We can view the fields $({\bf S}, {\bf p}, {\bf n}, {\boldsymbol{\nu}})$ as the restriction of some differential forms defined on $N^5$ on $M^4$ viewed as
a Cauchy slice $t = const$ for some choice of the time parameter. In this approach the minimal set of fields on $N^5$ would be $({\bf S}, {\bf P}, {\bf C}, {\bf N})$ - the $0\oplus 1 \oplus 3 \oplus 4$ forms. In fact, imposing the
moment map for ${\CalG}^{(1)}_{M^4}$ requires, as Lagrange multipliers, the fields 
\beq
{\CalV}_{t} dt \in {\sf Vect}(M^4) \otimes dt \oplus p_{t} dt \in {\sf C}^{\infty}(M^4) \otimes dt
\eeq
Thus, taking \eqref{eq:symf} and \eqref{eq:momg1} and \eqref{eq:momdiff4} as input, we can write the following action in $4+1$ dimensions:
\beq
\mathbb{S} = \int_{M^4 \times {\BR}} \, dt \wedge \left( {\boldsymbol{\nu}} \left( {\partial}_{t} {\bf S} + L_{{\CalV}_{t}} {\bf S} \right) + {\bf n} \wedge \left( {\partial}_{t} {\bf p} - {\bf d} p_{t} + \iota_{{\CalV}_{t}} {\bf dp} \right) \right) 
\label{eq:topac}
\eeq
where we redefined $p_{t} \mapsto p_{t} + \iota_{{\CalV}_{t}} {\bf p}$ to reduce clutter. Now define
\beq
\begin{aligned}
 & {\tilde{\bf n}} = \iota_{{\CalV}_{t}} {\boldsymbol{\nu}}\, , \ {\bf N} = {\boldsymbol{\nu}} - dt \wedge {\tilde{\bf n}} \in {\Omega}^{4}(N^5) \, , \\
 & {\tilde b} = \iota_{{\CalV}_{t}} {\bf n} \, , \ 
 {\bf C} = {\bf n} - dt \wedge {\tilde b}  \in {\Omega}^{3}(N^5)
 \end{aligned}
 \label{eq:5dfields}
\eeq
and
\beq
{\bf P} = p_{t} dt + {\bf p} \in {\Omega}^{1}(N^5)
\eeq
and \eqref{eq:topac} assumes an almost respectable form:
\beq
{\mathbb{S}} = \int_{N^5} {\bf N} \wedge D{\bf S} + {\bf C} \wedge D{\bf P} + k {\bf P} \wedge D {\bf P} \wedge D{\bf P}
\eeq
with $D$ the de Rham differential on $N^5$, and we added the Chern-Simons term for generality. 

The only trouble with this formulation is that the five-dimensional fields
${\bf C}$ and ${\bf N}$ are not independent: they share the same kernel, proportional to the vector field
\beq
{\CalV} = {\partial}_{t} - {\CalV}_{t}
\eeq
The way to formulate this condition algebraically is to recall that the $4$-form ${\bf N}$ can be identified with the ${\Omega}^{5}(N^5)$-valued vector field ${\bf N}^{\vee}$. We demand that the contraction of this vector field with ${\bf C}$ vanish:
\beq
\iota_{{\bf N}^{\vee}} {\bf C} = 0 \in {\Omega}^{2}(N^5) \otimes {\Omega}^{5}(N^5)
\label{eq:constr}
\eeq
To impose \eqref{eq:constr} at the level of a five dimensional action, we can introduce a Lagrange multiplier $\pi \in {\Gamma}\left( {\Lambda}^{2}(TN^5) \right)$, a bi-vector field, and replace $\mathbb{S}$ by $\mathbb{S} + \int_{N^5} \iota_{\pi} \iota_{{\bf N}^{\vee}} {\bf C}$. 
This action has two secondary gauge symmetries, coming from the shifts ${\pi} \mapsto {\pi} + \iota_{{\bf N}^{\vee}} {\zeta}$, ${\bf C} \mapsto {\bf C} + \iota_{v} {\bf N}$, with the parameters
$v \in {\Gamma}\left( TN^5 \right) = {\sf Vect}(N^5)$
and ${\zeta} \in {\Gamma} \left( {\Lambda}^{4}(TN^5) \right)$. We are thus led to extending the space of fields by polyvector fields on $N^5$ in addition to differential forms on $N^5$. This suggests a possible AKSZ-type formulation \cite{Alexandrov:1995kv}, although we do not establish it here. 

Of course, the $0\oplus 1 \oplus 3 \oplus 4$ degree differential forms on $N^5$ look more natural as the decomposition of a pair consisting of a $1$-form and a $4$-form on a $6$-manifold $W^6 = N^5 \times {\BR}$
\beq
\int_{W^6} {\mathsf{P}} \mathsf{D} {\mathsf{N}}
\label{eq:bf6}
\eeq
with $\mathsf{P} = {\bf S} du + {\bf P} \in {\Omega}^{1}(W^6)$, $\mathsf{N} = {\bf N} + du \wedge {\bf C} \in {\Omega}^{5} (W^6)$, and $u$ is the coordinate along ${\BR}$ in the local decomposition
$W^6 = N^5 \times {\BR}$. The constraint \eqref{eq:constr} now becomes a quadratic condition
on $\mathsf{N}$, 
\beq
\mathsf{N}^{\vee} \wedge \mathsf{N}^{\vee} = 0 \in {\Omega}^{4}(W^6) \otimes \left( {\Omega}^{6}(W^6) \right)^{{\otimes} 2}\, ,
\label{eq:puren}
\eeq
somewhat similar to the pure spinor condition
in Berkovits's approach to the covariant formulation of superstring theory \cite{Berkovits:2000fe}.

The six dimensional theory only represents the ${\CalG}^{(1)}_{M^4}$ or ${\CalG}_{N^5}$-constraints. The equation of state and its geometric realization via ${\CalL}_{{\ve}, {\bf g}, {\bf a}}$ should probably appear as
a boundary condition. 

It may very well be that the six-dimensional theory 
\eqref{eq:bf6} with \eqref{eq:puren} enforced
is equivalent by a clever gauge choice to the six-dimensional theory modeled on a Poisson sigma model on ${\mathscr{P}}_{M^4}$.

\subsection{Wavefronts and Einstein equations}

{}So far we have discussed the relation of $4$- or $3+1$-dimensional hydrodynamics to a topological theory in five dimensions, or to a hybrid theory in six dimensions. 

Let us make some remarks generalizing our intersection theory approach. The problem of finding
the specific points of intersection of two varieties, e.g., ${\CalC}$ and ${\CalL}$ is often accompanied by a probabilistic version, where one or both varieties are replaced by a probability measure, or, a topological version, where one is interested in the weighted count of the points of intersection.  

Given the classical mechanical setting of our intersection problem, it is natural to ask about its quantum mechanical analogues. 

One idea comes from interpreting the Lagrangian submanifold as a WKB quasiclassical state ${\bf\Psi}$ in the formal Hilbert space
associated with quantizing ${\mathscr{P}}_{M^4}$ with its symplectic form \eqref{eq:symf} for simplicity (the case of \eqref{eq:symf} with the correction \eqref{eq:bhydro} can be treated analogously). In the $({\bf S}, {\bf n})$-polarization, the WKB quantum state associated with the fluid equation of state and spacetime metric $\bf g$ has the wavefunction, cf. \eqref{eq:genes}: 
\beq
{\bf\Psi}_{{\ve}, {\bf g}} ({\bf S}, {\bf n} ) \sim e^{\frac{\ii}{\hbar} {\CalA}_{e, {\bf g}}[ {\bf S}, {\bf n} ]}
\eeq
But in quantum mechanics, we are allowed to take linear superpositions of states. For example, we can study the wave packet, associated to some family of metrics $\bf g$. If the measure ${\mu}[ {\bf g}]$ on that family is itself WKB-like, 
\beq
{\mu} [{\bf g}] \sim e^{\frac{\ii}{\hbar} L[ {\bf g}]}
\eeq
with some local functional, e.g., the Einstein-Hilbert action, the $\hbar \to 0$ limit of the wavepacket will correspond to taking an extremum with respect to $\bf g$, e.g., solving the Einstein equations with the fluid stress-tensor as a source:
\beq
{\bf A}[ {\bf S}, {\bf n} ] = {\rm Crit}_{\bf g} \, \left(  L[ {\bf g}] + {\CalA}_{e, {\bf g}}[{\bf S}, {|bf n}] \right)
\eeq
Mathematically, this means passing from a Lagrangian submanifold ${\CalL}$ to a wavefront of a family. 

It remains to understand the relation of our formalism to the
membrane paradigm and its recent developments, e.g.,\cite{Bhattacharyya:2007vjd}, \cite{Bredberg:2011jq}.

\subsection{Classical Liouville and analytic Langlands as intersection problem} 
\label{sec:BPSCFT}

The classical Liouville equation
\beq
{\partial}_{z}{\bar\partial}_{\bar z} {\phi} + e^{2\phi} = 0
\eeq
defined on a two-dimensional Riemann surface $\Sigma$ with local holomorphic coordinate $z$, describes the constant (negative) curvature metric
\beq
e^{2\phi} dzd{\bar z}\, .
\eeq
Here, the unknown is the conformal factor $\phi$. In the local coordinate patch ${\phi}(z, {\bar z})$ is just a function, but the transition $z \mapsto {\tilde z}(z)$ from one patch to another transforms ${\phi}$ not as a function but as a more sophisticated gadget:
\beq
{\phi} \mapsto {\tilde\phi} = {\phi} - {\rm log} | {\tilde z}^{\prime}(z) |
\eeq
As observed by H.~Poincare, the \emph{classical stress-tensor}
\beq
T_{zz} = \left( {\partial}_{z}{\phi} \right)^{2} - {\partial}^{2}_{zz} {\phi}
\label{eq:lst}
\eeq
obeys
\beq
{\bar\partial}_{\bar z} T_{zz} = 0\, , 
\eeq
$T_{zz}dz^2$ is not a $2$-differential: under the map $z \mapsto {\tilde z}(z)$
it transforms inhomogeneously
\beq
T_{zz} \mapsto {\tilde T}_{{\tilde z}{\tilde z}}  =  T_{zz}/({\tilde z}^{\prime})^2 + \frac 12
\{ {\tilde z} ; z \}
\eeq
with the Schwarzian derivative $\{ {\tilde z} ; z \} = {\tilde z}'''/{\tilde z}' - 3/2 ({\tilde z}''/{\tilde z}')^2$. In other words, the global meaning of \eqref{eq:lst} is that of a holomorphic projective connection
\beq
{\CalD} = -{\partial}_{z}^{2} + T_{zz} \ , 
\label{eq:cd2}
\eeq
More precisely, one should view the second-order differential operator ${\CalD}$ as acting on $(-\frac 12)$-differentials mapping them to $\frac 32$-differentials. 
Locally, the scalar differential operator of second order can be viewed as the first-order differential operator acting on two-component vectors, so that the horizontal sections of the former can be mapped to the horizontal sections of the latter and vice versa: 
\beq
0 = {\CalD}{\psi}_{-\frac 12} \Leftrightarrow {\partial}_{z} \left( \begin{matrix} 
{\psi}_{\frac 12} \\
 {\tilde\psi}_{-\frac 12} \end{matrix} \right) = \left( \begin{matrix} 0 
& 1 \\
T_{zz} & 0 \end{matrix} \right) \cdot \left( \begin{matrix} 
{\psi}_{\frac 12} \\
{\tilde\psi}_{-\frac 12} \end{matrix} \right) 
\eeq
Globally, 
\beq
 \left( \begin{matrix} 
{\psi}_{\frac 12} \\
{\tilde\psi}_{-\frac 12} \end{matrix} \right) 
\eeq
is a section of a rank-two complex vector bundle ${\mathscr E}$ over $\Sigma$ with trivial determinant
${\rm det}({\mathscr E}) \approx {\CalO}$, by the usual Wronskian considerations. Thus \eqref{eq:cd2} defines a flat connection on ${\mathscr E}$. 
Given a complex structure $\tau$ on $\Sigma$, the space of all holomorphic projective connections, also known as $SL_{2}$-opers thanks to their role in the geometric Langlands program \cite{BD1, BD2}, is a $3g-3$-dimensional affine complex variety, naturally viewed as a Lagrangian submanifold ${\CalL}_{\tau}$ of the moduli space ${\CalM}_{SL_{2}({\BC})}[{\Sigma}^{\rm top}]$ of 
flat $SL_{2}({\BC})$-connections on $\Sigma$. The latter is independent of the complex structure of $\Sigma$ and can be defined purely in topological terms, hence the superscript ``{\rm top}'' in the notation. 

Now, not every $\CalD$ corresponds to a solution of the Liouville equation. In fact, given $\tau$, one expects a unique solution, in agreement with the uniformization program. What is so special
about the Poincare stress-tensor \eqref{eq:lst}? For one thing, the Liouville field must be recoverable from $T_{zz}$: 
\beq
{\CalD}  \left( e^{-\phi} \right) = 0
\eeq
and its complex conjugate
\beq
{\bar\CalD} \left( e^{-\phi} \right) = 0
\eeq
However, given the holomorphy of $T_{zz}$ we could look for local holomorphic solutions 
\beq
{\CalD} {\psi} = 0
\eeq
which form a two-dimensional vector space: 
\beq
{\psi}(z) = a_{1} {\psi}_{1} (z) + a_{2} {\psi}_{2}(z)
\eeq
with some constants $a_1, a_2$. 
The Liouville field, therefore, is expressed as a linear combination:
\beq
e^{-\phi}  = {\ii} \left( {\psi}_{1}(z) {\psi}_{2}^{*}({\bar z}) - {\psi}_{2}(z) {\psi}_{1}^{*}({\bar z}) \right)\, , 
\label{eq:phipsi}
\eeq
with this form fixed by the reality of $\phi$ and the impossibility of ${\CalD}$ having $SU(2)$ monodromy. 
In going from one patch to another, and, perhaps, returning to where we started from, the
original basis of solutions transforms by an $SL_{2}({\BC})$ transformation, the monodromy:
\beq
 \left( \begin{matrix} 
{\psi}_{1} \\
{\psi}_{2} \end{matrix} \right) \mapsto \left( \begin{matrix} a & b \\ c & d \end{matrix} \right)
 \left( \begin{matrix} 
{\psi}_{1} \\
{\psi}_{2} \end{matrix} \right)
\eeq
with complex $a,b,c,d$ obeying
\beq
a d- bc = 1
\eeq
However, the single-valuedness of \eqref{eq:phipsi} implies that the $SL_{2}({\BC})$-matrix
$\left( \begin{matrix} a & b \\ c & d \end{matrix} \right)$ belongs to $SL(2, {\BR})$:
\beq
a,b,c, d \in {\BR}
\eeq 
$$ (a{\psi}_{1} + b{\psi}_{2} ) ( c {\psi}^{*}_{1} + d {\psi}_{2}^{*}) - ( c {\psi}_{1} + d{\psi}_{2})( a{\psi}_{1}^{*} + b{\psi}_{2}^{*} ) = {\psi}_{1} {\psi}_{2}^{*} - {\psi}_{2} {\psi}_{1}^{*} $$
Thus, 
\beq
{\phi} \in {\CalL}_{\tau} \cap {\CalM}_{SL(2, {\BR})} \subset {\CalM}_{SL_{2}({\BC})}[{\Sigma}^{\rm top}]
\label{eq:intlm}
\eeq
Actually, the set of intersection points  of the variety ${\CalL}_{\tau}$ of opers
and the locus of $SL(2, {\BR})$-flat connections  is infinite. Only one of these points corresponds to the smooth hyperbolic metric. But all intersection points play a role in quantum Liouville theory \cite{ZZ}. Recently, the self-adjoint version of the $SL_2$ Gaudin system 
was studied \cite{Teschner:2017djr,Etingof:2019pni}. It is easy to recognize in the description of \cite{Etingof:2021eub} the same
intersection \eqref{eq:intlm}, or, equivalently \cite{Gaiotto:2024tpl}
\beq
{\phi} \in {\CalL}_{\tau} \cap {\CalL}_{\bar\tau}\subset {\CalM}_{SL_{2}({\BC})}[{\Sigma}^{\rm top}]
\label{eq:intll}
\eeq

\subsubsection*{Quantization parameters and BPS/CFT correspondence}

Classical Liouville theory is but an approximation to quantum Liouville theory. The  correlation
functions admit a continuous conformal block decomposition \cite{BPZ,FZ,ZZ}. For example, for the $n$-point function on a
sphere:
\begin{multline}
\langle V_{\Delta_{1}}(z_{1}, {\bar z}_{1}) \ldots V_{{\Delta}_{n}}(z_{n}, {\bar z}_{n}) \rangle = \\
\int
d{\alpha}_{1} \ldots d{\alpha}_{n-3} \Biggl\vert {\bf\Psi}_{b} \left( z_{1}, \ldots , z_{n} ; {\Delta}_{1}, \ldots , {\Delta}_{n} ; {\alpha}_{1}, \ldots, {\alpha}_{n-3} \right) \Biggr\vert^2
\label{eq:npoint}
\end{multline}
where we have included the DOZZ three-point functions in the definition of conformal blocks, and left unspecified the discrete data: the fusion channels, the contour of integration over the momenta $\vec\alpha$ defining the dimensions
of intermediate channels, etc. The parameter $b$ defines the central charge and enters the dimensions of the primary fields:
\beq
{\Delta}_{\alpha} = {\alpha} (Q - {\alpha})\, , \ Q = b + b^{-1}\, , \ c = 1+ 6Q^2
\eeq
The BPS/CFT correspondence \cite{NN2004,NekBPSCFT} relates these correlation functions, specifically, the conformal blocks, 
to partition functions of $\Omega$-deformed \cite{Nekrasov:2002qd} supersymmetric gauge theories in four dimensions. Specifically, the $n$-point functions \eqref{eq:npoint} in Liouville theory, in a specific channel , are found \cite{Alday:2009aq, Gaiotto:2009we,Drukker:2009id} to be associated with a linear quiver gauge theory, with the gauge group $SU(2)^{n-3}$, with 
gauge couplings related to the cross-ratios $(z_{1}:z_{i+1}:z_{i+2}:z_{n})$, the momenta $\alpha_i$ to the vevs
of the vector multiplet scalars. Most interestingly, the ratio $b^2 = {\ve}_{1}/{\ve}_{2}$ of the $\Omega$-deformation
parameters controls the quantum parameter of Liouville theory. The conformal blocks ${\bf\Psi}_{b}$
are the quantum analogues of the variety ${\CalL}_{\bf z}$ of opers, its generating function ${\CalA}_{\bf z}({\alpha})$ in the NRS Darboux coordinates \cite{NRS} being given by the ${\ve}_{2}\to 0$ asymptotics 
\beq
{\bf\Psi}_{b} ({\bf z}; {\bf\Delta}; {\boldsymbol{\alpha}}) \sim e^{b^2 {\CalA}_{\bf z}({\alpha})}
\eeq
also known as Zamolodchikov's classical conformal block, studied e.g.,  in \cite{Litvinov:2013sxa}. 
It would be interesting to identify an analogue of the $({\ve}_{1}, {\ve}_{2})$-deformation for ${\CalC}$
and to clarify its possible relation to the two viscosities. 

An important tool in Liouville theory is its connection to CFT based on an $SU(2)$ current algebra \cite{FZ}. It has its gauge theory analogue \cite{NN2004,Frenkel:2015rda}, where the surface defects in gauge theory obey Knizhnik-Zamolodchikov \cite{KZ} equations
as non-perturbative Dyson-Schwinger equations \cite{NekBPSCFT}. It would be interesting to find an analogue of this correspondence in the geometry of relativistic hydrodynamics, possibly through the dynamics of vortex spacetime surfaces.

\subsection{Viscosity}
\label{sec:visc}

The inclusion of viscosity and dissipation is physically important. The stress-tensor approach of \cite{LL}
adds the terms
\beq
{\Theta}_{\mu\nu}
=
\begin{aligned}
&\eta\left(
D_{\mu}u_{\nu}+D_{\nu}u_{\mu}
+u_{\mu}u^{\lambda}D_{\lambda}u_{\nu}
+u_{\nu}u^{\lambda}D_{\lambda}u_{\mu}
\right)\\
&{}+
\left(\zeta-\frac{2}{3}\eta\right)
\left(g_{\mu\nu}+u_{\mu}u_{\nu}\right)D_{\lambda}u^{\lambda}\, ,
\end{aligned}
\label{eq:visc}
\eeq
responsible for the shear and bulk viscosity, respectively. 
Equivalently, if
$$
{\Delta}_{\mu\nu}=g_{\mu\nu}+u_{\mu}u_{\nu},\qquad
\vartheta=D_{\lambda}u^{\lambda},
$$
then
$$
{\Theta}_{\mu\nu}
=
\eta \left(L_{\bf u}{\Delta}\right)_{\mu\nu}
+
\left(\zeta-\frac{2}{3}\eta\right)
{\Delta}_{\mu\nu}\vartheta .
$$
A separated form of \eqref{eq:visc}, adapted to the ${\CalL} \cap {\CalC}$ picture developed in the main body of the paper, remains to be found. This raises the possibility of relating the two-parametric deformation by $({\eta}, {\zeta})$ to the
two-parametric deformation by $({\ve}_{1}, {\ve}_{2})$ of the BPS/CFT correspondence reviewed in the Appendix \ref{sec:BPSCFT}. 

\subsection{Spinning tops and two-dimensional Yang-Mills}
\label{sec:2dym}

If ${\CalX} = \mathfrak{g}^{*}$, the example discussed in \eqref{eq:kkf}, \eqref{eq:kirillov-kostant}, 
the ambient symplectic manifold ${\mathscr{M}}_{\CalX}$, the subvariety ${\CalC} $ and the space
of leaves ${\CalM}_{{\CalX}, {\pi}}$ can be given a gauge-theoretic interpretation in two-dimensional
Yang-Mills theory:
\beq
{\mathscr{M}}_{\mathfrak{g}^{*}} = T^{*} \Biggl\{ {\rm space}\, {\rm of}\, G-{\rm connections}\, {\rm on} \, {\BR} \, \Biggr\}
\eeq
where we view ${\boldsymbol{\xi}}$, a $1$-form on $\BR$ valued in $\mathfrak{g}$ as the connection 
form ${\bf A} = (A_{s}(s)ds)$, with $A_{s}(s) \in \mathfrak{g}$, while $\bf x$, a path in $\mathfrak{g}^{*}$, is the electric field ${\bf E} = (E(s))$. To avoid confusion, we changed the notation for the parameter along the path from $t$ to $s$. In this presentation, the parameter $s$ along the path is the \emph{spatial} coordinate in the gauge-theory
viewpoint. 

The subvariety ${\CalC}  \subset {\mathscr{M}}_{\mathfrak{g}^{*}}$, associated with the Lie-Poisson structure \eqref{eq:kkf} is nothing but the locus of $({\bf E},{\bf A})$ obeying the usual Gauss law:
\beq
{\partial}_{s} E(s) - {\rm ad}^{*}_{A_{s}(s)} (E(s)) = 0
\label{eq:gauss}
\eeq
The leaves of the foliation ${\CalF}_{\pi}$ are simply the orbits of the gauge group action:
\beq
g(s) : \left( E(s), A(s) \right) \mapsto \left( Ad^{*}_{g(s)}E(s) \,  , \, {\partial}_{s} g g^{-1} + Ad_{g(s)} A(s) \right)
\eeq
To map this presentation to the notations of Section \ref{sec:epa}, use ${\bf A} = {\bf Q} ds$, ${\bf E} = {\bf P}$. 

At this point, gauge theory and spinning tops
diverge: Yang-Mills theory has its own proper time $t$,  different from the spatial direction $s$, it has a Hamiltonian, given by a quadratic Casimir $\int_{\BR}\, ds\, c_{2}\left( E(s) \right)$, integrated over the space $\BR$ against a measure $ds$. 
The physical phase space of the two-dimensional Yang-Mills theory is the quotient ${\CalM}_{\mathfrak{g}^{*}} = {\CalC} /{\CalF}_{\pi}$. Specifically, how big or how small ${\CalM}_{\mathfrak{g}^{*}}$ is depends on the details of the boundary conditions. The simplest setting is that of periodic boundary conditions, i.e., where the domain is the circle $S^1$ as opposed to the real line $\BR$. In that case the conjugacy class of the monodromy $P{\rm exp}\oint_{S^1} {\bf A}$ around the circle is gauge-invariant, making ${\CalM}_{\mathfrak{g}^{*}}^{S^1} = (T^{*}G)/G$, with $G$ acting on $G$ via adjoint action. 
On the real line $\BR$ one can formally gauge $\bf A$ away, and solve for $\bf E$ in terms of $E(0) \in {\mathfrak{g}}^{*}$, so that ${\CalC} $ is described
as the quotient $\left( {\sf Maps} ({\BR}, G) \times {\mathfrak{g}}^{*} \right)/G$, with $G$ acting on the first factor by right multiplication (a symmetry of $({\partial}_{s} g(s) )g(s)^{-1}$) and on the second factor by the coadjoint action. Alternatively, we can describe ${\CalC} $
as a fibration over the stack $\mathfrak{g}^{*}/G$ of the space of conjugacy classes (space of coadjoint orbits), the fiber over $[{\CalO}_{\xi}] \in \mathfrak{g}^{*}/G$  being the space
of all paths ${\sf Maps} ({\BR}, {\CalO}_{\xi})$ in the corresponding orbit ${\CalO}_{\xi}$. 

The corresponding generating function, as a function of the pair 
$$ ({\xi}, {\sf path}\,  {\gamma}: I \to {\CalO}_{\xi}) $$ is given by
\eqref{eq:geomac} for the Kirillov-Kostant form on the coadjoint orbit ${\CalO}_{\xi}$.

Yang-Mills theory also has interesting non-local observables, given by the Wilson loops and lines. Especially interesting are the time-like Wilson lines. Their insertion changes the phase space. One needs to specify a collection ${\CalO}_{1}, \ldots {\CalO}_{n} \subset \mathfrak{g}^{*}$ of coadjoint orbits of $G$, and a collection $s_1 , \ldots , s_n$ of points on $\BR$.  The space $\mathscr{M}_{\mathfrak{g}^{*}}$ is generalized to
\beq
\mathscr{M}_{\mathfrak{g}^{*}; {\CalO}_{1}, s_{1}; \ldots ; {\CalO}_{n}, s_{n}} = 
\mathscr{M}_{\mathfrak{g}^{*}} \times {\CalO}_{1} \times \ldots \times {\CalO}_{n}
\eeq
while the 
Gauss law constraint \eqref{eq:gauss} is modified to
\beq
{\partial}_{s} E(s) - {\rm ad}^{*}_{A_s(s)} (E(s)) = \sum_{i=1}^{n} J_{i} {\delta} (s - s_{i})
\label{eq:gauss2}
\eeq
where $J_i$ is the moment map/embedding ${\CalO}_{i} \to {\mathfrak{g}^{*}}$ (more pedantically, it is the composition of the projection map $p_i : \mathscr{M}_{\mathfrak{g}^{*}; {\CalO}_{1}, s_{1}; \ldots ; {\CalO}_{n}, s_{n}} \longrightarrow {\CalO}_{i}$ on the $i$th factor and the embedding $\iota_{i} : {\CalO}_{i} \longrightarrow \mathfrak{g}^{*}$). 

The Yang-Mills dynamics becomes quite non-trivial in the presence of the sources as in \eqref{eq:gauss2}, containing many interesting (integrable) systems, cf. \cite{Gorsky:1993pe}.

On the spinning top side, the dynamics is unraveling in the $s$-direction, by imposing the equation of state, e.g., the ${\bf\Omega}$-map: ${\bf A} = {\bf\Omega}({\bf E})$, thus defining a Lagrangian submanifold ${\CalL}_{\bf\Omega} \subset {\mathscr{M}}_{\mathfrak{g}^{*}}$. Since $\bf A$ is a one-form while $\bf E$ is a scalar, the map involves a choice of the metric $ds^2$ on the domain of the paths, and some data on $\mathfrak{g}, \mathfrak{g}^{*}$, such as a non-degenerate (but not necessarily $G$-invariant!) quadratic form on $\mathfrak{g}^{*}$.

The generalization \eqref{eq:gauss2} describes the spinning top with occasional quenches, at the moments of time $s_1, s_2, \ldots, s_n$. It would be interesting to relate this to the intersections of vortex surfaces discussed in Section 
\ref{sec:vs}. 

\subsection{Helicoidal Lagrangians for nontrivial cohomology}
What can we do if $\CalB$ discussed in Section \ref{sec:anomf} is closed but not exact?
In this case, the $\CalA$ such that ${\CalB} = {\delta}{\CalA}$ is defined only locally. Given an open cover
${\mathscr{X}} = \cup_{\alpha} {\mathscr{X}}_{\alpha}$
with contractible ${\mathscr{X}}_{\alpha}$, define ${\CalA}_{\alpha} \in {\Omega}^{1}({\mathscr{X}}_{\alpha})$, s.t. ${\CalB} = d{\CalA}_{\alpha}$. 
On intersections of the open sets 
\beq
{\CalA}_{\alpha} - {\CalA}_{\beta} = {\delta}{\phi}_{\alpha\beta}\,, \ {\phi}_{\alpha\beta} \in {\sf C}^{\infty} \left( {\mathscr{X}}_{\alpha} \cap {\mathscr{X}}_{\beta} \right)
\label{eq:overlap}
\eeq
We could define the local patches ${\CalL}_{S, \alpha}$ of ${\CalL}_{S}$ by 
\beq
{\CalL}_{S, \alpha} = \{ \, (x,p) \, | \, x \in {\mathscr{X}}_{\alpha}\, , \, p = - A_{\alpha} + {\delta}S \, \}
\eeq
but these patches miss each other over the intersections ${\mathscr{X}}_{\alpha} \cap {\mathscr{X}}_{\beta}$, cf. 
\eqref{eq:overlap}. 

{}
To proceed further, consider a simple example. Let $\mathscr{P} =  {\BT}^4$ be the four-torus with the symplectic form ${\omega}_{0} = dp_1 \wedge dx^1 + dp_2 \wedge dx^2$, with 
$(x^1, x^2, p_1, p_2)$ the periodic coordinates 
\beq
\begin{aligned}
& x^i \sim x^i + 2\pi\, , \\ 
& p_i \sim p_i + 2{\pi}{\ell}_{i}\, , \ i = 1, 2
\\
\end{aligned}
\label{eq:xpper}
\eeq
with some periods ${\ell}_{1}, {\ell}_{2} \in {\BR}_{+}$. In the limit ${\ell}_{1}, {\ell}_{2} \to \infty$ the symplectic manifold $\mathscr{P}$
approaches the cotangent bundle $T^{*}{\BT}^{2}$. 

\begin{figure}
    \centering
    \includegraphics[width=4cm]{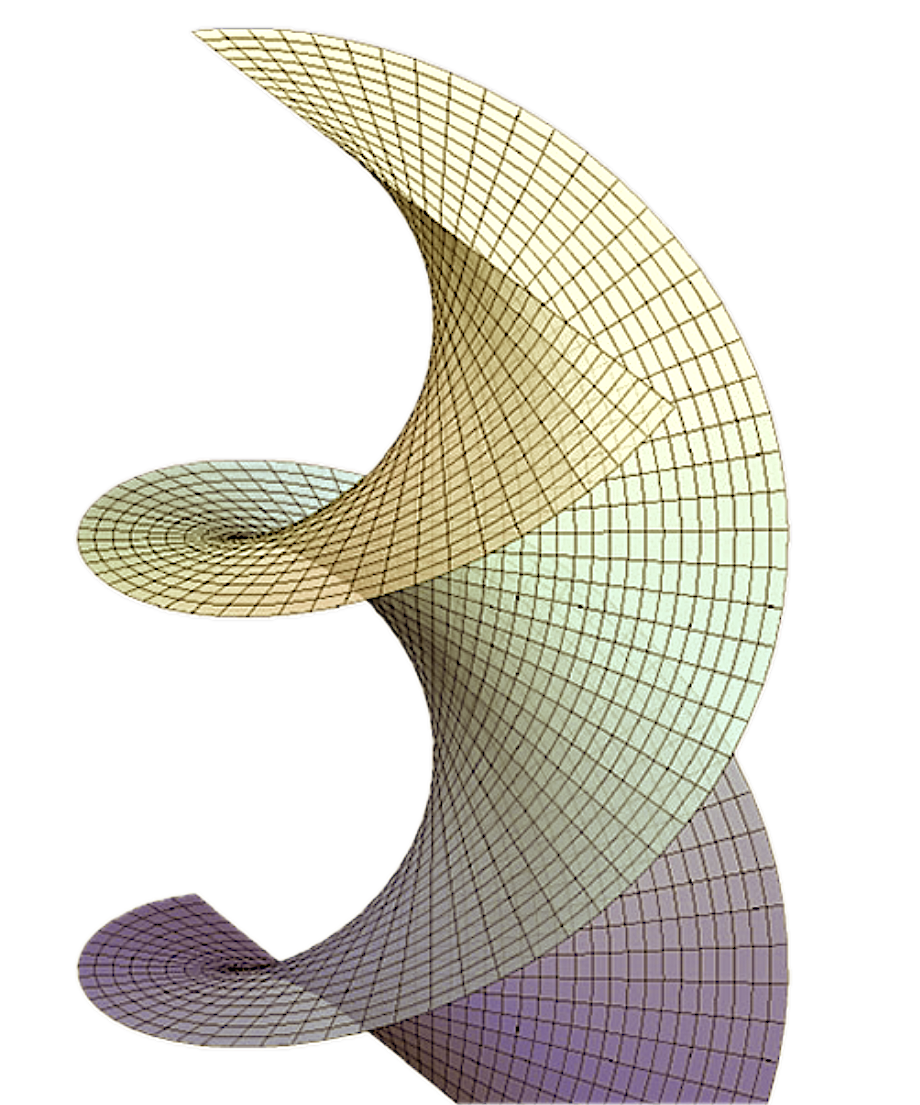}
    \caption{${\widetilde{\CalL}}_{S}$}
    \label{fig:helicolagrangian}
\end{figure}

{}A quasi-periodic function $S(x^1, x^2)$ obeying
\begin{multline}
S(x^1, x^2) = \\
S (x^1 + 2\pi, x^2) - 2\pi a {\ell}_1 x^1 - 2\pi b {\ell}_{2} x^2 = \\
S(x^1, x^2 + 2\pi)- 2\pi c {\ell}_1 x^1 - 2\pi d {\ell}_{2} x^2
\end{multline}
with some $a,b,c,d \in {\BZ}$ defines a Lagrangian submanifold ${\CalL}_{S}$ in the familiar fashion:
\beq
p_i = {\partial}_i S \, , \ i = 1,2
\label{eq:pdx}
\eeq
Now let us deform ${\omega}_{0} \to {\omega}_{k} =  {\omega}_{0} + k dx^1 \wedge dx^2$ with constant $k \in \BR$. The submanifold \eqref{eq:pdx} is no longer Lagrangian, ${\omega}_{k} \vert_{{\CalL}_{S}} = k dx^1 \wedge dx^2 \neq 0$. Deforming \eqref{eq:pdx} to
\beq
p_1 = {\xi} x^2 + {\partial}_{1} S \, , \ p_{2} = ({\xi}- k) x^1 + {\partial}_{2} S
\label{eq:pdxl}
\eeq
is not compatible with \eqref{eq:xpper} unless $\xi = p {\ell}_{1}$, 
${\xi} - k = q {\ell}_{2}$, for some $p,q \in {\BZ}$.
The resolution is that ${\CalL}_{S}$ becomes a non-compact Lagrangian submanifold ${\widetilde{\CalL}}_{S}$, which projects to ${\CalL}_{S}$ with an infinite number of branches as in Fig.~\ref{fig:helicolagrangian}.

The same helicoidal nature of typically noncompact Lagrangians is expected
in the more general situation:
\beq
{\widetilde{\CalL}}_{S}^{({\CalB})} \subset T^{*}{\mathscr{X}}
\label{eq:helicolag}
\eeq

\subsection{Theory of self-dual fields}

This section stands somewhat apart from the rest of the paper, but is included to illustrate the applicability of the intersection-theoretic view of dynamics to a wider class of problems.

Let $M$ be a spacetime of dimension $4l+2$. Define ${\mathscr P}_{l} = {\Omega}^{2l+1}(M)$ to be the space of middle-dimensional forms. This is a well-known symplectic vector space, with symplectic form
\beq
{\Omega}_{l} = \int_{M} {\delta}{\bf A} \wedge {\delta} {\bf A}
\eeq
for ${\bf A} \in {\Omega}^{2l+1}(M)$. Define ${\CalG}_{l} = {\Omega}^{2l}(M)/Z^{2l}(M)$ to be the space
of $2l$-forms modulo closed $2l$-forms, viewed as an abelian additive group. The group ${\CalG}_{l}$ acts on ${\mathscr P}_{l}$
in the obvious way:
\beq
{\bf A} \mapsto {\bf A} + {\bf dB}\, , \ {\bf B} \in {\CalG}_{l}
\eeq
The corresponding coisotropic submanifold ${\CalC}_{l}$ is the zero locus of the moment map
\beq
{\boldsymbol{\mu}} = {\bf dA}\,  
\eeq
namely, ${\CalC}_{l}={\boldsymbol{\mu}}^{-1}(0)$ is the space $Z^{2l+1}(M)$ of closed $2l+1$-forms. The quotient
\beq
{\CalC}_{l}/{\CalG}_{l} = H^{2l+1}(M, {\BR})
\eeq
is finite-dimensional.

Now choose a conformal/split structure on $M$, and define $\CalL$ to be the space of self-dual forms,
\beq
{\CalL} = \{ \, {\bf A}\, | \, {\bf A} = \star {\bf A} \, \}
\eeq
The intersection ${\CalC}_{l} \cap {\CalL} \subset {\mathscr P}_{l}$ is the space of self-dual closed forms. Writing
\beq
{\bf A} = {\bf A}_{0} + {\bf dB}\, , 
\eeq
where ${\bf A}_{0}$ is the harmonic representative of the corresponding class in $H^{2l+1}(M, {\BR})$, with ${\bf d \star A}_{0} = 0\, , \, {\bf dA}_{0} = 0$,
we see that $\bf B$ obeys the wave equation together with the chirality constraint.

The cases $l=0,1,2$ are particularly important in string theory: $l=0$ corresponds to chiral bosons in $1+1$ dimensions, namely on the string worldsheet; $l=1$ corresponds to the tensor field of the M5-brane theory in six dimensions; and $l=2$ corresponds to the $4$-form of IIB supergravity.

\section*{Statements and declarations}

\paragraph{Funding}

The research of NN was supported by NSF PHY Award 2310279 and by the Simons Collaboration ``Probabilistic Paths to QFT''. The work of PW was supported by the NSF under Grant NSF DMR-1949963.

\paragraph{Competing interests}

The authors declare that they have no relevant financial or non-financial interests to disclose.

\paragraph{Data availability}

No datasets were generated or analyzed during the current study.

\paragraph{Ethics approval}

Not applicable.

\paragraph{Consent to participate}

Not applicable.

\paragraph{Consent for publication}

Not applicable.

\paragraph{Code availability}

Not applicable.

\paragraph{Authors' contributions}

Both authors contributed to the conception and development of the work, the mathematical analysis, and the writing and revision of the manuscript. Both authors read and approved the final manuscript.

\end{document}